\documentclass[preprint, 12pt, number]{elsarticle}

\usepackage{changepage} 
\usepackage[mathscr]{euscript}
\usepackage{amssymb}
\usepackage{amsmath}
\usepackage{stackengine}
\usepackage{hyperref}
\usepackage{esint}
\usepackage{graphicx}
\usepackage{multirow}
\usepackage{multirow}
\usepackage{latexsym}
\usepackage{subcaption}
\usepackage{caption}
\usepackage{lipsum} 
\stackMath
\setcitestyle{square}
\usepackage{accents}
\usepackage{xcolor}
\usepackage{epstopdf}
\usepackage{bm}
\usepackage{booktabs}
\usepackage{float}
\usepackage[a4paper, margin=1in]{geometry} 

\begin{document}

\begin{frontmatter}

\title{Two-dimensional framework for modeling load transfer in nanocomposites with thin, stiff, and prestressed reinforcements}

\author[umn]{Rohit Satish Patil\corref{cor1}}
\ead{patil168@umn.edu}
\author[dhu]{Zhilin Han}
\author[umn]{Sofia G. Mogilevskaya}
\cortext[cor1]{Corresponding author.}

\address[umn]{Department of Civil, Environmental, and Geo-Engineering, University of Minnesota, 500 Pillsbury Drive S.E, 55414, Minneapolis, MN, USA}
\address[dhu]{College of Physics, Donghua University, Shanghai, 201620, China}

\begin{abstract}

This paper presents a novel framework for simulating load-transfer mechanisms in nanocomposites reinforced by thin, stiff, and prestressed reinforcements, such as, e.g., nanosheets or nanoplatelets. The reinforcements are modeled as open material surfaces treated as membranes endowed with their own elastic properties and residual surface tension; the matrix is assumed to be isotropic elastic. In the two-dimensional plane-strain setting considered here, each material surface is represented by a sufficiently smooth curve of otherwise arbitrary shape. The theoretical part of the framework is based on the Gurtin--Murdoch theory in which the surfaces are represented by traction jumps across them. The elastic fields in the matrix are exactly expressed by the integral representations written in terms of the unknown traction jumps that are functions of the second-order derivatives of the displacements. The computational part is based on the isogeometric boundary element method (IGABEM) that utilizes non-uniform rational B-splines (NURBS). The unknown traction jumps are found from the numerical solutions of the coupled system of singular boundary integral equations, after which the elastic fields in the matrix are evaluated. The procedure of extracting the values of the stress intensity factors (SIFs) at the tips of the surfaces is described. The proposed algorithm is validated against available benchmark solutions for multiple surfaces located along either straight segments or circular arcs. Selected parametric studies are performed to examine the influence of surface curvatures, relative spacing, orientation, and interactions. An example involving a graphene nanoplatelet-reinforced epoxy is presented.

    \begin{keyword}
    Nanocomposite; Gurtin--Murdoch material surfaces theory; Isogeometric boundary element method; Load transfer; Stress intensity factors
    \end{keyword}

\end{abstract}

\end{frontmatter}


\section{Introduction}\label{Section:1}
Two-dimensional thin, stiff, and prestressed reinforcements such as nanoplatelets or nanosheets are widely used to enhance the mechanical, thermal, and electrical properties of polymer-matrix composites \cite{Liu2019, Ramanathan2008}. Made of, e.g., graphene, boron nitride, metal oxides, such reinforcements are characterized by a high surface-to-volume ratio, high in-plane stiffness and strength, out-of-plane flexibility, and possess many other excellent properties, making nanosheet-reinforced composites attractive for structural, electronic, and energy applications \cite{Qadir2021, Akinwande2017, Tan2017}. In general, reinforcements can be randomly distributed and oriented, curved, wrinkled, or stacked \cite{Androulidakis2017, Potts2011}. Residual stresses can also develop in the manufacturing process, \cite{ChanHwang2023, Vassaux2022, Anagnostopoulos2015}, affecting local load-transfer mechanisms and stress fields. Therefore, those features should be taken into account in the realistic modeling of the local and overall mechanical behavior of the nanosheet-reinforced composites. 

Driven by advances in manufacturing and emerging applications, several studies have investigated the mechanical response of such composites. They can be organized into three main groups. The first group focuses on the characterization of the constituent properties through experiments and atomistic or coarse-grained simulations, providing information on reinforcement stiffness, strength, and the effects of structural and chemical modification \cite{Singh2024, Meng2017, Cao2014}. The second group addresses the measurement and prediction of the overall properties of composites using mechanical tests and simplified effective medium theories \cite{Akter2024, Rouway2021, Feng2018}. The last group investigates reinforcements–matrix interactions, interfacial and interlayer load transfer through experimental, analytical, and computational approaches. 

Some papers in the latter group describe multiscale formulations that connect molecular-scale descriptions with continuum mechanics models. In such formulations, reinforcement is represented by a molecular-structural network, while the polymer matrix is modeled as a continuum; the two are coupled through interaction springs, effective interface elements, or a continuum interphase \cite{Spanos2016, Giannopoulos2014}. Other studies adopt sequential multiscale approaches, in which molecular simulations provide the effective response of a reinforcement together with its very surrounding polymer (interphase), for subsequent continuum finite-element or micromechanical homogenization \cite{Caliskan2023, Barakat2023, AlMahmud2021, Hadden2015, Montazeri2011}. 

It has also been suggested, see, e.g. \cite{Papageorgiou2017}, that for multi-layered/stacked or oxidized (oxide-) reinforcements, load transfer mechanisms in graphene-reinforced polymer composites can be modeled using continuum mechanics approaches only.  For example,  in \cite{Young2018, Papageorgiou2020}, shear-lag theory was used to study stress transfer from the matrix to a reinforcement through shear stress at the interface. In \cite{Elmasry2021, Kilikeviius2021}, the three-dimensional elastic model of representative volume elements is used in which reinforcements are modeled as finite-thickness ellipsoidal inclusions perfectly attached to the matrix. The problem was solved using the Finite Element Method.

In the above studies, various simplifying assumptions were made about reinforcement shapes, spatial arrangements, and interactions. In addition, prestress or residual stresses have not been included in the models and their influence on equilibrium and subsequent mechanical response was not addressed. Moreover, explicitly resolving the thickness of ultra-thin reinforcements can be computationally demanding.

To avoid explicitly resolving the finite thickness, while keeping mechanical properties accounted for and including the effects of residual stresses, it was proposed in \cite{Mogilevskaya2021_AMR, Mogilevskaya2021_CST} to use the continuum material surface theories of Gurtin--Murdoch, \cite{Gurtin1978, Gurtin1975} and Steigmann--Ogden, \cite{Steigmann1999, Steigmann1997}. In the theories, the surface (e.g., interphase or boundary of the bulk material) is treated as a zero-thickness membrane or shell attached to the bulk without slipping. It is characterized by its own mechanical properties and surface tension. The equilibrium conditions for boundary value problems of those theories require the continuity of displacements and jumps in tractions across a surface. These jumps are expressed in terms of displacements via higher-order differential operators. Thus, the approximations for the related fields have to satisfy the elevated smoothness requirements. 

Later, see \cite{Mogilevskaya2021_AMR, Han2018, Zemlyanova2018_IJSS}, it was shown that, in the absence of surface tension, the jump conditions in the Gurtin--Murdoch and Steigmann--Ogden theories represent special cases of conditions that appear across thin and stiff interphase layers and can be obtained using asymptotic analysis, \cite{Benveniste2001}. The elastic problems of those types were rigorously studied, and comprehensive reviews of relevant theoretical, computational, and experimental results can be found, e.g., in \cite{Bessoud2011, Vilchevskaya1992} and the references therein. 

The Gurtin--Murdoch and Steigmann--Ogden theories were first used to model interfaces or coatings in fiber- or particle-reinforced composites, see \cite{Firooz2021, Zemlyanova2018_JAM, Kushch2011, He2006, Lim2006} and the references therein, i.e., the surfaces were located along closed curves representing the boundaries of fibers or particles. In some publications, see \cite{Parvanova2024}, the theories were used in static or elastodynamic problems that included semi-infinite solids. In most of the studies, the surfaces had canonical shapes, e.g., circular, spherical, or infinite straight lines. For arbitrarily shaped closed surfaces, the boundary element method (BEM) was used \cite{Dong2012, Dong2011}, while the boundary perturbation method was used in problems involving nanosized films with rough coatings or surface asperities, see \cite{Kostyrko2022, Kostyrko2021} and the references therein. 

Early applications of both theories to reinforcements located along curves of finite lengths considered isolated reinforcement located along a straight segment \cite{Zemlyanova2025, Zemlyanova2023_PhysicaD, Mogilevskaya2021_CST} or a circular arc \cite{Han2024_IJSS, Han2024_IJES}. In those studies, the unknown scalar components of the surface stress tensor were approximated by a series of Chebyshev polynomials or trigonometric functions multiplied by square root weight functions in order to capture the near-tip asymptotic behavior. These approximations also provided the required smoothness conditions for the unknowns and automatically satisfied tip conditions. The studies were also extended to include multiple interacting reinforcements located along straight segments or along the arcs of the same circle \cite{Han2024_JAM, Patil2024}. 

However, the series expansion-based global approximations, below referred to as (GAM), cannot be used in problems involving reinforcements of arbitrary shapes. Thus, in \cite{Patil2026}, we proposed using the Isogeometric boundary element method (IGABEM) \cite{Simpson2013, Simpson2012}, where the Non-Uniform Rational B-Splines functions (NURBS), see \cite{Farin1999, PieglLesandTiller1997}, were used to represent curved geometries and approximate unknown field variables. This method combines the advantages of Isogeometric Analysis (accurate geometry representation and higher-order inter-elements continuity), \cite{Cottrell2009, Hughes2005}, with those of BEM (exact integral representations for the elastic fields, boundary-only discretizations), see \cite{CrouchMogilevskaya2024}. 

The IGABEM-based algorithm described in \cite{Patil2026} was created only for problems that involved a single reinforcement of an arbitrary but sufficiently smooth shape. However, it could not be used to model multiple reinforcements and those located along curves of mixed curvatures (curves whose curvatures change sign along their paths at the inflection points, where they change from locally convex to locally concave, or vise versa). In addition, the procedure for extracting stress intensity factors (SIFs) at the tips of a surface has not been developed. Thus, here we generalize the algorithm proposed in \cite{Patil2026} to develop a framework capable of handling multiple interacting reinforcements of arbitrary shapes with arbitrary varying curvatures to evaluate local fields everywhere in the material system, including SIFs. 

The structure of the paper is as follows. The formulation of the problem is given in Section~\ref{Section:2}, while the governing equations of the Gurtin–Murdoch model in a plane strain setting are reviewed in Section~\ref{sec:GM_plane_strain}. The integral representations of the elastic fields are presented in Section~\ref{sec:integral_representations}, with the details provided in \ref{app:singular_formulation}. Section~\ref{Section:5} describes the main steps of the IGABEM technique. The procedure for evaluating stress intensity factors at the tips of reinforcements is described in Section~\ref{Section:SIF}. In Section~\ref{Sub-Section:6.1}, the proposed framework is validated against benchmark solutions for multiple straight and circular-arc reinforcements. Section~\ref{Section:7} presents several numerical examples designed to study the influence of selected governing parameters, as well as the example illustrating the capability of the approach to model the graphene nanoplatelet-reinforced epoxy composite. The concluding remarks are presented in Section~\ref{Section:8}.

\section{Problem formulation}\label{Section:2}

\begin{figure}[!htbp]
    \centering
    \includegraphics[width=0.75\linewidth]  {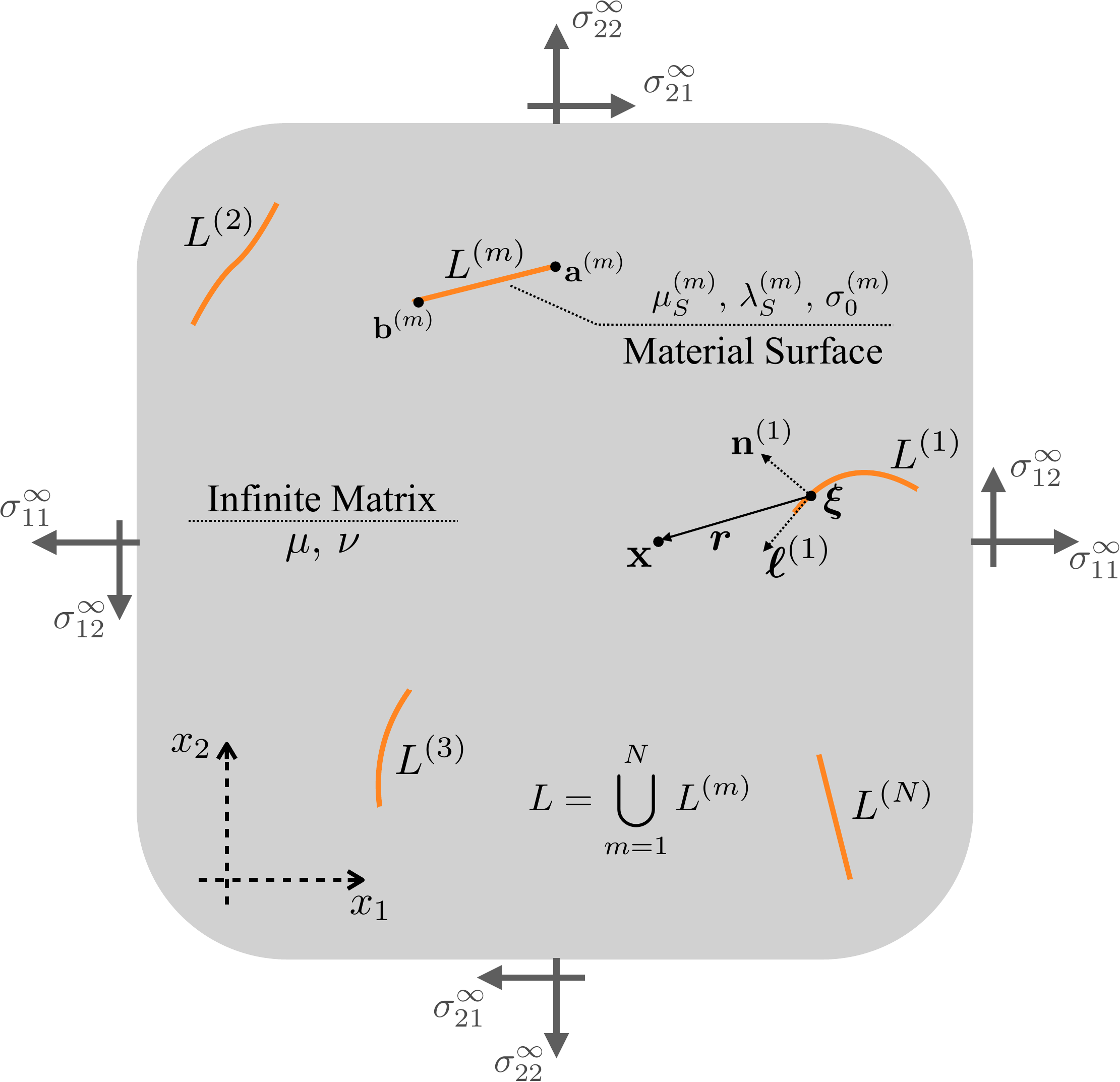}
    \caption{Problem formulation.}
    \label{fig:problem-formulation}
\end{figure}

Consider the two-dimensional plane strain problem for an infinite, isotropic, linearly elastic matrix subjected to uniform far-field stress $\boldsymbol{\sigma}^{\infty}$, with Cartesian components $\sigma^{\infty}_{11}$, $\sigma^{\infty}_{12}$, $\sigma^{\infty}_{22}$. The matrix contains $N$ open sufficiently smooth material surfaces $L^{(m)}$, $m = 1,\ldots,N$ of arbitrary shapes, see Fig.~\ref{fig:problem-formulation}. 
The tips of the $m$-th material surface are denoted by 
$\mathbf{a}^{(m)}$ and $\mathbf{b}^{(m)}$, and the totality of all boundaries is defined as
\[
L = \bigcup_{m=1}^{N} L^{(m)} .
\]
The material surfaces, which are assumed to be separated from each other by finite distances, may have different positions, orientations, curvatures, and lengths.


The Gurtin--Murdoch theory is adopted in which every surface is modeled as an elastic membrane of negligible thickness, endowed with its own elastic properties and residual surface tension, perfectly bonded to the surrounding bulk matrix. The latter assumptions imply displacement continuity across each surface and traction jump across it. 

The properties of the $m$-th surface are described by the surface shear modulus $\mu_{S}^{(m)}$, the surface Lamé parameter $\lambda_{S}^{(m)}$, and the residual surface tension $\sigma_{0}^{(m)}$, all of dimensions $\mathrm{N/m}$. Generally, these properties can be different for different surfaces. The surrounding isotropic elastic matrix is characterized by the shear modulus $\mu$, with dimensions of $\mathrm{N/m^{2}}$, and the Poisson ratio $\nu$.

The objectives are to evaluate the elastic fields inside all components of the material system and to evaluate SIFs at the tips $\mathbf{a}^{(m)}$ and $\mathbf{b}^{(m)}$, $m = 1,\ldots,N$.


\section{System of governing equations of the Gurtin--Murdoch model in plane strain setting}
\label{sec:GM_plane_strain}

The system was presented in a plane strain setting in ~\cite{Mogilevskaya2021_AMR, Mogilevskaya2021_CST}. It consists of the classical equilibrium and constitutive equations in the bulk supplemented by the conditions of displacement continuity and traction jump across each \(L^{(m)}\), as well as constitutive relations and tip conditions at \(L^{(m)}\).

In the absence of body forces, the bulk equilibrium equation is
\begin{equation}
\nabla\cdot\boldsymbol{\sigma}=\mathbf{0}
,
\label{eq:bulk_equilibrium_GM}
\end{equation}
where \(\boldsymbol{\sigma}\) is the bulk Cauchy stress tensor. For an
isotropic linearly elastic material,
\begin{equation}
\boldsymbol{\sigma}
=
2\mu\boldsymbol{\varepsilon}
+
\lambda\,\operatorname{tr}(\boldsymbol{\varepsilon})\mathbf{I},
\qquad
\boldsymbol{\varepsilon}
=
\frac{1}{2}
\left(
\nabla\mathbf{u}
+
\nabla\mathbf{u}^{\mathsf T}
\right),
\label{eq:bulk_constitutive_GM}
\end{equation}
where \(\boldsymbol{\varepsilon}\) is the bulk strain tensor, \(\mathbf{u}\) is the displacement vector in the Cartesian coordinate system and \(\mathbf{I}\) is the second-order identity tensor. The operator \(\operatorname{tr}(\cdot)\) denotes the tensor trace. Under the assumption of plane strain, all out-of-plane displacement and strain components vanish, and all in-plane fields depend only on Cartesian coordinates \(x_1\) and \(x_2\).

For each \(L^{(m)}\), \(s^{(m)}\) denotes the arc-length measured from the tip \(\mathbf{a}^{(m)}\), while \(\boldsymbol{\ell}^{(m)}\) and \(\mathbf{n}^{(m)}\) denote the local unit tangent and normal vectors, respectively. The signed local radius of curvature is denoted by \(R^{(m)}=R^{(m)}(s^{(m)})\).

The condition of continuity of the displacement vector across \(L^{(m)}\) is:
\begin{equation}
\mathbf{u}^{(m)+}
=
\mathbf{u}^{(m)-}
=
\mathbf{u}^{(m)}
,
\qquad  m = 1, \dots, N, 
\label{eq:displacement_continuity_GM}
\end{equation}
in which superscripts \(+\) and \(-\) denote the limiting values of a field as \(L^{(m)}\) is approached from the direction of \(\mathbf{n}^{(m)}\) and from the opposite direction, respectively.

The normal and tangential components of the bulk traction on \(L^{(m)}\), \(m=1,\ldots,N\), in the local coordinate system, Fig.~\ref{fig:problem-formulation},  are denoted by \(t_n^{(m)}\) and \(t_{\ell}^{(m)}\). The surface-equilibrium conditions relate the traction jumps across \(L^{(m)}\)\ to the surface stress components at \(L^{(m)}\) as
\begin{equation}
\Delta t_{\ell}^{(m)}
=
t_{\ell}^{(m){+}} - t_{\ell}^{(m){-}}
=
\frac{\partial \sigma^{S(m)}}{\partial s^{(m)}}
+
\sigma_{0}^{(m)}\frac{\omega^{S(m)}}{R^{(m)}},
\label{eq:tangential_traction_jump_GM}
\end{equation}
\begin{equation}
\Delta t_{n}^{(m)}
=
t_{n}^{(m){+}}-t_{n}^{(m){-}}
=
-\frac{\sigma^{S(m)}}{R^{(m)}}
+
\sigma_{0}^{(m)}\frac{\partial\omega^{S(m)}}{\partial s^{(m)}},
\label{eq:normal_traction_jump_GM}
\end{equation}
where the tangential component \(\sigma^{S(m)}\)  of the surface stress tensor at \(L^{(m)}\) is characterized by the following constitutive relation:
\begin{equation}
\sigma^{S(m)}
=
\sigma_{0}^{(m)}
+
\left(
\lambda_{S}^{(m)}+2\mu_{S}^{(m)}\right)\varepsilon^{S(m)},
\label{eq:surface_constitutive_scalar_GM}
\end{equation}
in which \(\varepsilon^{S(m)}\) is the only nonvanishing component of the surface strain tensor at \(L^{(m)}\) expressed in terms of the local displacement components as
\begin{equation}
\varepsilon^{S(m)}
=
\frac{u_n^{(m)}}{R^{(m)}}
+
\frac{\partial u_{\ell}^{(m)}}{\partial s^{(m)}},
\label{eq:surface_strain_GM}
\end{equation}
with $u_n^{(m)}$ and $u_{\ell}^{(m)}$ being the normal and tangential components of the displacement vector in the local coordinate system on \(L^{(m)}\). The normal component of the surface stress tensor on \(L^{(m)}\) is \(\sigma_{0}^{(m)}\omega^{S(m)}\), where
\begin{equation}
\omega^{S(m)}
=
-\frac{u_{\ell}^{(m)}}{R^{(m)}}
+
\frac{\partial u_n^{(m)}}{\partial s^{(m)}}.
\label{eq:surface_rotation_GM}
\end{equation}

If the surface \(L^{(m)}\) has tips, the following conditions must be satisfied:
\begin{equation}
\sigma^{S(m)} = 0,
\qquad
\sigma_{0}^{(m)}\omega^{S(m)} = 0.
\qquad
\text{at } \mathbf{a}^{(m)},\mathbf{b}^{(m)}.
\label{eq:GM_tip_conditions}
\end{equation}
When \(\sigma_{0}^{(m)}\neq 0\), the second condition requires that \(\omega^{S(m)}=0\) at the tips, while when \(\sigma_{0}^{(m)}=0\), it is satisfied identically.

Eqs.~\eqref{eq:bulk_equilibrium_GM}--\eqref{eq:GM_tip_conditions}
define the boundary-value problem for an infinite elastic matrix containing \(N\) open Gurtin--Murdoch material surfaces of arbitrary but sufficiently smooth shapes.

It has been shown in ~\cite{Mogilevskaya2021_AMR, Mogilevskaya2021_CST} that the Gurtin--Murdoch theory with vanishing residual surface tension describes the membrane-type interface conditions associated with a thin, stiff interphase layer of finite thickness \(h\), \cite{Benveniste2001}, provided that
\begin{equation}
\frac{\lambda_I}{\lambda+2\mu}\frac{h}{H}
\sim \Theta(1),
\qquad
\frac{\mu_I}{\lambda+2\mu}\frac{h}{H}
\sim \Theta(1),
\label{eq:membrane_scaling}
\end{equation}
as
\begin{equation}
\frac{\mu_I}{\mu}\to\infty,
\qquad
\frac{h}{H}\to0,
\label{eq:membrane_asymptotic_regime}
\end{equation}
in which  \(H\) is a characteristic length scale of the problem, \(\lambda_I\) and \(\mu_I\) are the Lam\'e parameter and shear modulus of the layer, respectively. In a such a case, the elastic parameters of the surface are related to the properties of the finite-thickness interphase layer by
\begin{equation}
\lambda_S
=
\frac{2\mu_I\lambda_I}{\lambda_I+2\mu_I}\,h,
\qquad
\mu_S
=
\mu_I h.
\label{eq:surface_moduli_from_layer}
\end{equation}

\section{Integral representations of the elastic fields in the material system}
\label{sec:integral_representations}

The boundary-value problem formulated in the preceding section requires the displacement field to satisfy the Navier equations in the bulk and to remain continuous across each material surface, while the bulk traction vector should undergo a jump governed by surface equilibrium. That motivated us to use a single-layer elastic potential, which satisfies the required conditions for the displacement and traction fields.

A single-layer elastic potential is an integral representation of the displacements generated by a continuous distribution of concentrated forces along a curve. Its kernel \(G_{kj}(\mathbf{x},\boldsymbol{\xi})\), the so-called Kelvin fundamental solution (Green's function), represents the displacement component \(u_k\) at the point \(\mathbf{x}=(x_1,x_2)\), Fig.~\ref{fig:problem-formulation}, produced by a unit concentrated force applied at the point \(\boldsymbol{\xi}=(\xi_1,\xi_2)\) in the \(j\)-th direction. The solution satisfies the homogeneous Navier equation everywhere in the plane except at the point of force application, where it is singular. 

Under a prescribed uniform far-field load, the total displacement field can be obtained by the standard superposition procedure and represented as
\begin{equation}
u_k(\mathbf{x})
=
u_k^{\infty}(\mathbf{x})
+
\sum_{m=1}^{N}
\int_{L^{(m)}}
G_{kj}(\mathbf{x},\boldsymbol{\xi})
\rho_j^{(m)}(\boldsymbol{\xi})
\,\mathrm{d}s_{\boldsymbol{\xi}},
\qquad k,j=1,2,
\label{eq:SLEP_displacement_general}
\end{equation}
where repeated Cartesian indices imply summation, \(\mathbf{x} \notin L\), \(\boldsymbol{\xi} \in L\),  \(\rho_j^{(m)}(\boldsymbol{\xi})\) is the sufficiently smooth Cartesian component of the force density at the point  \(\boldsymbol{\xi} \in L^{(m)}\), and \(\mathrm{d}s_{\boldsymbol{\xi}}\) is the length of the arc at \(\boldsymbol{\xi}\).
Furthermore, \(u_k^{\infty}\) represents the displacement component in the homogeneous matrix (which does not contain $L$) that is subjected to the prescribed far-field load. Under plane-strain conditions,  \(G_{kj}(\mathbf{x},\boldsymbol{\xi})\) has the following form:
\begin{equation}
G_{kj}(\mathbf{x},\boldsymbol{\xi})
=
\frac{1}{2\pi\mu(\kappa+1)}
\left[
-\kappa\delta_{kj}\ln r
+
r_{,k}r_{,j}
\right],
\label{eq:Kelvin_displacement_kernel}
\end{equation}
where
\begin{equation}
\kappa=3-4\nu,
\qquad
r=\lVert\mathbf{x}-\boldsymbol{\xi}\rVert,
\qquad
r_{,k}
=
\frac{\partial r}{\partial x_k}
=
\frac{x_k-\xi_k}{r},
\label{eq:Kelvin_kernel_notation}
\end{equation}
and \(\delta_{kj}\) is the Kronecker delta.

The displacement representation, Eq.~\eqref{eq:SLEP_displacement_general}, is regular for
\(\mathbf{x}\notin L\) and its value at the point \(\mathbf{x}_0\in L^{(q)}\) can be obtained for each \(q\in\{1,\ldots,N\}\) by taking the limit
\(\mathbf{x}\to\mathbf{x}_0\) from either side of \(L^{(q)}\). Since the Kelvin displacement kernel has only a weak logarithmic singularity, the limiting displacement values from the two sides coincide, so
\begin{equation}
u_k^{+}(\mathbf{x}_0)
=
u_k^{-}(\mathbf{x}_0)
=
u_k^{\infty}(\mathbf{x}_0)
+
\sum_{m=1}^{N}
\int_{L^{(m)}}
G_{kj}(\mathbf{x}_0,\boldsymbol{\xi})
\rho_j^{(m)}(\boldsymbol{\xi})
\,\mathrm{d}s_{\boldsymbol{\xi}}.
\label{eq:SLEP_displacement_limit}
\end{equation}
Therefore, as required, the displacement vector is continuous across $L$.


Using the strain--displacement relations together with Hooke's law, the traction vector \(\mathbf{t}(\mathbf{x})\) at the field point \(\mathbf{x}\), located inside the matrix material on a line characterized by the unit normal \(\mathbf{n}(\mathbf{x})\), can be expressed as follows:
\begin{equation}
t_k(\mathbf{x})
=
t_k^{\infty}(\mathbf{x})
+
\sum_{m=1}^{N}
\int_{L^{(m)}}
T_{kj}(\mathbf{x},\boldsymbol{\xi})
\rho_j^{(m)}(\boldsymbol{\xi})
\,\mathrm{d}s_{\boldsymbol{\xi}},
\qquad
\mathbf{x}\notin L,
\label{eq:SLEP_traction_general}
\end{equation}
where \(T_{kj}\) is the traction kernel associated with the Kelvin fundamental solution, see \ref{Appendix:Kelvin_traction_kernel}.

Unlike the displacement representation of Eq.~\eqref{eq:SLEP_displacement_general}, the traction representation of Eq.~\eqref{eq:SLEP_traction_general} possesses different limiting values from the two sides of each \(L^{(q)}\), $q=1,\ldots,N$, due to the strong singularity of the traction kernel. The corresponding limiting values of the tractions at \(\mathbf{x}_0\in L^{(q)}\) are given by
\begin{align}
t_k^{\pm}(\mathbf{x}_0)
={}&
t_k^{\infty}(\mathbf{x}_0)
\pm
\frac{1}{2}\rho_k^{(q)}(\mathbf{x}_0)
\nonumber\\
&+
\fint_{L^{(q)}}
T_{kj}(\mathbf{x}_0,\boldsymbol{\xi})
\rho_j^{(q)}(\boldsymbol{\xi})
\,\mathrm{d}s_{\boldsymbol{\xi}}
+
\sum_{\substack{m=1\\m\neq q}}^{N}
\int_{L^{(m)}}
T_{kj}(\mathbf{x}_0,\boldsymbol{\xi})
\rho_j^{(m)}(\boldsymbol{\xi})
\,\mathrm{d}s_{\boldsymbol{\xi}},
\label{eq:SLEP_traction_limit}
\end{align}
where \( \fint \) denotes the Cauchy principal value. Subtracting the two limiting values, one gets
\begin{equation}
\rho_k^{(q)}(\boldsymbol{\xi})
=
t_k^{(q)+}(\boldsymbol{\xi})
-
t_k^{(q)-}(\boldsymbol{\xi})
=
\Delta t_k^{(q)}(\boldsymbol{\xi}),
\qquad
\boldsymbol{\xi}\in L^{(q)}.
\label{eq:SLEP_density_identification}
\end{equation}
Thus, the density of the single-layer elastic potential has the physical meaning of the Cartesian traction jump. The displacement field can therefore be written in terms of traction jumps as
\begin{equation}
u_k(\mathbf{x})
=
u_k^{\infty}(\mathbf{x})
+
\sum_{m=1}^{N}
\int_{L^{(m)}}
G_{kj}(\mathbf{x},\boldsymbol{\xi})
\Delta t_j^{(m)}(\boldsymbol{\xi})
\,\mathrm{d}s_{\boldsymbol{\xi}}.
\label{eq:displacement_traction_jump_representation}
\end{equation}

The traction jumps in Eqs.~\eqref{eq:tangential_traction_jump_GM}--~\eqref{eq:normal_traction_jump_GM} are expressed in the local coordinate system associated with each surface, whereas the traction jumps appearing in Eq.~\eqref{eq:displacement_traction_jump_representation} are given in Cartesian components. Similarly, the surface constitutive relations for \(\sigma^{S(m)}\) and \(\omega^{S(m)}\), given in Eqs.~\eqref{eq:surface_constitutive_scalar_GM}--\eqref{eq:surface_rotation_GM}, are formulated in terms of the local displacement components and their tangential derivatives. Therefore, appropriate transformations between the Cartesian and local coordinate systems are required. Substituting the transformed displacement representations and traction jumps into
Eqs.~\eqref{eq:surface_constitutive_scalar_GM}--\eqref{eq:surface_rotation_GM} yields a system of coupled singular boundary integral equations for the unknown surface fields \(\sigma^{S(m)}\) and \(\omega^{S(m)}\),
\(m=1,\ldots,N\).

The resulting system, presented in \ref{app:coupled-singular-bies}, is supplemented by the tip conditions of Eq.~\eqref{eq:GM_tip_conditions}. Once the system has been solved for the unknown surface fields \(\sigma^{S(m)}\) and \(\omega^{S(m)}\), \(m=1,\ldots,N\), the elastic fields at any point \(\mathbf{x}\notin L\) can be recovered from the corresponding integral representations. In particular, the displacement field is evaluated from the single-layer potential in
Eq.~\eqref{eq:SLEP_displacement_general}, whereas the tractions on a line characterized by the unit normal \(\mathbf{n}(\mathbf{x})\) is
obtained from Eq.~\eqref{eq:SLEP_traction_general}. By appropriately selecting the unit normal, the Cartesian components of the bulk stress
tensor can be recovered from the traction representation. The corresponding strain components \(\varepsilon_{ij}\) are then evaluated from the strain-stress constitutive relation as
\begin{equation}
\varepsilon_{ij}(\mathbf{x})
=
\frac{1}{2\mu}
\left[
\sigma_{ij}(\mathbf{x})
-
\frac{3-\kappa}{4}
\sigma_{kk}(\mathbf{x})\delta_{ij}
\right],
\qquad i,j,k=1,2,
\label{eq:inverse_plane_strain_constitutive_relation}
\end{equation}
where \(\sigma_{ij}\) are the Cartesian components of the bulk stress tensors, and \(\sigma_{kk}\) denotes the trace of the in-plane stress tensor.


\section{Isogeometric Boundary Element Method}\label{Section:5}

The coupled boundary integral equations derived in the previous section are defined for surfaces of arbitrary shape and involve \(\sigma^{S(m)}\), \(\omega^{S(m)}\) and their tangential derivatives. Their numerical solution therefore requires a discretization that can represent curved geometries accurately, while providing sufficiently smooth approximations of the unknown surface fields. These requirements motivate the use of the isogeometric boundary element method (IGABEM), in which the same basis functions are used to represent the geometry of the surfaces and to approximate the unknown surface quantities.

In the present formulation, each surface is represented exactly or to high geometric accuracy by a non-uniform rational B-spline (NURBS) curve. The unknown fields \(\sigma^{S(m)}\) and \(\omega^{S(m)}\) are approximated using the same NURBS basis functions, and the resulting boundary integral equations are evaluated at a set of collocation points. Because the governing equations involve the first tangential derivatives of the unknown fields, quadratic NURBS basis functions with simple interior knots, i.e., interior knots of multiplicity one, are employed. The resulting approximation is smooth within each element and \(C^{1}\)-continuous across adjacent elements of each surface. A special numerical treatment, detailed in Ref.~\cite{Patil2026}, is employed for the singular integrals, whereas the interaction integrals between distinct material surfaces are evaluated using standard Gaussian quadrature.

\subsection{Representation of the surfaces using NURBS}
\label{subsec:NURBS_geometry}

In this work, the surfaces \(L^{(m)}\) are represented parametrically using NURBS curves. The fundamentals of B-splines and NURBS, including the construction of the basis functions and their parametric derivatives, are described in detail in our previous work~\cite{Patil2026}; see also \cite{Farin1999, Cottrell2009}. Only the relations required for the present formulation are summarized below.

The geometry of the \(m\)-th surface is represented as
\begin{equation}
\mathbf{x}^{(m)}(\xi)
=
\sum_{A=1}^{n^{(m)}}
R_{A,p}^{(m)}(\xi)\mathbf{P}_{A}^{(m)},
\qquad
\xi\in
\left[\xi_1^{(m)},\xi_{n^{(m)}+p+1}^{(m)}\right].
\label{eq:NURBS_geometry}
\end{equation}
where \(n^{(m)}\) is the number of NURBS basis functions associated with \(L^{(m)}\), \(p\) is the degree of basis functions,
\(\mathbf{P}_{A}^{(m)}\) denotes the \(A\)-th control point, and \(R_{A,p}^{(m)}(\xi)\) is the corresponding NURBS basis function. The basis functions are constructed from the knot vector
\[
\boldsymbol{\Xi}^{(m)}
=
\left[
\xi_1^{(m)},\xi_2^{(m)},\ldots,
\xi_{n^{(m)}+p+1}^{(m)}
\right]
\]
and the associated weights \(w_A^{(m)}\). The NURBS basis functions of degree \(p\), defined by
\begin{equation}
R_{A,p}^{(m)}(\xi)
=
\frac{
N_{A,p}^{(m)}(\xi)w_A^{(m)}
}{
\displaystyle
\sum_{B=1}^{n^{(m)}}
N_{B,p}^{(m)}(\xi)w_B^{(m)}
},
\label{eq:NURBS_basis}
\end{equation}
where \(N_{A,p}^{(m)}\) denote the corresponding B-spline basis functions.

The first- and second- order parametric derivatives of the surface geometry are obtained directly from the derivatives of the NURBS basis functions:
\begin{equation}
\frac{\mathrm{d}\mathbf{x}^{(m)}(\xi)}{\mathrm{d}\xi}
=
\sum_{A=1}^{n^{(m)}}
\frac{\mathrm{d}R_{A,p}^{(m)}(\xi)}{\mathrm{d}\xi}
\mathbf{P}_{A}^{(m)},
\qquad
\frac{\mathrm{d}^{2}\mathbf{x}^{(m)}(\xi)}{\mathrm{d}\xi^{2}}
=
\sum_{A=1}^{n^{(m)}}
\frac{\mathrm{d}^{2}R_{A,p}^{(m)}(\xi)}{\mathrm{d}\xi^{2}}
\mathbf{P}_{A}^{(m)}.
\label{eq:NURBS_geometry_derivatives}
\end{equation}
The geometric Jacobian is
\begin{equation}
J^{(m)}(\xi)
=
\left\|
\mathbf{x}_{,\xi}^{(m)}(\xi)
\right\|,
\label{eq:NURBS_geometric_Jacobian}
\end{equation}
so that
\begin{equation}
\mathrm{d}s^{(m)}
=
J^{(m)}(\xi)\,\mathrm{d}\xi,
\qquad
\frac{\partial}{\partial s^{(m)}}
=
\frac{1}{J^{(m)}(\xi)}
\frac{\partial}{\partial \xi}.
\label{eq:parametric_arc_length_relation}
\end{equation}
 
For \(\mathbf{x}^{(m)}=(x_1^{(m)},x_2^{(m)})\), the signed curvature is computed as
\begin{equation}
\mathcal{K}^{(m)}(\xi)
=
\frac{
x_{1,\xi}^{(m)}x_{2,\xi\xi}^{(m)}
-
x_{2,\xi}^{(m)}x_{1,\xi\xi}^{(m)}
}{
\left[J^{(m)}(\xi)\right]^3
}.
\label{eq:NURBS_signed_curvature}
\end{equation}

\subsection{Approximation of the surface unknowns}
\label{subsec:NURBS_unknowns}

According to the isogeometric concept, \(\sigma^{S(m)}\) and \(\omega^{S(m)}\) are approximated using the same NURBS basis functions used to represent the geometry of each surface. For the \(m\)-th surface, the approximations are 
\begin{equation}
\sigma^{S(m)}(\xi)
=
\sum_{A=1}^{n^{(m)}}
R_{A,p}^{(m)}(\xi)\,
\widehat{\sigma}_{A}^{S(m)},
\qquad
\omega^{S(m)}(\xi)
=
\sum_{A=1}^{n^{(m)}}
R_{A,p}^{(m)}(\xi)\,
\widehat{\omega}_{A}^{S(m)},
\label{eq:NURBS_approximation_surface_unknowns}
\end{equation}
where, \(\widehat{\sigma}_{A}^{S(m)}\) and \(\widehat{\omega}_{A}^{S(m)}\) are the unknown coefficients associated with the \(A\)-th NURBS basis function.

Although the geometry and the unknown surface fields are represented globally by NURBS expansions, the boundary integrals are evaluated numerically over individual knot spans. Consequently, the parametric domain of each surface is partitioned into elements defined by nonzero knot spans. Owing to the local-support property of the NURBS basis, only a limited number of basis functions are nonzero over a given element. For a basis of degree \(p\), this number is \(p+1\) for a regular knot span. The local basis functions in element \(e\) are related to the corresponding global NURBS basis functions through the following connectivity mapping:
\begin{equation}
N_l^{e(m)}(\xi)
=
R_{A,p}^{(m)}(\xi),
\qquad
A=\operatorname{conn}^{(m)}(e,l),
\qquad
l=1,\ldots,p+1,
\label{eq:NURBS_connectivity}
\end{equation}
where \(\operatorname{conn}^{(m)}(e,l)\) identifies the global basis function associated with the \(l\)-th local basis function of element \(e\).

Consequently, on the element \(e\), the unknown geometry and surface unknowns are represented as
\begin{equation}
\begin{aligned}
\mathbf{x}^{(m)}(\xi)
&=
\sum_{l=1}^{p+1}
N_l^{e(m)}(\xi)\,
\mathbf{P}_l^{e(m)},
\\[6pt]
\sigma^{S(m)}(\xi)
&=
\sum_{l=1}^{p+1}
N_l^{e(m)}(\xi)\,
\widehat{\sigma}_l^{S,e(m)},
\\[6pt]
\omega^{S(m)}(\xi)
&=
\sum_{l=1}^{p+1}
N_l^{e(m)}(\xi)\,
\widehat{\omega}_l^{S,e(m)}.
\end{aligned}
\label{eq:elementwise_NURBS_approximations}
\end{equation}
Here, \(\mathbf{P}_l^{e(m)}\) denotes the control point associated with the \(l\)-th local basis function, while \(\widehat{\sigma}_l^{S,e(m)}\) and \(\widehat{\omega}_l^{S,e(m)}\) denote the corresponding local representations of the global unknown coefficients.

The element-wise approximations are substituted into the coupled boundary integral equations, and the resulting equations are evaluated at a set of collocation points on each surface. In the present formulation, the collocation points are chosen as the physical images of the Greville abscissae associated with the NURBS basis functions. For a basis of degree \(p\), the Greville abscissae for \(L^{(m)}\) are
\begin{equation}
\bar{\xi}_{a}^{(m)} = \frac{1}{p} \sum_{j=1}^{p} \xi_{a+j}^{(m)},
\qquad
a=1,\ldots,n^{(m)},
\label{eq:Greville_abscissae}
\end{equation}
and the corresponding physical collocation point is
\begin{equation}
\mathbf{x}_{a}^{(m)} = \mathbf{x}^{(m)}
\left(\bar{\xi}_{a}^{(m)}\right).
\label{eq:physical_collocation_points}
\end{equation}

Enforcing the system of coupled boundary integral equations at all collocation points yields a linear algebraic system of the form
\begin{equation}
\mathbf{C} \bm{x}
=
\mathbf{d},
\label{eq:global_IGABEM_system}
\end{equation}
where \(\bm{x}\) collects the unknown coefficients associated with \(\sigma^{S(m)}\) and \(\omega^{S(m)}\) on all surfaces, \(\mathbf{C}\) contains the self- and inter-elements contributions, and \(\mathbf{d}\) contains the terms associated with the prescribed far-field load and residual surface tension. The tip conditions in Eq.~\eqref{eq:GM_tip_conditions} are imposed together with the collocation equations to complete the discrete system.

The integrals over elements that do not contain collocation points are regular ones and are computed using Gaussian quadrature. The integrals over elements containing collocation points require special treatment; the singular-integration procedure adopted here follows that developed in our previous work~\cite{Patil2026}. 

Once the resulting linear system is solved, the \(\sigma^{S(m)}\) and \(\omega^{S(m)}\) are reconstructed from their NURBS approximations. The elastic fields inside all components of the material system can then be evaluated, as explained in Section \ref{sec:integral_representations}.

\section{Evaluation of SIFs at the tips}
\label{Section:SIF}

Here, the values of \(\sigma^{S(m)}\) and \(\omega^{S(m)}\), obtained from the NURBS approximations,  will be used to evaluate SIFs \(K_1^{(m,\eta)}\) and \(K_2^{(m,\eta)}\) at the tips of each surface, with \(\eta = a\) at the tip \(\mathbf{a}^{(m)}\) and \(\eta = b\) at the tip \(\mathbf{b}^{(m)}\). We will obtain these SIFs using the complex variables theory and the well-known near-tip behavior of the complex Cauchy type integrals. 

\begin{figure}[!htbp]
    \centering   \includegraphics[width=0.65\linewidth]  {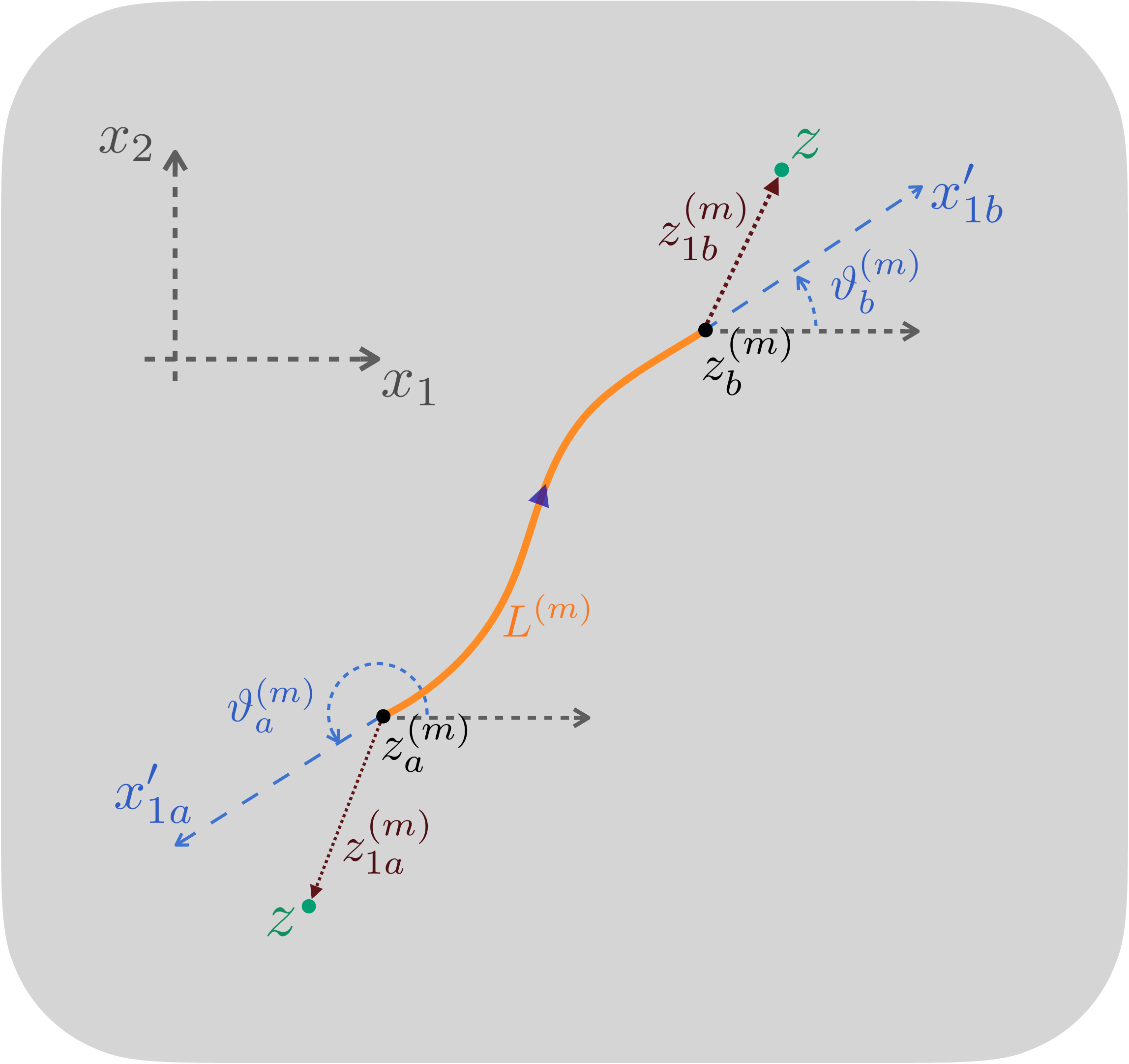}
    \caption{Schematic of the global and local coordinate systems at the tips of the surface \(L^{(m)}\), used for SIF evaluation.}
    \label{fig:SIF}
\end{figure}

The complex coordinates of the field and source points are represented by \(z=x_1+ \mathrm{i} x_2 \notin L\) and \(\tau=\tau_1+ \mathrm{i} \tau_2 \in L\) respectively, while 
the complex coordinates of the tips \(\mathbf{a}^{(m)}\) and \(\mathbf{b}^{(m)}\) are denoted by \(z_a^{(m)}\) and \(z_b^{(m)}\). As shown  in Fig.~\ref{fig:SIF}, the complex coordinates in the local coordinate systems with the centers at the two tips are defined as
\begin{equation}
z
=
z_{\eta}^{(m)}
+
z_{1\eta}^{(m)}
e^{\mathrm{i}\vartheta_{\eta}^{(m)}},
\qquad
\eta=a,b,
\label{eq:local_tip_coordinates}
\end{equation}
where \(z_{1\eta}^{(m)}\) is the complex coordinate measured from the corresponding tip \(z_{\eta}^{(m)}\), and
\(\vartheta_{\eta}^{(m)}\) denotes the orientation of the tangent at that tip directed away from the curve. Accordingly, the positive real axes of the local coordinate systems are aligned with the corresponding tangent directions.

As shown in ~\cite{Savruk1981, Han2024_JAM} complex SIFs can be obtained from the leading near-tip behavior of the Kolosov--Muskhelishvili potential \(\Phi(z)\) in the local coordinate system associated with each tip. Consequently, for the \(m\)-th surface,
\begin{equation}
K_1^{(m,\eta)}
-
\mathrm{i}K_2^{(m,\eta)}
=
\lim_{z_{1\eta}^{(m)}\to 0^{+}}
2\sqrt{2\pi z_{1\eta}^{(m)}}\,
\Phi\!\left(
z_{\eta}^{(m)}
+
z_{1\eta}^{(m)}
e^{\mathrm{i}\vartheta_{\eta}^{(m)}}
\right),
\qquad
\eta=a,b.
\label{eq:SIF_tip_from_Phi}
\end{equation}
In Eq.~\eqref{eq:SIF_tip_from_Phi}, the limits are taken along the positive real axes of the local coordinate systems. Thus, mode-I and mode-II SIFs can be obtained from the real and negative imaginary parts, respectively, of the limiting complex quantity.

It has been shown, see, e.g., ~\cite{Han2024_JAM}, that the Kolosov--Muskhelishvili potential for the problem under study can be expressed in terms of complex traction jumps as
\begin{equation}
\Phi(z)
=
\Phi^{\infty}(z)
+
\frac{1}{2\pi\mathrm{i}(\kappa+1)}
\sum_{q=1}^{N}
\int_{L^{(q)}}
\frac{\Delta \sigma^{(q)}(\tau)}
{\tau-z}
\,\mathrm{d}\tau,
\label{eq:KM_Phi_traction_jump}
\end{equation}
where \[\Phi^{\infty}(z) = \frac{\sigma^{\infty}_{11} + \sigma^{\infty}_{22}}{4} \,, \] 
and
\(\Delta\sigma^{(q)}= \Delta t_n^{(q)} +\mathrm{i}\Delta t_\ell^{(q)}\), see Eqs.~\eqref{eq:tangential_traction_jump_GM}-\eqref{eq:normal_traction_jump_GM}. 

Eq.~\eqref{eq:KM_Phi_traction_jump} can be rewritten in terms of the computed \(\sigma^{S(q)}\) and \(\omega^{S(q)}\) by using Eqs.~\eqref{eq:tangential_traction_jump_GM}--~\eqref{eq:normal_traction_jump_GM} and integrating
by parts. The endpoint contributions generated by the integration vanish because of the tip conditions of Eq.~\eqref{eq:GM_tip_conditions}. 

We introduce the complex field \(\chi^{(q)}(\tau)\) over each $q$-th surface as
\begin{equation}
\chi^{(q)}(\tau) =
\sigma_0^{(q)}\omega^{S(q)}(\tau) +
\mathrm{i}\sigma^{S(q)}(\tau),
\label{eq:complex_surface_field_combination}
\end{equation}
to obtain the following form  of $\Phi(z)$:
\begin{equation}
\Phi(z)
=
\Phi^{\infty}(z)
+
\frac{1}{2\pi\mathrm{i}(\kappa+1)}
\sum_{q=1}^{N}
\int_{L^{(q)}}
\frac{ \chi^{(q)}(\tau) e^{\mathrm{i}\alpha^{(q)}(\tau)}
}{
(\tau-z)^2
}
\,\mathrm{d}\tau,
\label{eq:KM_Phi_surface_fields}
\end{equation}
where \(\alpha^{(q)}(\tau)\) is the angle measured from the positive \(x_1\)-axis to the local tangent to \(L^{(q)}\) at the source point \(\tau\).

To explicitly account for the known square-root behavior of the surface fields near each tip, see ~\cite{Baranova2020_JE}, the complex quantity \(\chi^{(m)}\) is re-interpolated over the element adjacent to that tip using square-root-modified Lagrange polynomials. The interpolation points on the  tip-adjacent element of the \(m\)-th surface are denoted by \(\tau_j\), \(j=1,\ldots,n_p\), and the corresponding values of \(\chi^{(m)}\), evaluated from the NURBS approximations, are given by \[
\chi_j^{(m)}
=
\chi^{(m)}(\tau_j).
\] 
The parametric locations of these interpolation points are defined relative to the corresponding tip-element interval as
\[
\xi_{j,\eta}^{(m)}
=
\xi_{\mathrm{tip},\eta}^{(m)}
+
f_j
\left(
\xi_{\mathrm{e},\eta}^{(m)}
-
\xi_{\mathrm{tip},\eta}^{(m)}
\right),
\qquad
j=1,\ldots,n_p\,\,\,,
\]
where \(\xi_{\mathrm{tip},\eta}^{(m)}\) denotes the parametric coordinate
of tip \(\eta\), \(\xi_{\mathrm{e},\eta}^{(m)}\) denotes the parametric
coordinate of the opposite end of the corresponding tip element, and
\(f_j\) specifies the relative location of the \(j\)-th interpolation point
within that element.

Thus, similarly to what was done in~\cite{Mogilevskaya1998} for crack problems, $\chi^{(m)}(\tau)$ over the element adjacent to the tip of the \(m\)-th surface is represented as
\begin{equation}
\chi^{(m)}(\tau)
\approx
\sum_{j=1}^{n_p}
P_{j,\eta}^{(m)}(\tau)\,
\chi_j^{(m)},
\qquad
\tau\in L_{\mathrm{tip},\eta}^{(m)},
\qquad
\eta=a,b,
\label{eq:chi_reinterpolation_tip}
\end{equation}
where
\begin{equation}
P_{j,\eta}^{(m)}(\tau)
=
\sqrt{
\frac{
k_{\eta}\left(\tau-z_{\eta}^{(m)}\right)
}{
k_{\eta}\left(\tau_j-z_{\eta}^{(m)}\right)
}
}
\,Q_{j,\eta}^{(m)}(\tau),
\label{eq:modified_Lagrange_tip}
\end{equation}
with
\begin{equation}
k_{\eta}
=
\begin{cases}
1, & \eta=a,\\
-1, & \eta=b.
\end{cases}
\label{eq:tip_sign_parameter}
\end{equation}
and \(Q_{j,\eta}^{(m)}(\tau)\) are the standard Lagrange polynomials defined by
\begin{equation}
Q_{j,\eta}^{(m)}(\tau)
=
\prod_{\substack{s=1\\s\neq j}}^{n_p}
\frac{
\tau-\tau_{s,\eta}^{(m)}
}{
\tau_{j,\eta}^{(m)}-\tau_{s,\eta}^{(m)}
},
\qquad
\eta=a,b.
\label{eq:standard_Lagrange_functions}
\end{equation}

To evaluate the SIFs at each tip
of the \(m\)-th surface using Eq.~\eqref{eq:SIF_tip_from_Phi}, the field point is allowed to approach \(z_\eta^{(m)}\). In this limit, the integrals involved in the right-hand side of
Eq.~\eqref{eq:KM_Phi_surface_fields} from all surfaces
\(L^{(q)}\) with \(q\neq m\) remain regular and bonded because the corresponding elements are separated from the selected tip by finite distances. The far-field term \(\Phi^\infty\) and the integrals over the elements of \(L^{(m)}\) at the points that are not adjacent to \(z_\eta^{(m)}\) also remain bounded. After multiplication of \(\Phi(z)\) of Eq.~\eqref{eq:KM_Phi_surface_fields} by \(\sqrt{z_{1\eta}^{(m)}}\) and taking the limit when \({z_{1\eta}^{(m)}\to0^+}\), the products of multiplications corresponding to the bounded terms vanish. Therefore, Eq.~\eqref{eq:SIF_tip_from_Phi} reduces to
\begin{equation}
K_1^{(m,\eta)}-\mathrm{i}K_2^{(m,\eta)}
=
\lim_{z_{1\eta}^{(m)}\to0^+}
2\sqrt{2\pi z_{1\eta}^{(m)}}\,
\Phi_{\mathrm{tip},\eta}^{(m)}
\left(
z_\eta^{(m)}
+
z_{1\eta}^{(m)}
e^{\mathrm{i}\vartheta_\eta^{(m)}}
\right),
\qquad
\eta=a,b,
\label{eq:SIF_from_tip_element}
\end{equation}
where,
\begin{equation}
\Phi_{\mathrm{tip},\eta}^{(m)}(z)
=
\frac{1}{2\pi\mathrm{i}(\kappa+1)}
\sum_{j=1}^{n_p}
\chi_j^{(m)}
\int_{L_{\mathrm{tip},\eta}^{(m)}}
\frac{
P_{j,\eta}^{(m)}(\tau)
e^{\mathrm{i}\alpha^{(m)}(\tau)}
}{
(\tau-z)^2
}
\,\mathrm{d}\tau .
\label{eq:KM_tip_element_contribution}
\end{equation}

To evaluate the integral of Eq.~\eqref{eq:KM_tip_element_contribution}, \(e^{\mathrm{i}\alpha^{(m)}(\tau)}\) is expanded in a Taylor series about the selected tip. After substitution, the integral is reduced to linear combinations of integrals of the form
\begin{equation}
I_{j,\eta}^{(r)}(z)
=
\int_{L_{\mathrm{tip},\eta}^{(m)}}
\frac{P_{j,\eta}^{(m)}(\tau)}
{(\tau-z)^r}
\,\mathrm{d}\tau,
\qquad
r=1,2,
\label{eq:tip_element_integrals}
\end{equation}
which are evaluated analytically by following the procedure developed in ~\cite{Mogilevskaya1996}. Substitution of the resulting expressions into Eq.~\eqref{eq:SIF_from_tip_element}, followed by dimensionalization by $\sqrt{l^{(m)}}$, with \(l^{(m)}\) being the semi-length of the $m$-th surface, yields
\begin{align}
K_1^{(m,a)}-\mathrm{i}K_2^{(m,a)}
&=
\frac{\mathrm{i}\sqrt{2\pi}}
{2 \mu (\kappa+1) \sqrt{l^{(m)}}}
\exp\left(
\frac{\mathrm{i}\vartheta_a^{(m)}}{2}
\right)
\sum_{j=1}^{n_p}
\frac{
\chi_{j,a}^{(m)}
Q_{j,a}^{(m)}\!\left(z_a^{(m)}\right)
}{
\sqrt{
\tau_{j,a}^{(m)}-z_a^{(m)}
}
},
\label{eq:SIF_tip_a_final}
\\
K_1^{(m,b)}-\mathrm{i}K_2^{(m,b)}
&=
-\frac{\mathrm{i}\sqrt{2\pi}}
{2 \mu (\kappa+1) \sqrt{l^{(m)}}}
\exp\left(
\frac{\mathrm{i}\vartheta_b^{(m)}}{2}
\right)
\sum_{j=1}^{n_p}
\frac{
\chi_{j,b}^{(m)}
Q_{j,b}^{(m)}\!\left(z_b^{(m)}\right)
}{
\sqrt{
z_b^{(m)}-\tau_{j,b}^{(m)}
}
},
\label{eq:SIF_tip_b_final}
\end{align}
where \(\chi_{j,a}^{(m)}\) and \(\chi_{j,b}^{(m)}\) denote the values of \(\chi^{(m)}\) at the interpolation points located on the elements adjacent to tips \(\mathbf{a}^{(m)}\) and
\(\mathbf{b}^{(m)}\), respectively.

\section{Validation of the proposed approach}\label{Sub-Section:6.1}

To validate the approach, two configurations are considered. The first configuration involves a matrix containing multiple material surfaces located along straight segments. The benchmark solutions for this example are available in ~\cite{Patil2024}, where global approximations by the series of Chebyshev polynomials for the unknown scalar components of the surface stress tensor are used. In the second configuration, the embedded surfaces are assumed to be located along circular arcs; the benchmark solutions for that example are obtained in ~\cite{Han2024_JAM} using approximations by the series of trigonometric functions. Thus, the two examples provide tests of the proposed approach for geometries with zero and finite curvatures. In the IGABEM analysis for both studies, the surfaces were represented by a quadratic NURBS. The study is conducted using \(N_e=50\) elements and $75$ Gaussian quadrature points per element.

\subsection{Validation for the case of multiple straight surfaces}
\label{subsec:validation_straight_surfaces}

\begin{figure*}[!htbp]
    \centering

    \begin{subfigure}[t]{0.32\textwidth}
        \centering
        \includegraphics[width=\textwidth]{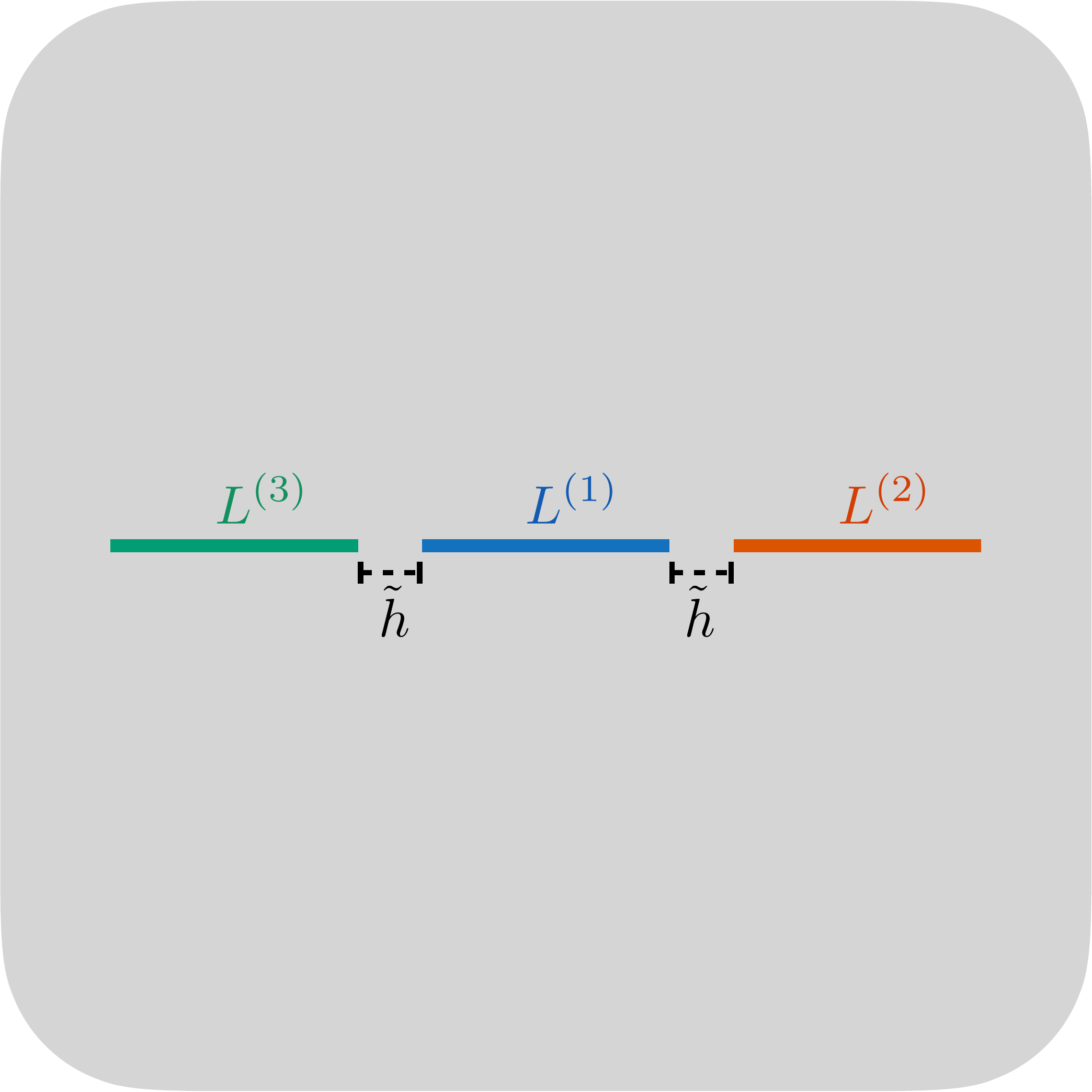}
        \caption{}
        \label{fig:collinear_1}
    \end{subfigure}
    \hfill
    \begin{subfigure}[t]{0.32\textwidth}
        \centering
        \includegraphics[width=\textwidth]{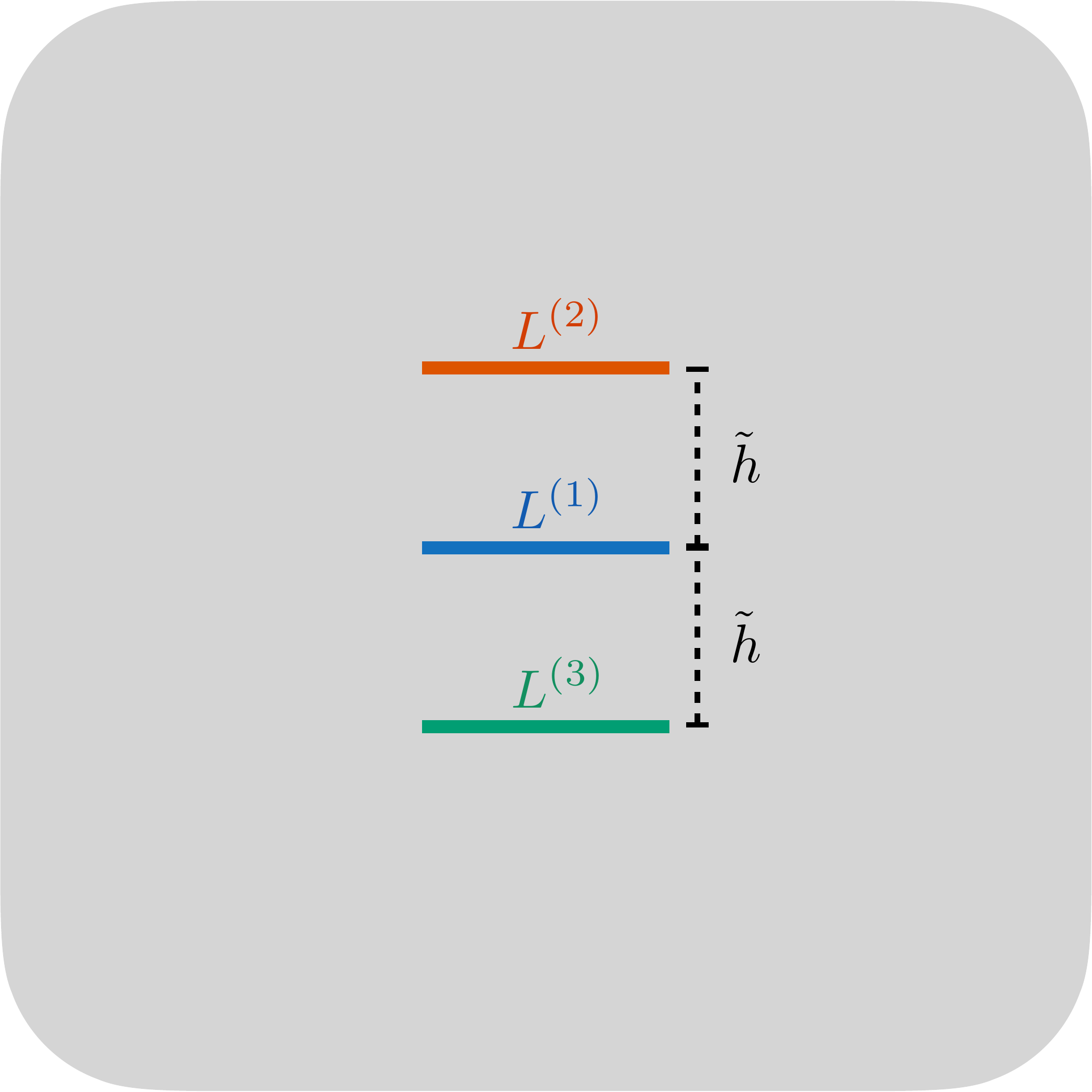}
        \caption{}
        \label{fig:collinear_2}
    \end{subfigure}
    \hfill
    \begin{subfigure}[t]{0.32\textwidth}
        \centering
        \includegraphics[width=\textwidth]{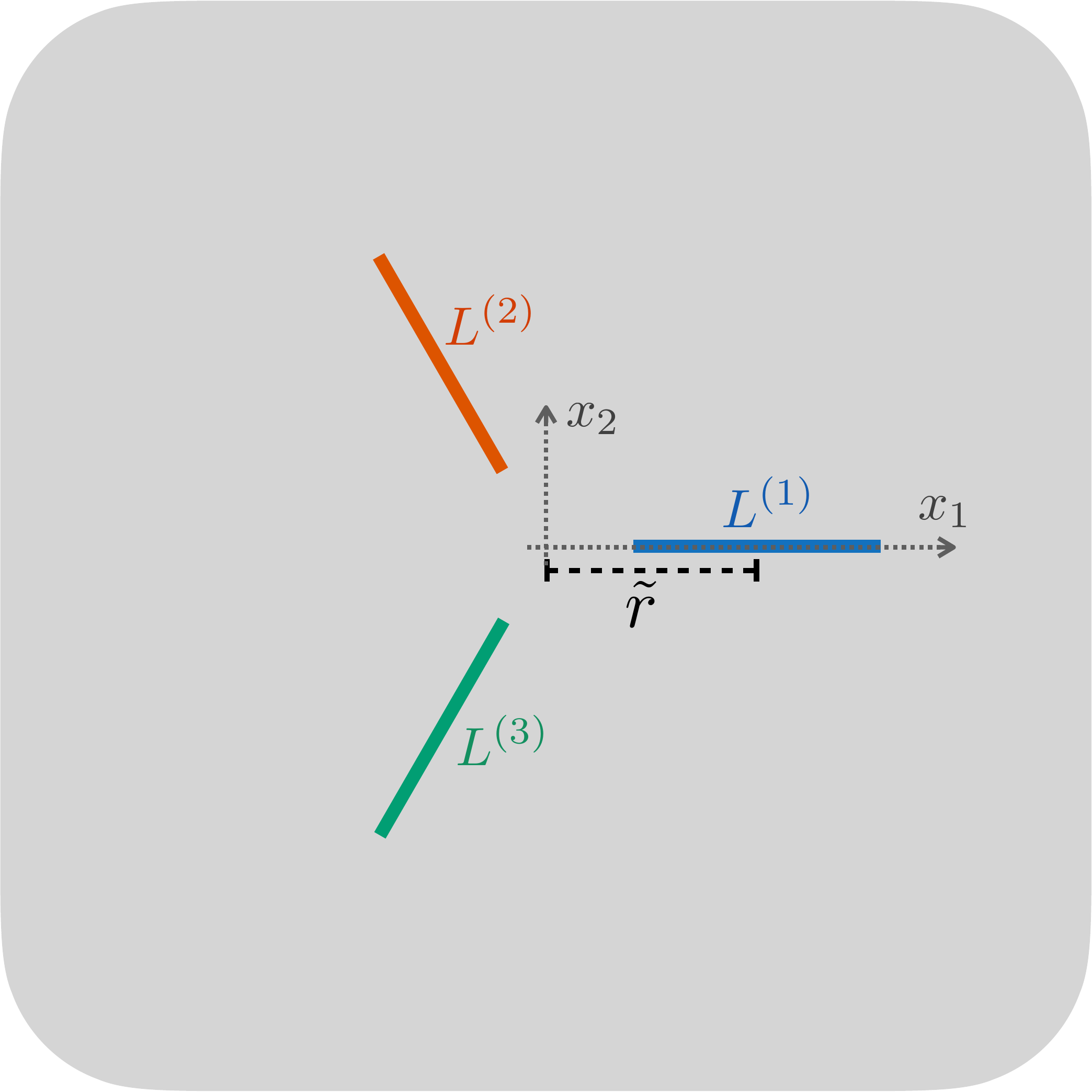}
        \caption{}
        \label{fig:collinear_3}
    \end{subfigure}

    \vspace{0.4cm}

    \begin{subfigure}[t]{0.325\textwidth}
        \centering
        \includegraphics[width=\textwidth]{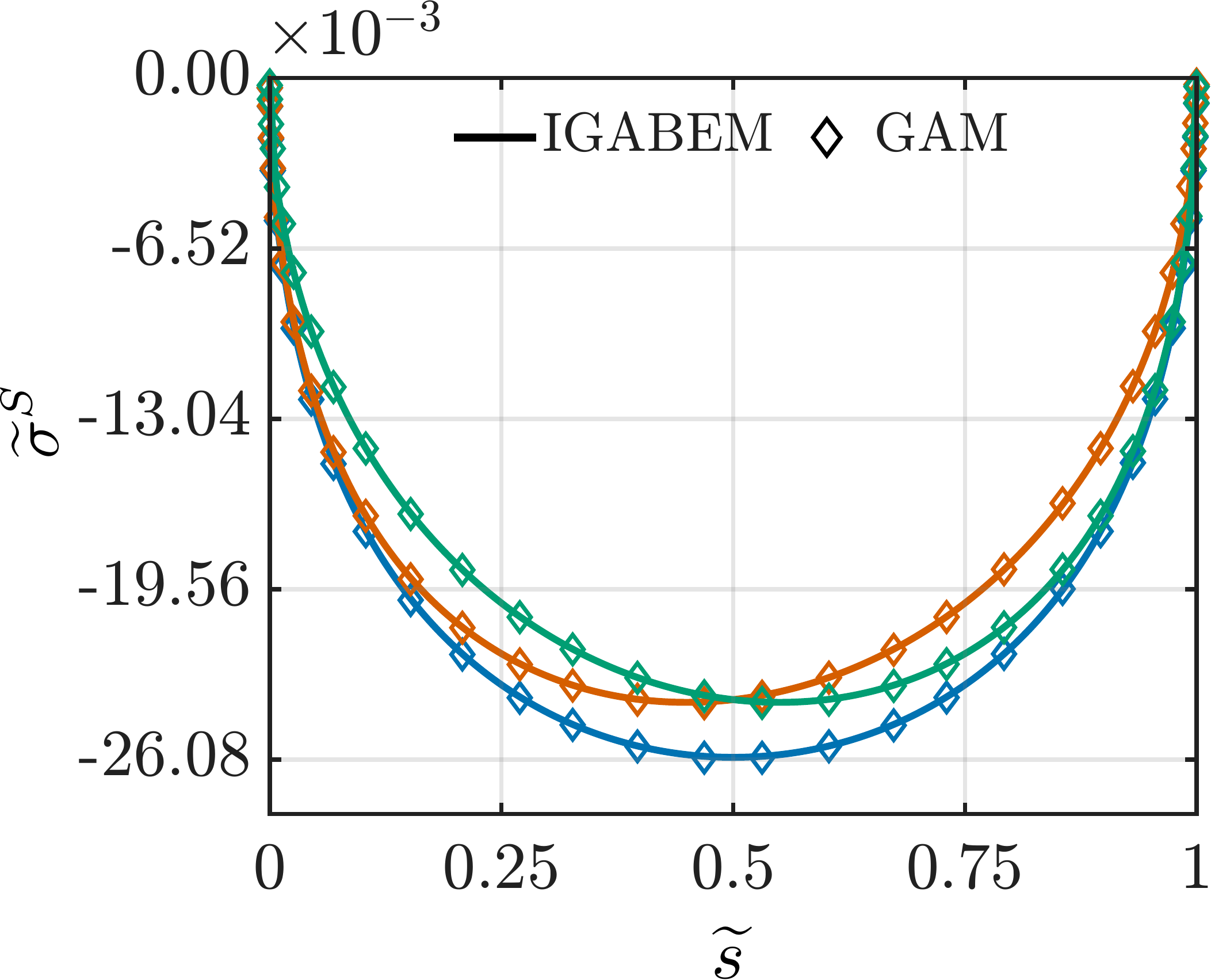}
        \caption{}
        \label{fig:parallel_1}
    \end{subfigure}
    \hfill
    \begin{subfigure}[t]{0.32\textwidth}
        \centering
        \includegraphics[width=\textwidth]{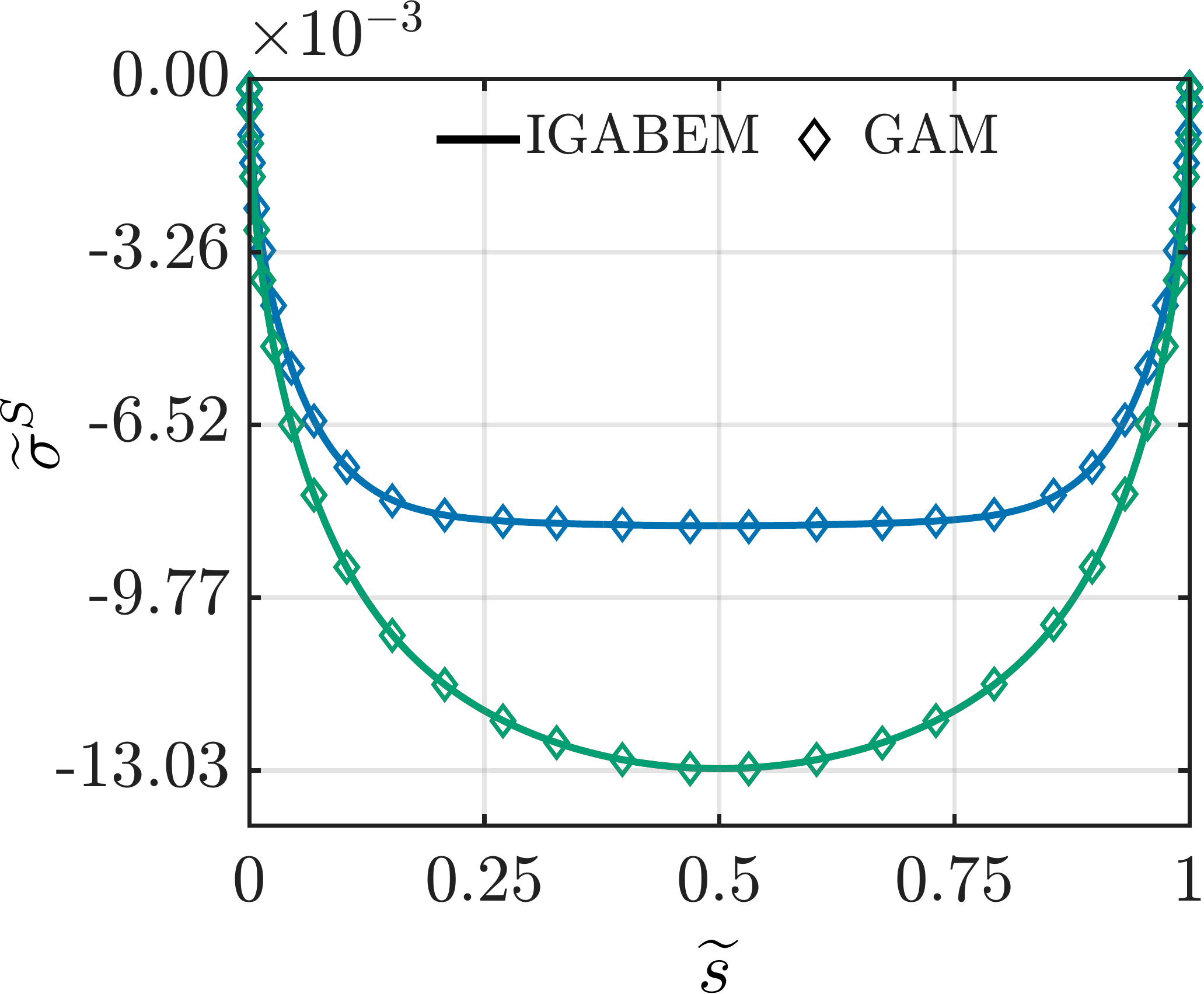}
        \caption{}
        \label{fig:parallel_2}
    \end{subfigure}
    \hfill
    \begin{subfigure}[t]{0.32\textwidth}
        \centering
        \includegraphics[width=\textwidth]{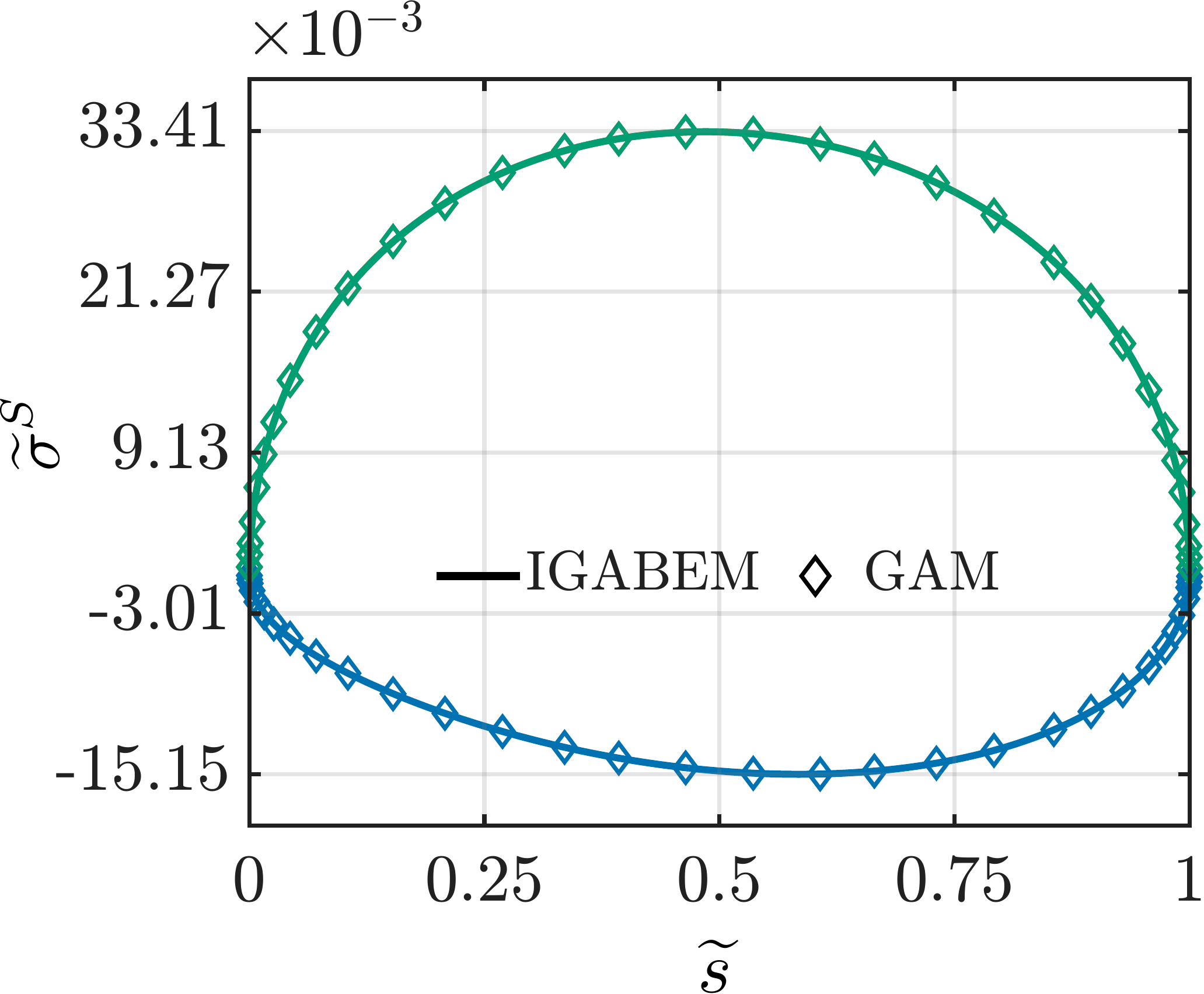}
        \caption{}
        \label{fig:parallel_3}
    \end{subfigure}

    \vspace{0.4cm}

    \begin{subfigure}[t]{0.325\textwidth}
        \centering
        \includegraphics[width=\textwidth]{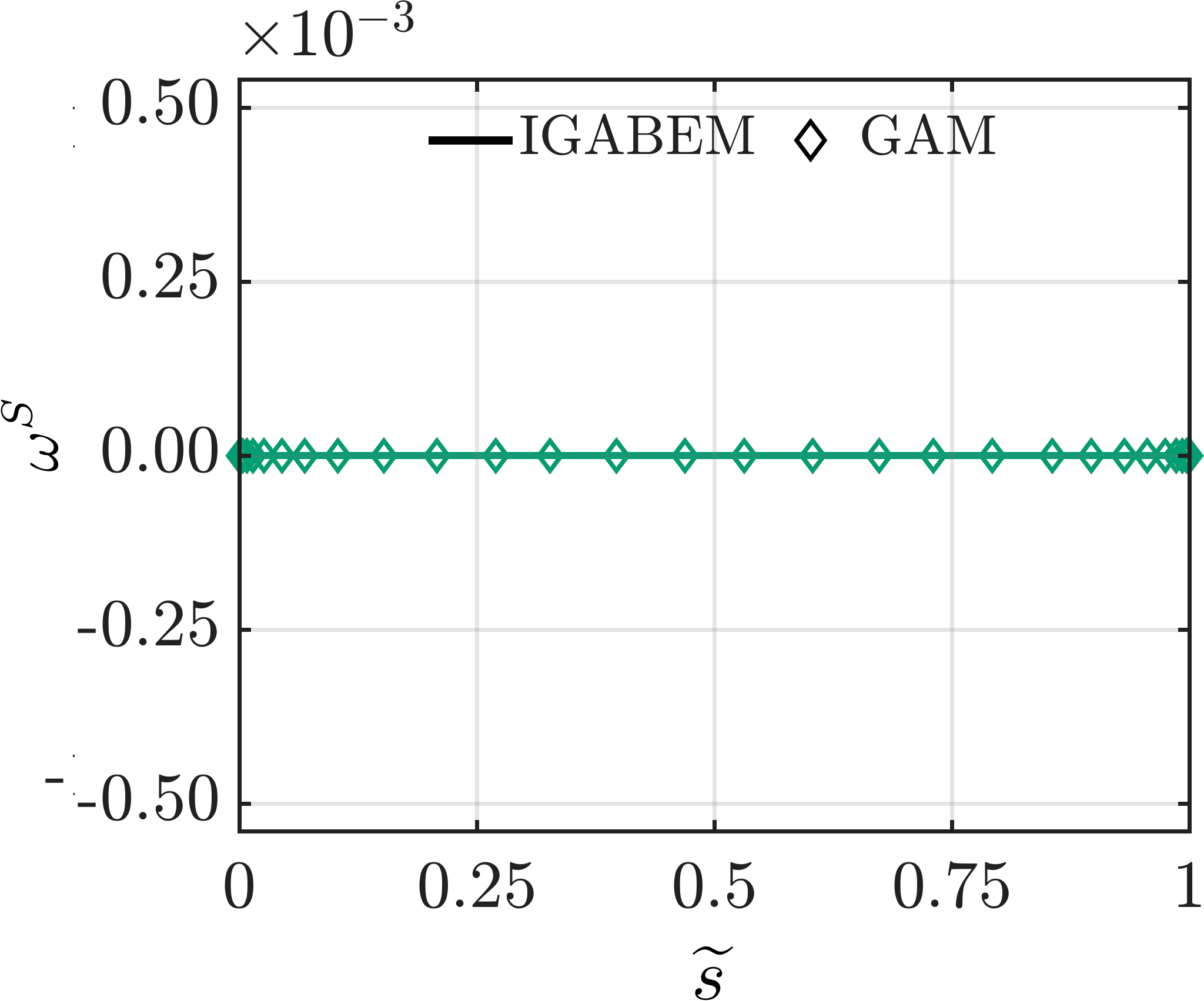}
        \caption{}
        \label{fig:radial_1}
    \end{subfigure}
    \hfill
    \begin{subfigure}[t]{0.325\textwidth}
        \centering
        \includegraphics[width=\textwidth]{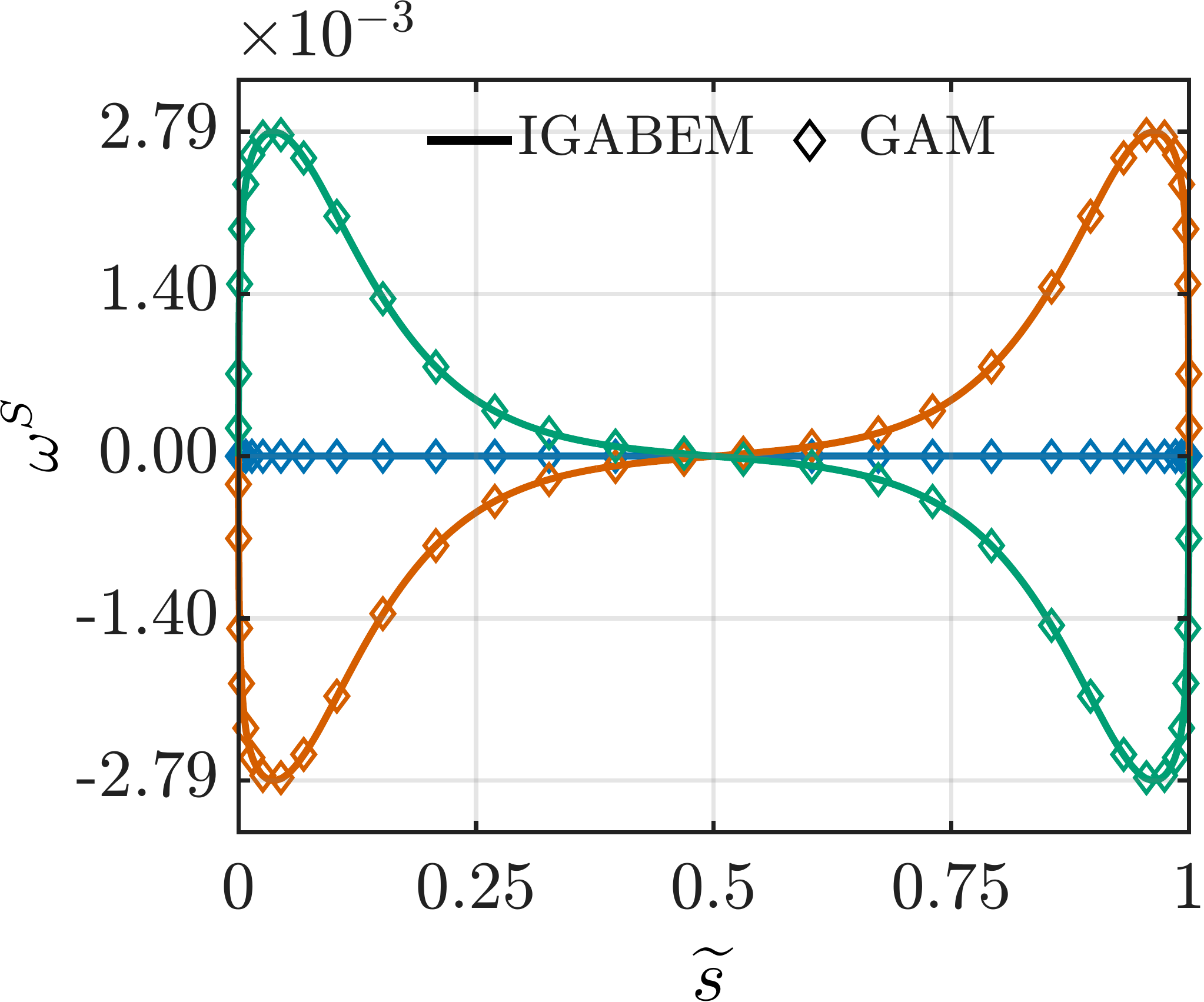}
        \caption{}
        \label{fig:radial_2}
    \end{subfigure}
    \hfill
    \begin{subfigure}[t]{0.325\textwidth}
        \centering
        \includegraphics[width=\textwidth]{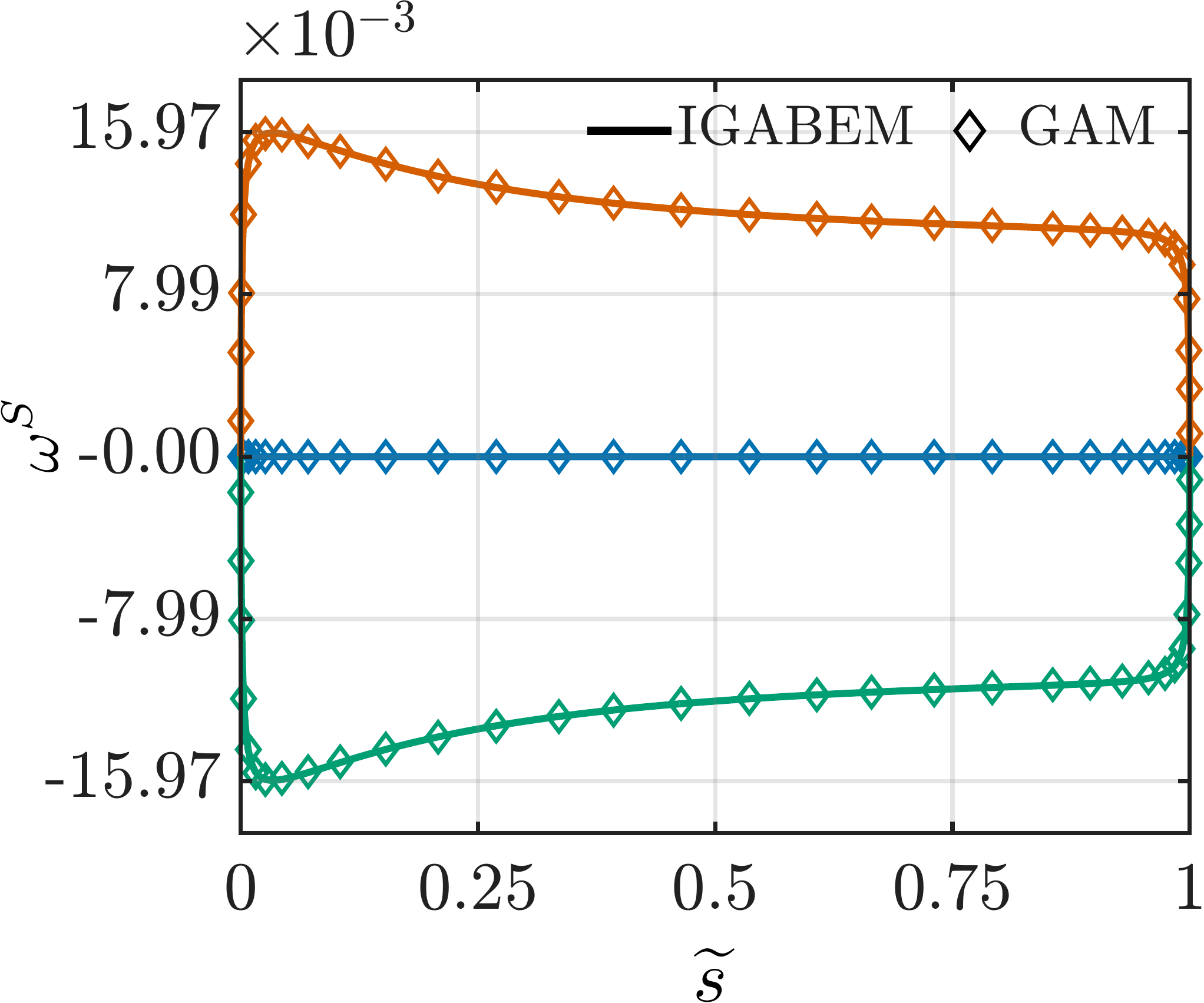}
        \caption{}
        \label{fig:radial_3}
    \end{subfigure}

    \caption{Validation of the IGABEM results (solid curves) against the GAM benchmark solutions (diamond markers) for three straight-surface arrangements. Top row: geometries of the (a) collinear, (b) parallel, and (c) radial arrangements. Middle row: distributions of \(\widetilde{\sigma}^{S}\) for (d) collinear, (e) parallel, and (f) radial arrangements. Bottom row: distributions of \(\omega^{S}\) for the (g) collinear, (h) parallel, and (i) radial arrangements. Within each column, the colors of the surface-field distributions correspond to those of the respective surfaces shown in the top-row geometry.}
    \label{fig:interaction_configurations}
\end{figure*}

As in ~\cite{Patil2024}, three representative surface arrangements are considered: collinear, parallel, and radial, as shown in Figs.~\ref{fig:collinear_1}--\ref{fig:collinear_3}. The material properties and loading conditions are chosen to match the benchmark solutions of ~\cite{Patil2024}. Thus, it is assumed that the matrix is made of epoxy, with $\mu = 2~\mathrm{GPa}$ and $\nu = 0.35$ and each surface $L^{(m)}$ is located along a straight segment of length $2l$ with $l=5~\mathrm{nm}$. The dimensionless surface parameters and the residual surface tension for $L^{(m)}$ are defined as
\begin{equation}
\widetilde{\sigma}^{S(m)}=\frac{\sigma^{S(m)}}{\mu l^{(m)}},
\qquad
\gamma^{(m)}
=
\frac{2\mu l^{(m)}}{2\mu_S^{(m)}+\lambda_S^{(m)}},
\qquad
\widetilde{\sigma}_{0}^{(m)}
=
\frac{\sigma_{0}^{(m)}}{\mu l^{(m)}} ,
\qquad m=1,2,3,
\label{eq:surface_parameters_each_straight_surface}
\end{equation}
while the dimensionless elastic stress fields in the matrix are introduced as
\begin{equation}
\widetilde{\sigma}_{ij}=\frac{\sigma_{ij}}{\mu},
\qquad m=1,2,3.
\label{eq:nondimensional_quantities_straight_validation}
\end{equation}

In the examples presented here, all three surfaces are assigned the following identical parameters:
\begin{equation}
\gamma^{(1)}=\gamma^{(2)}=\gamma^{(3)}=0.12,
\qquad
\widetilde{\sigma}_{0}^{(1)}
=
\widetilde{\sigma}_{0}^{(2)}
=
\widetilde{\sigma}_{0}^{(3)}
=
0.025, 
\label{eq:identical_surface_parameters_straight_validation}
\end{equation}
and the only nonzero normalized far-field stress component is
\begin{equation}
\widetilde{\sigma}^{\infty}_{22}=0.05 .
\label{eq:far_field_loading_straight_validation}
\end{equation}

For the collinear and parallel arrangements shown in Figs.~\ref{fig:collinear_1} and~\ref{fig:collinear_2}, respectively, the normalized separation distance is defined as
\begin{equation}
\widetilde{h}=\frac{h}{l},
\label{eq:normalized_spacing_straight}
\end{equation}
where \(h\) is the clear distance between the neighboring surfaces. Here,  we assume that
\begin{equation}
\widetilde{h}=0.25 .
\label{eq:h_straight_validation}
\end{equation}
For the arrangement shown in Fig.~\ref{fig:collinear_3}, the segments are arranged radially with respect to a prescribed center point, and the normalized radial distance between that point and the midpoint of the \(j\)th surface is denoted as
\begin{equation}
\widetilde{r}_{j}=\frac{r_j}{l},
\qquad j=1,2,3.
\label{eq:normalized_radial_distance_straight}
\end{equation}
In this case, we choose $\widetilde{r}_{j}$ as
\begin{equation}
\widetilde{r}_{j}=\widetilde{r}=1.25,
\qquad j=1,2,3 .
\label{eq:radial_distance_straight_validation}
\end{equation}
The angle between adjacent surfaces in the radial arrangement is defined as
\begin{equation}
\phi = \frac{2\pi}{N},
\label{eq:radial_angular_spacing}
\end{equation}
where \(N\) is the number of surfaces. Here, we assume that \(N=3\), and hence \(\phi = 2\pi/3\).

The geometries of the three surface arrangements and the plots of the corresponding distributions of \(\widetilde{\sigma}^{S}\) and \(\omega^{S}\) are presented in Fig.~\ref{fig:interaction_configurations}. The solid curves denote the IGABEM results, while the diamond markers denote the benchmark solutions obtained with global approximations and identified as GAM. As the results for $L^{(2)}$ and $L^{(3)}$  of Figs.~\ref{fig:collinear_2}, \ref{fig:collinear_3} are identical due to symmetry, the corresponding plots of Figs.~\ref{fig:parallel_2}, \ref{fig:parallel_3} overlap.  Excellent agreement is observed between the two methods for all three arrangements and for all surfaces. 

To isolate the effect of the orientation of the surfaces, \(L^{(1)}\) is kept fixed and horizontal in all three configurations, while \(L^{(2)}\) and \(L^{(3)}\) are repositioned from the collinear arrangement to the parallel and radial arrangements. The resulting changes in the fields on \(L^{(1)}\) therefore reflect only the interaction with the two surrounding surfaces. 

As shown in Figs.~\ref{fig:parallel_1}, \ref{fig:parallel_2}, and~\ref{fig:parallel_3},
\(L^{(1)}\) develops the largest compressive surface stress \(\widetilde{\sigma}^{S}\) in the collinear arrangement. When \(L^{(2)}\) and \(L^{(3)}\) are moved to the parallel arrangement, the magnitude of the compressive stress \(\widetilde{\sigma}^{S}\) on \(L^{(1)}\) decreases substantially.
In the radial arrangement, the compressive stress \(\widetilde{\sigma}^{S}\) on \(L^{(1)}\) increases, compared to the case of the parallel arrangement, but remains smaller than that obtained for the collinear arrangement. Additionally, the inclined arrangement of \(L^{(2)}\) and \(L^{(3)}\) breaks the symmetry of the \(\widetilde{\sigma}^{S}\) distribution on \(L^{(1)}\) about its midpoint. As observed in Figs.~\ref{fig:radial_1},
\ref{fig:radial_2} and~\ref{fig:radial_3}, \(\omega^{S}\) on \(L^{(1)}\) remains zero in all three arrangements.

The fields in \(L^{(2)}\) and \(L^{(3)}\) show a pronounced dependence on the relative configuration of the surfaces. In the
collinear arrangement, all three surfaces develop compressive
distributions of \(\widetilde{\sigma}^{S}\), while
\(\omega^{S}\) vanishes identically. In parallel arrangement, the outer surfaces \(L^{(2)}\) and \(L^{(3)}\) also develop compressive \(\widetilde{\sigma}^{S}\) distributions, although their magnitudes are smaller than those obtained in the collinear arrangement. The respective values of \(\omega^{S}\) are nonzero, equal in magnitude and opposite in sign. Moreover, the values  of \(\omega^{S}\) change sign at the midpoint of each surface, leading to antisymmetric distributions. This behavior reflects the opposite rotational responses induced by the symmetric placement of \(L^{(2)}\) and \(L^{(3)}\) above and below \(L^{(1)}\).

In the radial arrangement, the inclined segments \(L^{(2)}\) and \(L^{(3)}\) develop tensile distributions of \(\widetilde{\sigma}^{S}\), in contrast to the compressive response of \(L^{(1)}\) and to the compressive distributions observed at all three surfaces in the collinear and parallel arrangements. The corresponding values of \(\omega^{S}\) again have opposite signs; however, unlike in the parallel arrangement, they retain their respective signs over the entire lengths of the inclined surfaces and do not change sign at the midpoints. This qualitative change highlights the sensitivity of the \(\omega^{S}\) response to the relative orientation of the surfaces. These results show that the distributions of \(\widetilde{\sigma}^{S}\) and \(\omega^{S}\) are strongly affected by both the relative positions and orientations of the neighboring surfaces.

Using the procedure described in Section~\ref{Section:SIF}, the normalized
SIFs are evaluated at the tips of each material surface for all three arrangements. For the problems of straight-surfaces considered here, the relative locations of the interpolation points, which are described in Section~\ref{Section:SIF}, within each tip-adjacent element are specified by
\[
f_j=
\left\{
\frac{1}{13},\,
\frac{1}{2},\,
\frac{12}{13}
\right\}.
\]

\begin{table}[!htbp]
    \centering
    \caption{Comparison of the normalized SIFs
    (\(\times 10^{-2}\)) at the tip \(\eta = a\) of the three straight
    surfaces for the collinear and parallel arrangements.}
    \label{tab:straight_surface_validation}
    \begin{tabular}{cccccc}
        \toprule
        & & \multicolumn{2}{c}{IGABEM}
        & \multicolumn{2}{c}{Ref.~\cite{Patil2024}} \\
        \cmidrule(lr){3-4}
        \cmidrule(lr){5-6}
        Arrangement
        & Surface
        & \(K_{1}^{(m,\eta)}\)
        & \(K_{2}^{(m,\eta)}\)
        & \(K_{1}^{(m,\eta)}\)
        & \(K_{2}^{(m,\eta)}\) \\
        \midrule

        \multirow{3}{*}{{Fig.~\ref{fig:collinear_1}}}
        & \(L^{(1)}\) & 0.64 & 0.00 & 0.64 & 0.00 \\
        & \(L^{(2)}\) & 0.63 & 0.00 & 0.62 & 0.00 \\
        & \(L^{(3)}\) & 0.49 & 0.00 & 0.48 & 0.00 \\

        \midrule

        \multirow{3}{*}{{Fig.~\ref{fig:collinear_2}}}
        & \(L^{(1)}\) & 0.29 & 0.00 & 0.28 & 0.00 \\
        & \(L^{(2)}\) & 0.33 & 0.01 & 0.33 & 0.01 \\
        & \(L^{(3)}\) & 0.33 & -0.01 & 0.33 & -0.01 \\
        
        \bottomrule
    \end{tabular}
\end{table}
For the collinear configuration, the symmetry of the geometry, material, and loading leads to a symmetric Mode-I response, while the Mode-II normalized SIFs vanish at all tips. For parallel configuration, symmetry-related tips have identical normalized Mode-I SIFs, whereas the corresponding Mode-II SIFs are equal in magnitude and opposite in sign.
Consequently, for collinear and parallel arrangements, only one symmetry-independent set of normalized SIFs is reported in Table~\ref{tab:straight_surface_validation}, where the IGABEM results (presented here with more significant digits) are compared with the corresponding GAM results from ~\cite{Patil2024}.

\begin{table*}[!htbp]
    \centering
    \caption{Comparison of the normalized SIFs
    (\(\times10^{-2}\)) at both tips of the three straight surfaces for the radial arrangement of Fig.~\ref{fig:collinear_3}.}
    \label{tab:SIF_radial_comparison}

    \setlength{\tabcolsep}{5pt}
    \begin{tabular}{ccccccccc}
        \toprule
        \multirow{2}{*}{Surface}
        & \multicolumn{4}{c}{IGABEM}
        & \multicolumn{4}{c}{Ref.~\cite{Patil2024}} \\
        \cmidrule(lr){2-5}
        \cmidrule(lr){6-9}

        & \(K_1^{(m,a)}\)
        & \(K_2^{(m,a)}\)
        & \(K_1^{(m,b)}\)
        & \(K_2^{(m,b)}\)
        & \(K_1^{(m,a)}\)
        & \(K_2^{(m,a)}\)
        & \(K_1^{(m,b)}\)
        & \(K_2^{(m,b)}\) \\
        \midrule

        \(L^{(1)}\)
        &  0.25 &  0.00 &  0.37 &  0.00
        &  0.24   &  0.00   & 0.36      & 0.00 \\

        \(L^{(2)}\)
        & -0.75 & -0.08 & -0.71 & -0.05
        & -0.74   & -0.08   & -0.70      & -0.05 \\

        \(L^{(3)}\)
        & -0.75 &  0.08 & -0.71 &  0.05
        & -0.74   &  0.08   & -0.70      & 0.05 \\

        \bottomrule
    \end{tabular}
\end{table*}
For the radial arrangement, the symmetry relates the two inclined surfaces, \(L^{(2)}\) and \(L^{(3)}\), but the two tips of an individual surface are not symmetry-equivalent. Therefore, the normalized SIFs at both tips of all three surfaces are reported in Table~\ref{tab:SIF_radial_comparison} and compared with the corresponding GAM results from ~\cite{Patil2024}. In this configuration, the normalized Mode-II SIFs at the tips of surface \(L^{(1)}\), which lies along the symmetry axis, are practically
zero, while those associated with the corresponding tips of \(L^{(2)}\) and \(L^{(3)}\) have approximately equal magnitudes and opposite signs.

As can be seen from the tables, the results obtained with IGABEM agree well with GAM results reported in \cite{Patil2024}.

\subsection{Validation for the case of surfaces along circular arcs}
\label{subsec:validation_circular_arcs}

\begin{figure}[htb]
	\centering
	\includegraphics[width=0.5\textwidth]{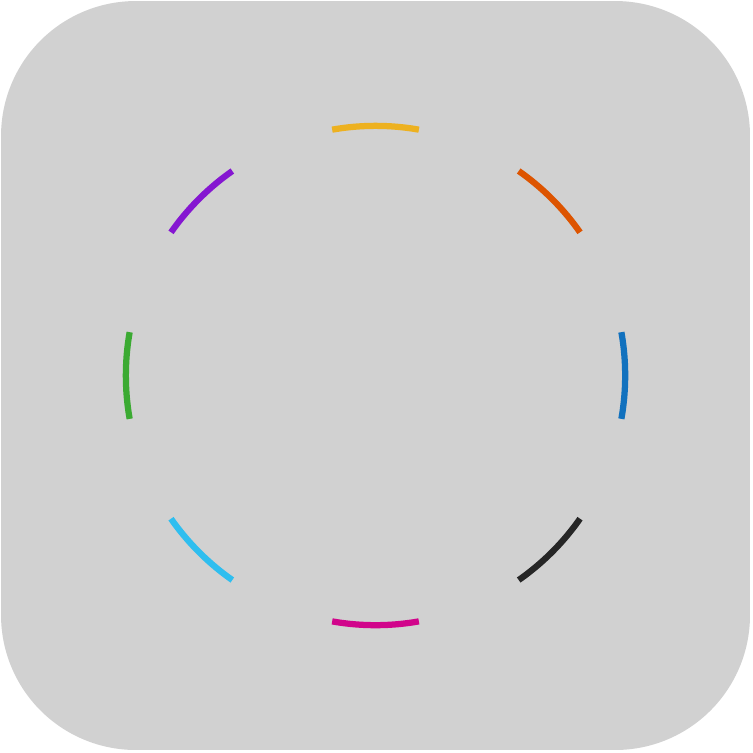}
    \caption{Arrangement of eight circular arcs along the same circle. The colors distinguish the individual surfaces and are used consistently in the corresponding surface-field distributions shown in Fig.~\ref{Fig:6}.}
    \label{Fig:5}
\end{figure}

For validation, we use the benchmark solutions reported ~\cite{Han2024_JAM}. The problem's geometry consists of eight circular-arc surfaces \(L^{(m)}\), \(m=1,\ldots,8\), symmetrically located along the same circle and embedded into an infinite isotropic elastic matrix, as shown in Fig.~\ref{Fig:5}. The epoxy matrix is again characterized by $\mu = 2~\mathrm{GPa}$ and $\nu=0.35$ and it is assumed that each arc has the same angular span
\begin{equation}
\theta_m = \frac{20\pi}{180},
\qquad m=1,\ldots,8 .
\label{eq:circular_arc_angle_validation}
\end{equation}

Also, as in the benchmark solution, the only nonzero normalized far-field stress component is chosen to be
\begin{equation}
\widetilde{\sigma}^{\infty}_{11}=0.1,
\label{eq:far_field_loading_circular_validation}
\end{equation}
and for each surface $L^{(m)}$, the dimensionless residual surface tension is taken to be
\begin{equation}
\widetilde{\sigma}_{0}^{(m)}=0,
\qquad m=1,\ldots,8,
\label{eq:residual_surface_tension_circular_validation}
\end{equation}
while the dimensionless surface stiffness parameter is prescribed as
\begin{equation}
\gamma^{(m)}=0.12,
\qquad m=1,\ldots,8 .
\label{eq:gamma_circular_validation}
\end{equation}
For the $q$-th surface, arc length is normalized with respect to the semi-arc length as
\begin{equation}
s_q=\frac{R_q\theta_q}{2},
\label{eq:semi_arc_length_circular_validation}
\end{equation}
where $R_q$ is the radius of the $q$-th arc and $\theta_q$ is its angular span defined as
\begin{equation}
\theta_q=\beta_2^{(q)}-\beta_1^{(q)},
\label{eq:theta_beta_circular_validation}
\end{equation}
where $\beta$ denotes the angle between the outward unit normal to the arc and the $x_1$-axis. Additionally, the following dimensionless quantities are introduced:
\begin{equation}
\gamma^{(q)}
=
\frac{\mu R_q \theta_q}{2\mu_S^{(q)}+\lambda_S^{(q)}},
\qquad
\widetilde{\sigma}^{S(q)}
=
\frac{2\sigma^{S(q)}}{\mu R_q \theta_q},
\qquad
\widetilde{\sigma}_{0}^{(q)}
=
\frac{2\sigma_{0}^{(q)}}{\mu R_q \theta_q},
\qquad
\widetilde{\sigma}_{ij}
=
\frac{\sigma_{ij}}{\mu}.
\label{eq:nondim_circular_arc_validation}
\end{equation}

The IGABEM solution is validated against the corresponding GAM solution presented in ~\cite{Han2024_JAM} by comparing the
distributions of the normalized \(\widetilde{\sigma}^{S}\) and \(\omega^{S}\) along the material surfaces. Owing to the symmetry of the geometry and loading, the results are presented only over the angular interval
\begin{equation}
\alpha\in[0,\pi].
\label{eq:angular_interval_circular_validation}
\end{equation}

\begin{figure}[H]
  \centering
  \begin{subfigure}[b]{0.49\textwidth}
      \centering
      \includegraphics[angle=0, width=1\textwidth]{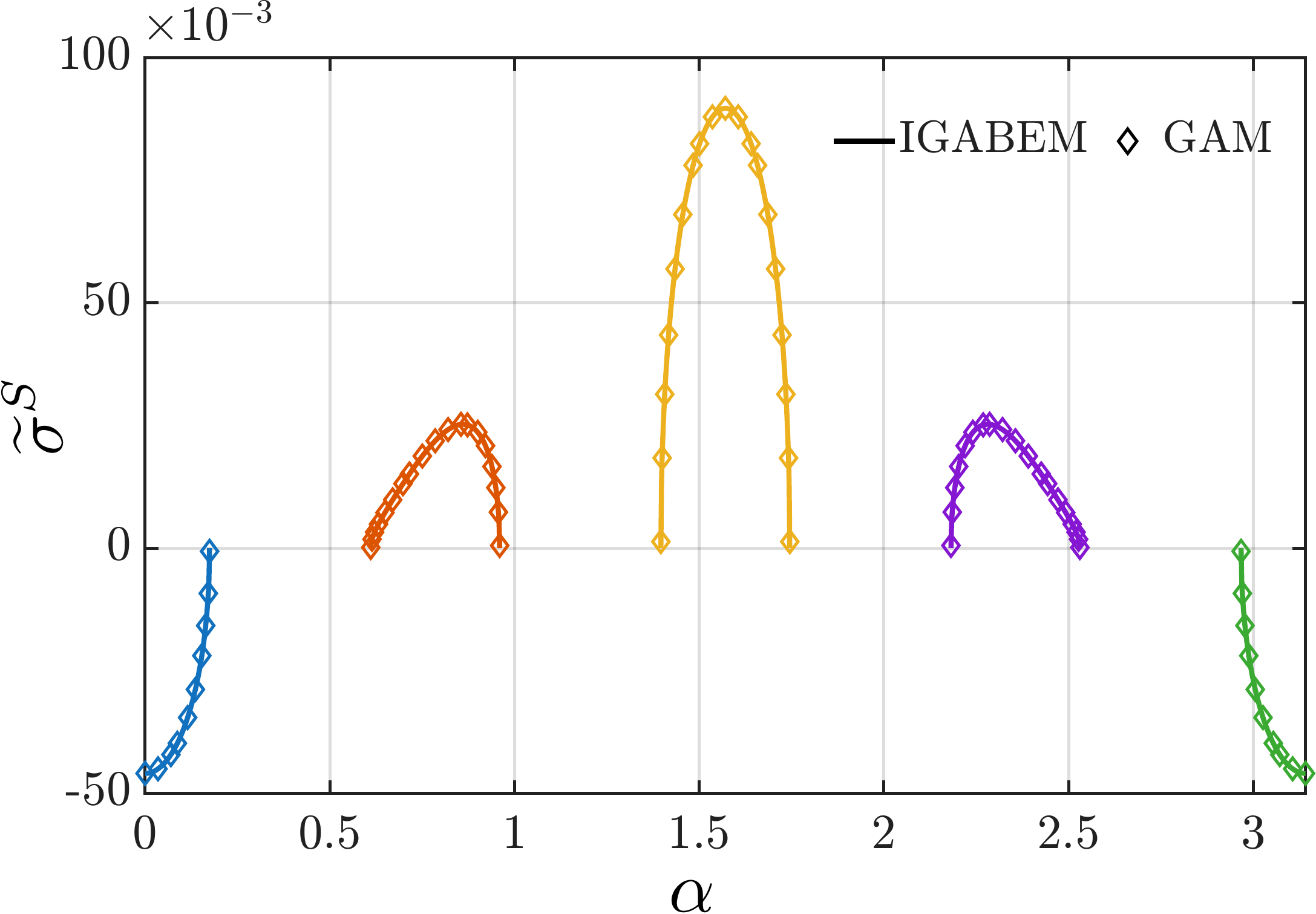}
      \subcaption{}
      \label{Fig:6a}
  \end{subfigure}
  \hfill
  \begin{subfigure}[b]{0.49\textwidth}
      \centering
      \includegraphics[angle=0, width=1\textwidth]{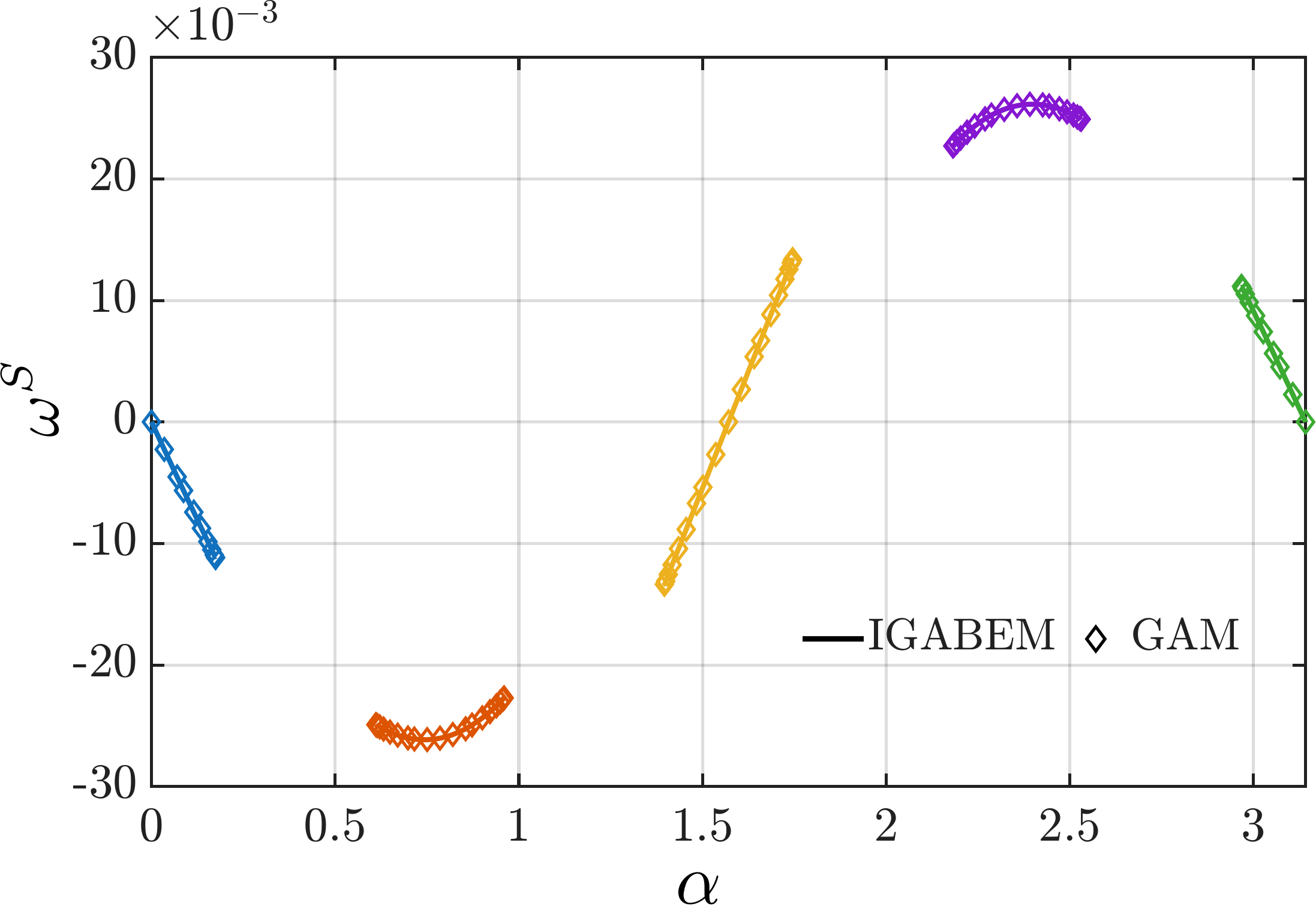}
      \subcaption{}
      \label{Fig:6b}
  \end{subfigure}
  \caption{Comparison of the IGABEM solutions (solid curves) and GAM solutions (diamond markers) for the eight circular arcs arrangement over the angular interval \(\alpha\in[0,\pi]\): distributions of (a) \(\widetilde{\sigma}^{S}\) and (b) \(\omega^{S}\). The color of each surface-field distribution corresponds to that of the respective circular surface shown in Fig.~\ref{Fig:5}.}
  \label{Fig:6}
\end{figure}

The comparison shown in Fig.~\ref{Fig:6} demonstrates close agreement between the IGABEM and GAM solutions over the interval, confirming the accuracy of the proposed formulation for interacting circular arc surfaces.

We also evaluated normalized SIFs at the tips of all circular-arc surfaces. For the circular-arcs problem, the interpolation points are defined as described in Section~\ref{Section:SIF}, with the relative locations within the tip-adjacent element specified by
\[
f_j=
\left\{
\frac{1}{10},\,
\frac{1}{2},\,
\frac{9}{10}
\right\}.
\]

\begin{table*}[!htbp]
    \centering
    \caption{Comparison of the normalized Mode-I SIFs
    (\(\times10^{-2}\)) at the eight tips of the four circular-arc
    material surfaces.}
    \label{tab:SIF_circular_arcs_comparison}

    \setlength{\tabcolsep}{7pt}
    \begin{tabular}{ccccc}
        \toprule
        \multirow{2}{*}{Surface}
        & \multicolumn{2}{c}{IGABEM}
        & \multicolumn{2}{c}{Ref.~\cite{Han2024_JAM}} \\
        \cmidrule(lr){2-3}
        \cmidrule(lr){4-5}

        & \(K_1^{(m,a)}\)
        & \(K_1^{(m,b)}\)
        & \(K_1^{(m,a)}\)
        & \(K_1^{(m,b)}\) \\
        \midrule

        \(L^{(1)}\)
        &  1.05 & -1.05
        &  1.05 & -1.05 \\

        \(L^{(2)}\)
        & -0.20 &  0.84
        & -0.19 &  0.84 \\

        \(L^{(3)}\)
        & -2.10 &  2.10
        & -2.10 &  2.10 \\

        \(L^{(4)}\)
        & -0.84 &  0.20
        & -0.84 &  0.19 \\

        \bottomrule
    \end{tabular}
\end{table*}
Due to the symmetry of the configuration, Table~\ref{tab:SIF_circular_arcs_comparison} reports the normalized Mode-I SIFs at the both tips of only the first four circular arcs, together with the corresponding GAM results reported in ~\cite{Han2024_JAM}. Due to the geometric and material symmetry of the configuration under the prescribed loading, the normalized Mode-II SIFs vanish at all tips. Consequently, the latter values are not presented in Table~\ref{tab:SIF_circular_arcs_comparison}.
The agreement of the normalized SIFs with the reference results validates the developed IGABEM formulation and the interpolation-based procedure employed to extract the SIFs for multiple curved material surfaces.

\section{Numerical Results}\label{Section:7}

The numerical results of this Section were designed to demonstrate capability of the present approach to (i) properly model surfaces with mixed curvatures, (ii) study various effects of interactions of multiple curved surfaces, (iii) model materials reinforced by wrinkled multilayer graphene sheets.

\subsection{Material surfaces of mixed curvatures}\label{Sub-Section:7.1}
\begin{figure*}[!htbp]
    \centering
    \captionsetup[subfigure]{skip=2pt}

    \begin{subfigure}[t]{0.235\textwidth}
        \centering
        \includegraphics[width=\linewidth]
        {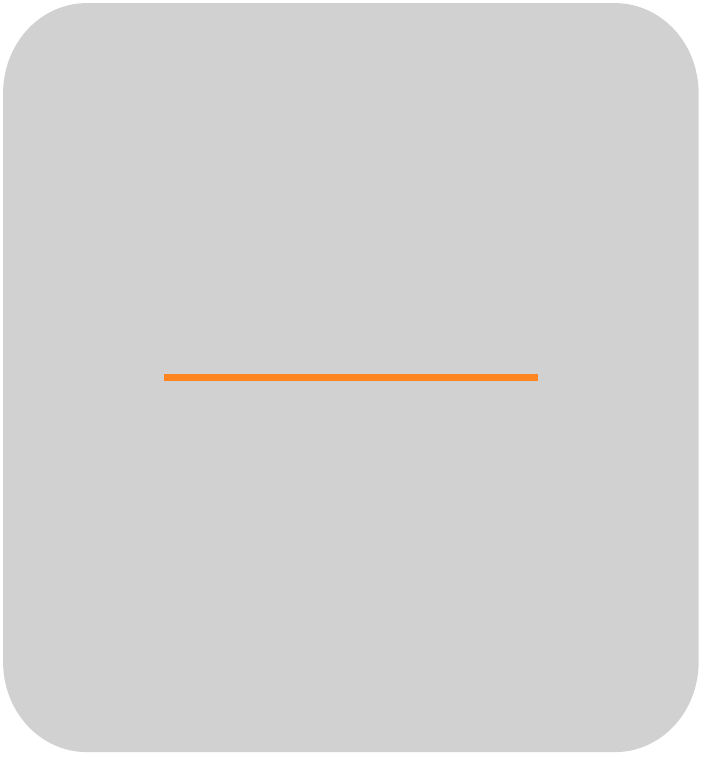}
        \caption{}
        \label{fig:geometry_1}
    \end{subfigure}
    \hfill
    \begin{subfigure}[t]{0.235\textwidth}
        \centering
        \includegraphics[width=\linewidth]
        {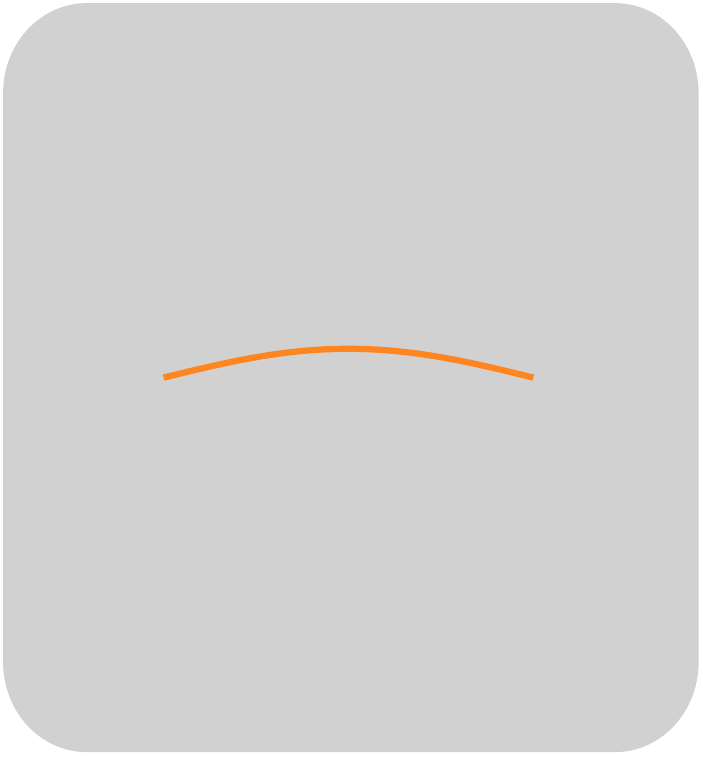}
        \caption{}
        \label{fig:geometry_2}
    \end{subfigure}
    \hfill
    \begin{subfigure}[t]{0.235\textwidth}
        \centering
        \includegraphics[width=\linewidth]
        {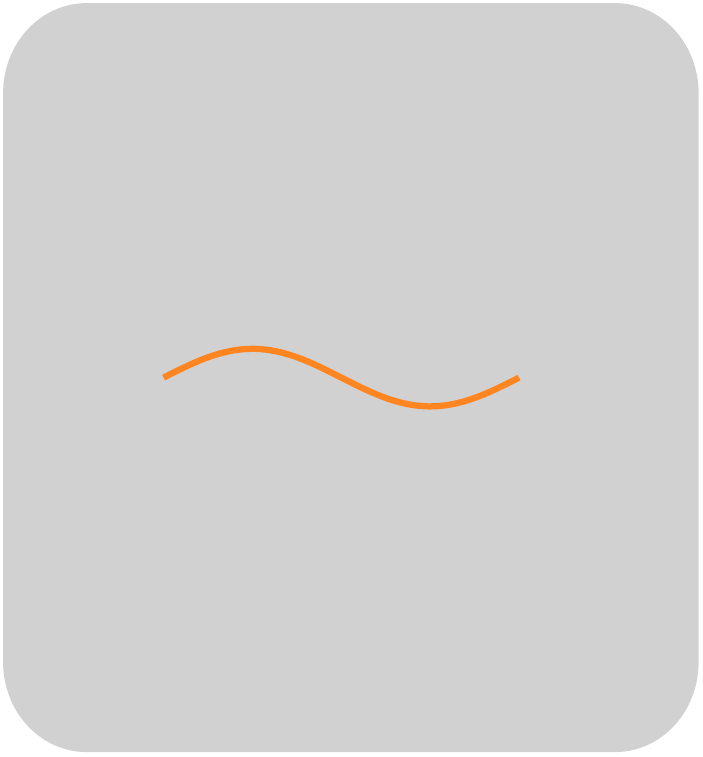}
        \caption{}
        \label{fig:geometry_3}
    \end{subfigure}
    \hfill
    \begin{subfigure}[t]{0.235\textwidth}
        \centering
        \includegraphics[width=\linewidth]
        {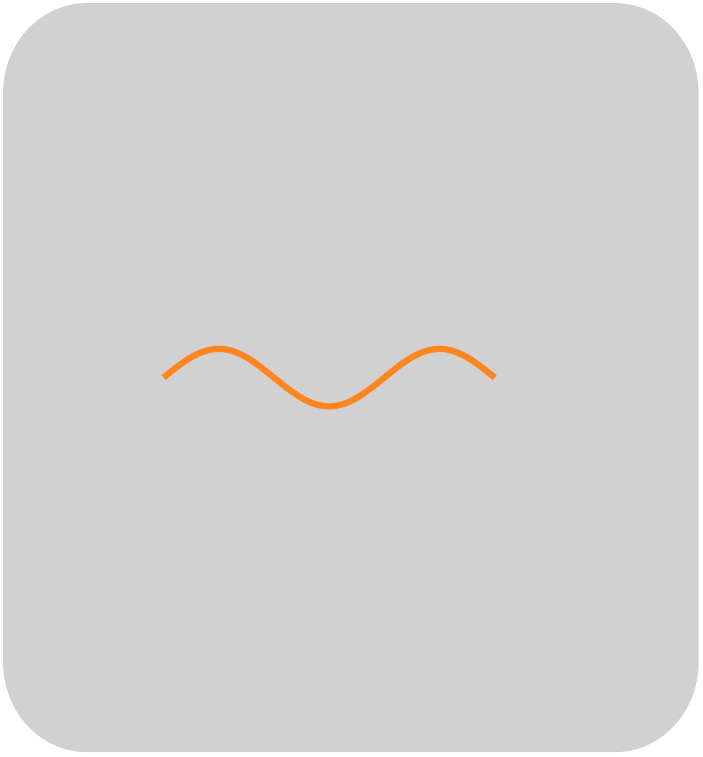}
        \caption{}
        \label{fig:geometry_4}
    \end{subfigure}

    \caption{
        Representative surface geometries: (a) straight surface; and sinusoidal surfaces - (b) without an inflection point, (c) with one inflection point, and
        (d) with two inflection points.
    }
    \label{fig:curvature_configurations}
\end{figure*}

To demonstrate that the proposed approach can properly handle surfaces of mixed curvatures and for the sake of comparison, the following four representative surface geometries are considered: (a) a straight surface, (b) a curved surface without inflection points, (c) a surface with one inflection point, and (d) a surface with two inflection points, see Fig.~\ref{fig:curvature_configurations}. The geometry of the \(n\)-th surface is non-dimensionalized with respect to the common half-arc length \(\ell\) and described as
\begin{equation}
    \widetilde{y}_n(\widetilde{x})
    =
    \widetilde{A}
    \sin\left[
        n\pi
        \left(
            \frac{\widetilde{x}}{\widetilde{d}_n}
            +\frac{1}{2}
        \right)
    \right],
    \qquad
    -\frac{\widetilde{d}_n}{2}
    \leq
    \widetilde{x}
    \leq
    \frac{\widetilde{d}_n}{2},
    \qquad
    n=0,1,2,3,
    \label{eq:surface_geometry}
\end{equation}
where \(\widetilde{x}=x/\ell\) and \(\widetilde{y}_n=y_n/\ell\) are the dimensionless coordinates, \(\widetilde{A}=A/\ell\) is the dimensionless amplitude, and \(\widetilde{d}_n=d_n/\ell\) is the dimensionless horizontal distance between the two tips. The values \(n=0,1,2,\) and \(3\) correspond to the surfaces located along  (a) a straight line and sinusoidal curves (b), (c), (d), respectively.

All four surfaces are assumed to have the same arc length \(2\ell\) with \(d_n\) chosen accordingly to satisfy this constraint.  For the \(n\)-th surface, dimensionless surface characteristics \(\widetilde{\sigma}^{S}_{n}\), \(\gamma\), \(\widetilde{\sigma}_{0}\) are introduced (similarly as in Eq.~\eqref{eq:surface_parameters_each_straight_surface}) as
\begin{equation}
    \widetilde{\sigma}^{S}_{n}
    =
    \frac{\sigma^{S}_{n}}{\mu\ell},
    \qquad
    \gamma
    =
    \frac{2\mu\ell}{2\mu_S+\lambda_S},
    \qquad
    \widetilde{\sigma}_{0}
    =
    \frac{\sigma_{0}}{\mu\ell}.
   \label{eq:surface_parameters_inflection_study}
\end{equation}
The corresponding Cauchy stress components in the surrounding matrix are normalized by the bulk shear modulus as in
Eq.~\eqref{eq:nondimensional_quantities_straight_validation}.

For this study, again we choose the epoxy matrix that is characterized by
\(\mu=2~\mathrm{GPa}\) and \(\nu=0.35\). The surface parameters (the same for all four surfaces) are set as \(\gamma=0.12\) and \(\widetilde{\sigma}_{0}=0.025\). The matrix is subjected to uniform biaxial far-field loading, with dimensionless components.
\begin{equation}
    \widetilde{\sigma}^{\infty}_{11}
    =
    \widetilde{\sigma}^{\infty}_{22}
    =
    0.025,
    \qquad
    \widetilde{\sigma}^{\infty}_{12}=0.
    \label{eq:remote_loading_inflection_study}
\end{equation}

\begin{figure*}[t]
    \centering
    \captionsetup[subfigure]{skip=2pt}

    \begin{subfigure}[t]{0.495\textwidth}
        \centering
        \includegraphics[width=\linewidth]
        {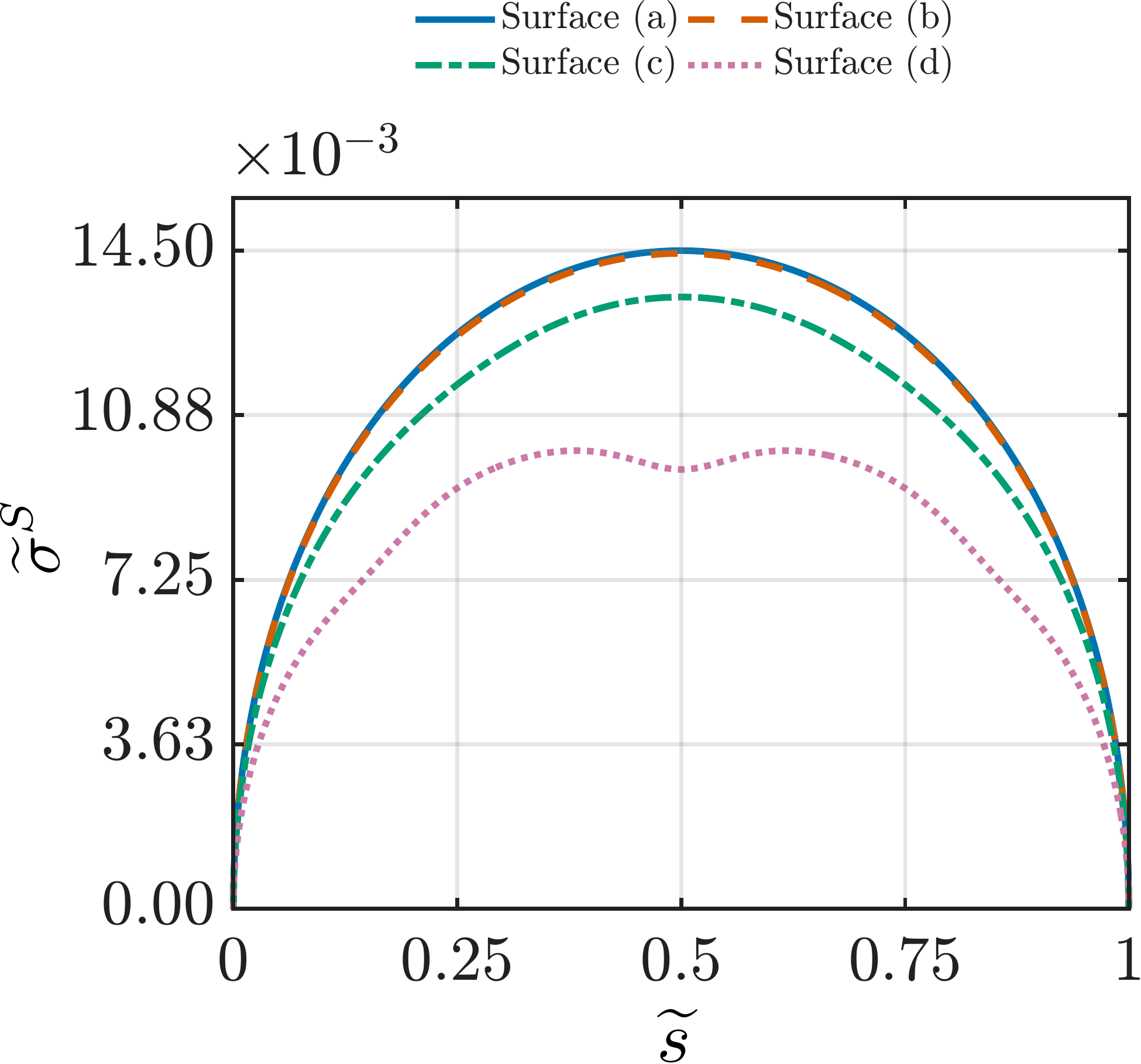}
        \caption{}
        \label{fig:sigmaS_combined}
    \end{subfigure}
    \hfill
    \begin{subfigure}[t]{0.495\textwidth}
        \centering
        \includegraphics[width=\linewidth]
        {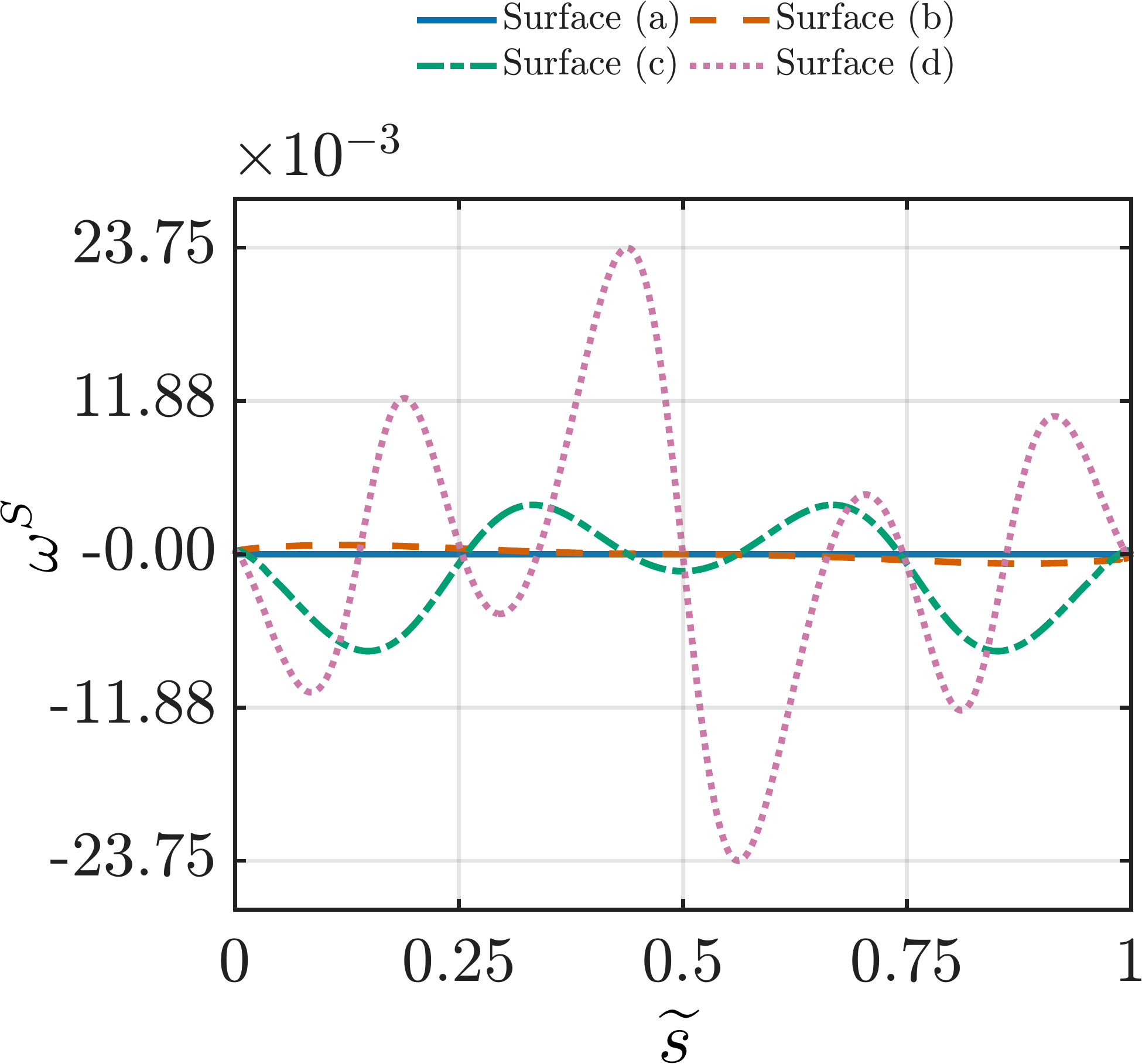}
        \caption{}
        \label{fig:omegaS_combined}
    \end{subfigure}

    \caption{
        Effect of surface geometry on the surface fields:
         distributions of (a)
        \(\widetilde{\sigma}^{S}\); and
        (b) \(\omega^{S}\)
        for the four representative surface geometries shown in
        Fig.~\ref{fig:curvature_configurations}.
    }
    \label{fig:curvature_surface_fields}
\end{figure*}

In Fig.~\ref{fig:curvature_surface_fields}, the distributions of \(\widetilde{\sigma}^{S}\)
and \(\omega^{S}\) are presented for the four surface geometries. As can be seen in Fig.~\ref{fig:sigmaS_combined}, the distributions of
\(\widetilde{\sigma}^{S}\) remain positive and vary smoothly along the surface in all four cases. The distributions for the surfaces (a)  \((n = 0)\) and (b) \((n = 1)\) , which do not contain an interior inflection point, have single extrema, with maximum values of
approximately \(14.5\times10^{-3}\). When the curvature changes sign once, i.e., surface (b)  \((n = 2)\), the distribution of \(\widetilde{\sigma}^{S}\) still has a single extrema point, with the maximum occurring near the central inflection point and decreasing slightly to
approximately \(13.0\times10^{-3}\). When the curvature changes sign twice, i.e., surface (c) \((n = 3)\), three local extrema develop, two of which near the corresponding inflection
points, and the maximum value decreases to approximately \(10.0\times10^{-3}\). Thus, the number and location of the inflection points directly influence the distribution of \(\widetilde{\sigma}^{S}\). It is also worth noting that the values of $\widetilde{\sigma}^{S}$  along all surfaces are smaller than that of surface tension, \(\widetilde{\sigma}_0=0.025\), i.e, the effect of the latter is reduced.

The distributions of \(\omega^{S}\) for the corresponding surfaces are shown in
Fig.~\ref{fig:omegaS_combined}. Under the prescribed loading, \(\omega^{S}\) vanishes along the surface (a); however, distributions of \(\omega^{S}\) for the curved surfaces (b)-(d) do not vanish. The maximum absolute value of \(\omega^{S}\) increases from approximately \(0.7\times10^{-3}\) for the surface without an inflection point to \(7.5\times10^{-3}\) and \(24\times10^{-3}\) for the surfaces with one and two inflection points, respectively. Thus, distribution of \(\omega^{S}\)  is highly sensitive to variations in curvatures and its behavior becomes more oscillatory with the increase in number of inflection points.


\begin{figure*}[t]
    \centering
    \includegraphics[width=1\textwidth]
    {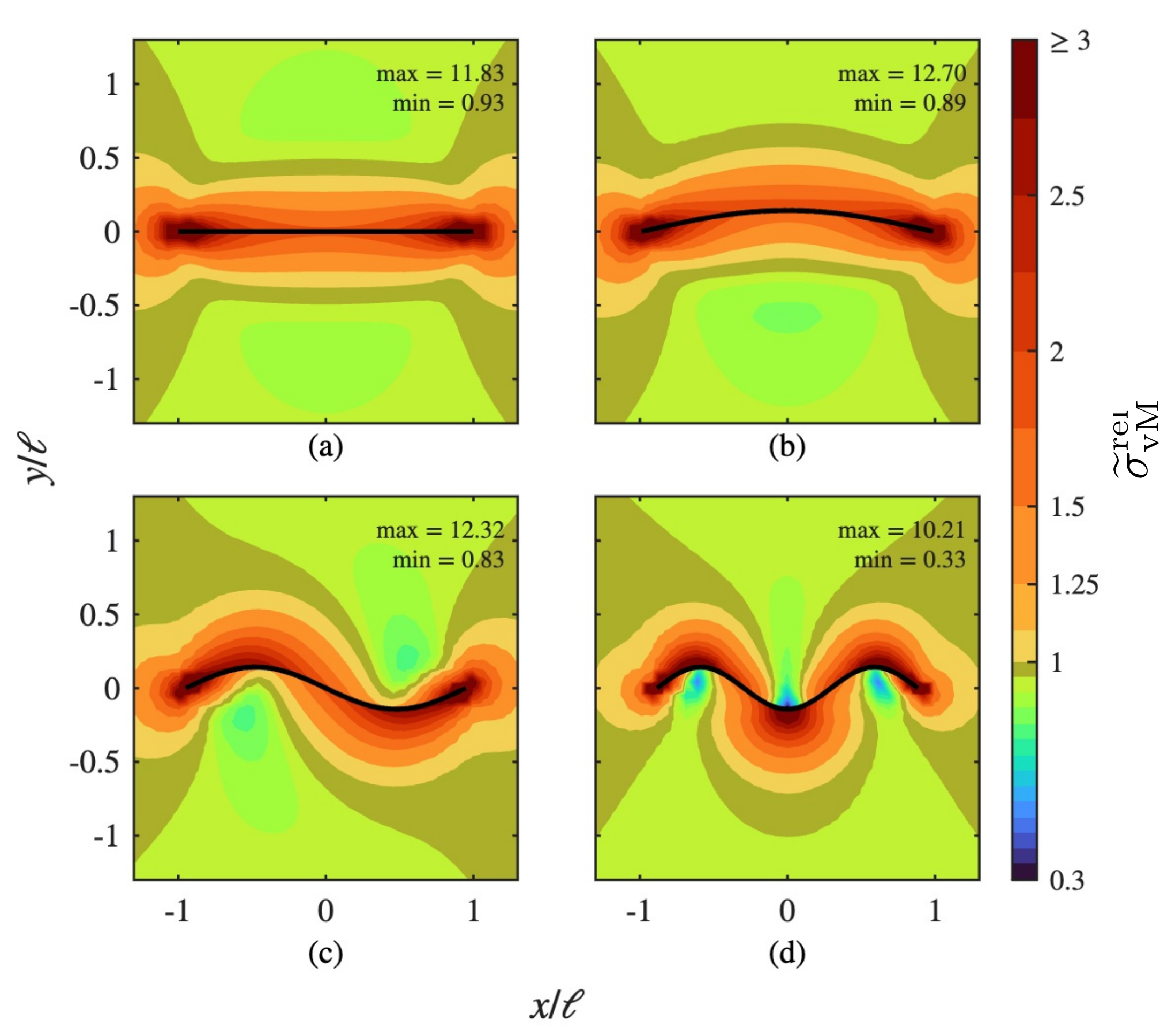}
    \caption{
        Distribution of the dimensionless relative von Mises stresses in the matrix surrounding the surfaces under study.
    }
    \label{fig:vm_ns_sin_curves}
\end{figure*}

The combination of decreasing values of \(\widetilde{\sigma}^{S}\) and increasing those of \(\omega^{S}\) with the increase in the number of inflection points indicates that the load transfer mechanism is profoundly affected  by the curvatures of the surfaces.

Furthermore, to better illustrate the mechanism of load redistributions in the matrix,  Fig.~\ref{fig:vm_ns_sin_curves} presents the contour plots of the dimensionless relative von Mises stress defined as
\begin{equation}
    \widetilde{\sigma}_{\mathrm{vM}}^{\mathrm{rel}}(\mathbf{x})
    =
    \frac{
        \widetilde{\sigma}_{\mathrm{vM}}(\mathbf{x})
    }{
        \widetilde{\sigma}_{\mathrm{vM}}^{\infty}
    },
    \label{eq:relative_von_mises_stress}
\end{equation}
where
\begin{equation}
\begin{aligned}
    \widetilde{\sigma}_{\mathrm{vM}}
    =
    \Bigg[
        &\frac{1}{2}
        \left(
            \widetilde{\sigma}_{11}
            -
            \widetilde{\sigma}_{22}
        \right)^2
        +
        \frac{1}{2}
        \left(
            \widetilde{\sigma}_{22}
            -
            \widetilde{\sigma}_{33}
        \right)^2
        \\
        &+
        \frac{1}{2}
        \left(
            \widetilde{\sigma}_{33}
            -
            \widetilde{\sigma}_{11}
        \right)^2
        +
        3(\widetilde{\sigma}_{12})^{\,2}
    \Bigg]^{1/2},
    \qquad
    \widetilde{\sigma}_{33}
    =
    \nu
    \left(
        \widetilde{\sigma}_{11}
        +
        \widetilde{\sigma}_{22}
    \right).
    \label{eq:von_mises_stress_plane_strain}
\end{aligned}
\end{equation}
and \(\widetilde{\sigma}_{\mathrm{vM}}^{\infty}\) is the von Mises stress in the homogeneous matrix subjected to the same far-field loading conditions \(\widetilde{\sigma}_{11}^{\infty}
=\widetilde{\sigma}_{22}^{\infty}=0.025\),
\(\widetilde{\sigma}_{12}^{\infty}=0\), and Poisson ratio \(\nu=0.35\), i.e.,
\[
    \widetilde{\sigma}_{\mathrm{vM}}^{\infty}=0.0075.
\]
Thus,
\(\widetilde{\sigma}_{\mathrm{vM}}^{\mathrm{rel}}=1\) identify the homogeneous matrix. The \(\widetilde{\sigma}_{\mathrm{vM}}^{\mathrm{rel}} > 1\) identify
stress-amplification zones, while the \(\widetilde{\sigma}_{\mathrm{vM}}^{\mathrm{rel}} < 1\) identify stress-relaxation zones. 

For all four surfaces, the relative von Mises stress approaches unity with increasing distance from the material surface, indicating recovery of the homogeneous matrix response under the prescribed far-field loading.

For the straight surface (a), the amplification zone forms a continuous
band along the surface, with strong localization near its tips. Two
broad relaxation regions develop symmetrically above and below the
surface. The rapid decrease of
\(\widetilde{\sigma}^S\) to zero at the tips produces the strong
tip-localized amplification, while the neighboring relaxation regions
reflect the corresponding self-equilibrated stress redistribution.

For the surface without an inflection point (b), the amplification band
follows the curved profile, and the field remains symmetric about the
vertical centerline. However, the distributions on the two sides of the
surface differ: a pronounced relaxation region develops on its concave side, while the magnitude of relaxation on the opposite side decreases. The minimum relative von Mises stress decreases from \(0.93\) for surface
(a) to \(0.893\) for surface (b), corresponding to increases in the
maximum relaxation from \(7.2\%\) to \(10.7\%\). Although the two
surfaces exhibit similar distributions of
\(\widetilde{\sigma}^S\), the curvature-induced normal load transfer in
surface (b) produces a different matrix response and increases the
maximum relative von Mises stress from \(11.95\) to \(12.91\).

For the surface with one inflection point (c), the amplification and
relaxation regions move to the regions near the inflection points. The
curvature-dependent normal load transfer reverses direction between the
two curved segments, producing relaxation on one side and amplification
on the other. Load transfer continues through the inflection point
because gradients of \(\widetilde{\sigma}^S\) and \(\omega^S\) remain nonzero there.
The minimum relative von Mises stress is approximately \(0.830\),
corresponding to a maximum relaxation of about \(17.0\%\), whereas the
maximum reaches \(13.39\), the largest among the four configurations.

For the surface with two inflection points (d), successive curvature
reversals generate several alternating amplification and relaxation
regions. Unlike the broad relaxation regions in the preceding cases,
these regions are compact and highly localized, consisting of two
small outer pockets and a narrow central region. The more oscillatory
distribution of \(\omega^S\) further increases the spatial variation of
the normal and tangential load transfer. The minimum relative von Mises
stress reaches approximately \(0.35\), corresponding to about \(65\%\)
local relaxation, while the maximum decreases to \(10.93\). Thus, the
additional curvature reversals produce deeper but more spatially
confined relaxation while distributing the matrix perturbation among
several localized regions.

For the geometries considered, additional inflection points do not
produce a monotonic increase in the maximum matrix stress. Instead, the
complete signed-curvature distribution and the associated
\(\widetilde{\sigma}^S\) and \(\omega^S\) fields control the locations,
intensities, and spatial extents of the amplification and relaxation
regions.

\subsection{Selected studies of interaction effects}
\label{subsec:surface_interaction_studies}

This subsection presents two parametric studies of interacting material surfaces with mixed curvatures. The first study examines the effect of surface separation by varying the distance between adjacent surfaces while keeping their surface properties fixed. The second study examines the effect of surface elasticity by varying the parameter \(\gamma\) while keeping the separation distances fixed. In both studies, the interaction response is quantified using SIFs at the surface tips, while the contours of the corresponding relative von Mises stress illustrate the regions of stress amplification and shielding in the surrounding matrix.

\subsubsection{Surface arrangements and loading conditions}
\label{subsubsec:surface_configurations_loading}

\begin{figure*}[!htbp]
    \centering
    \captionsetup[subfigure]{skip=2pt}

    \begin{subfigure}[t]{0.48\textwidth}
        \centering
        \includegraphics[width=\linewidth]
        {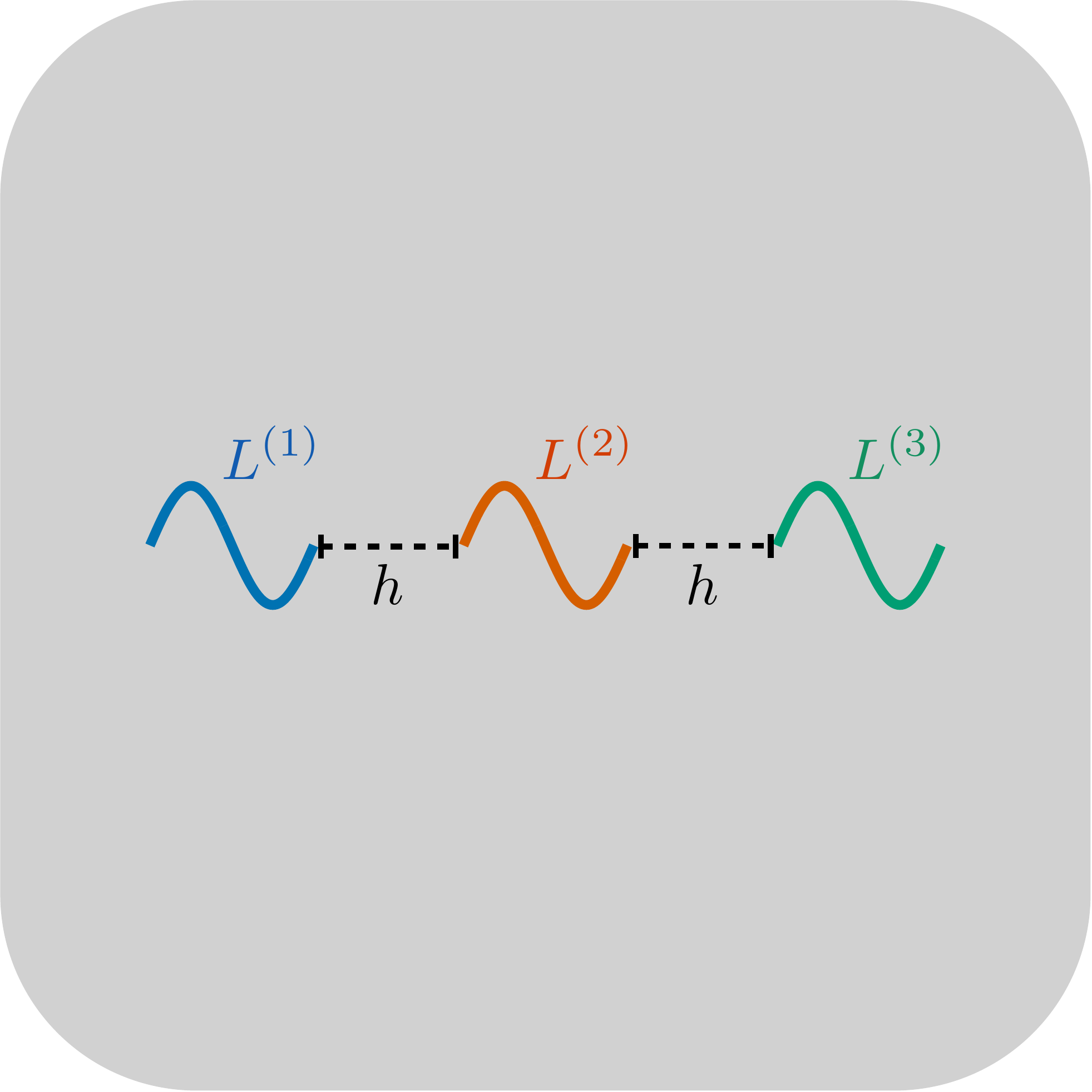}
        \caption{}
        \label{fig:collinear_configuration}
    \end{subfigure}
    \hfill
    \begin{subfigure}[t]{0.48\textwidth}
        \centering
        \includegraphics[width=\linewidth]
        {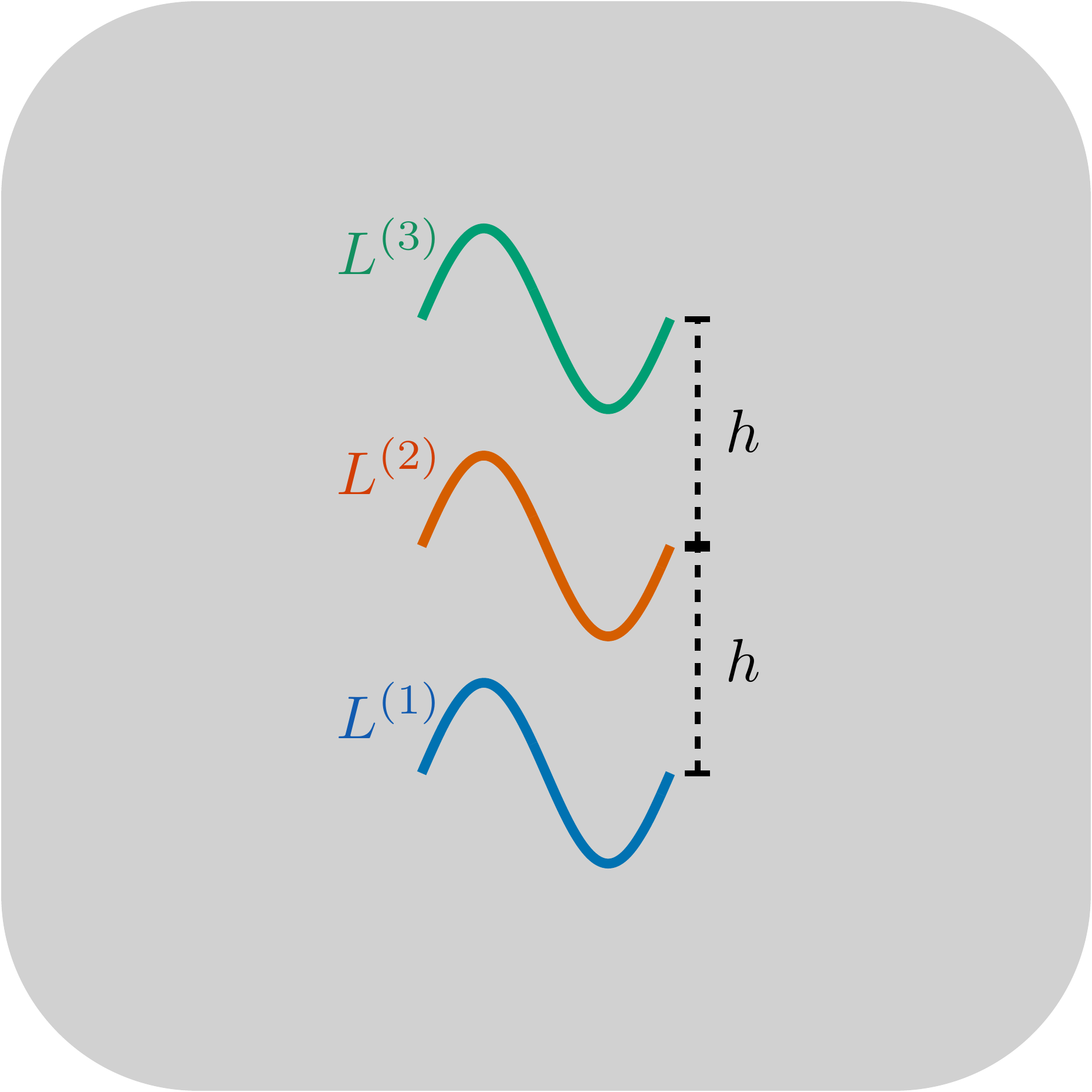}
        \caption{}
        \label{fig:parallel_configuration}
    \end{subfigure}

    \caption{
        Geometries of the three interacting sinusoidal surfaces: (a) collinear arrangement with tip-to-tip separation \(h\);
        and (b) parallel arrangement with vertical separation \(h\).
    }
    \label{fig:multiple_surface_configurations}
\end{figure*}

Three material surfaces of the same length and properties are considered in parallel and collinear arrangements as shown in Fig.~\ref{fig:multiple_surface_configurations}. Each surface is located along a sinusoidal curve and all geometric quantities are normalized by the length of the half-arc \(\ell\). Consequently, each surface has a dimensionless arc length of \(2\), a dimensionless amplitude of \(0.4\), and a dimensionless horizontal span of approximately \(1.082\). The \(m\)-th surface is parameterized as
\begin{equation}
    \begin{aligned}
        \widetilde{x}^{(m)}(t)
        &=
        \widetilde{x}_c^{(m)}
        +
        1.082\left(t-\frac{1}{2}\right),
        \\
        \widetilde{y}^{(m)}(t)
        &=
        \widetilde{y}_c^{(m)}
        +
        0.4\sin(2\pi t),
    \end{aligned}
    \qquad
    0\leq t\leq1,
    \qquad
    m=1,2,3,
    \label{eq:dimensionless_multiple_surface_parameterization}
\end{equation}
where, \(t\) is a dimensionless parametric coordinate that traces each surface from its left tip (\(t=0\)) to its right tip (\(t=1\)), and \(m\) represents the individual surface. The quantities \(\widetilde{x}_c^{(m)}\) and \(\widetilde{y}_c^{(m)}\) denote the dimensionless horizontal and vertical offsets, respectively, used to position the \(m\)-th surface within the matrix.

The dimensionless separation distance  is defined as
\begin{equation}
    \widetilde{h}=\frac{h}{\ell},
    \label{eq:dimensionless_surface_separation}
\end{equation}
where \(h\) denotes the corresponding distance between the tips of adjacent surfaces. For the parallel arrangement, the surfaces are placed one above another. They have the same dimensionless horizontal offset, while their dimensionless vertical offsets are separated by
\(\widetilde{h}\), i.e.,
\begin{equation}
    \widetilde{x}_c^{(1)}
    =
    \widetilde{x}_c^{(2)}
    =
    \widetilde{x}_c^{(3)}
    =
    0,
    \qquad
    \left(
        \widetilde{y}_c^{(1)},
        \widetilde{y}_c^{(2)},
        \widetilde{y}_c^{(3)}
    \right)
    =
    \left(
        -\widetilde{h},0,\widetilde{h}
    \right).
    \label{eq:dimensionless_parallel_surface_offsets}
\end{equation}

For the collinear arrangement, the surfaces are placed end to end. They have the same dimensionless vertical offset, while their
dimensionless horizontal offsets are prescribed as
\begin{equation}
    \left(
        \widetilde{x}_c^{(1)},
        \widetilde{x}_c^{(2)},
        \widetilde{x}_c^{(3)}
    \right)
    =
    \left(
        -(1.082+\widetilde{h}),
        0,
        1.082+\widetilde{h}
    \right),
    \qquad
    \widetilde{y}_c^{(1)}
    =
    \widetilde{y}_c^{(2)}
    =
    \widetilde{y}_c^{(3)}
    =
    0.
    \label{eq:dimensionless_collinear_surface_offsets}
\end{equation}

In both arrangements, the isotropic matrix is made from epoxy with \(\mu=2~\mathrm{GPa}\) and \(\nu=0.35\); the dimensionless parameter $\widetilde{\sigma}_0$ is chosen as \(\widetilde{\sigma}_0=0.025\) and the applied far-field tensile load has the following dimensionless components:
\begin{equation}
    \widetilde{\sigma}_{22}^{\infty}
    =
    0.05,
    \qquad
    \widetilde{\sigma}_{11}^{\infty} = \widetilde{\sigma}_{12}^{\infty}=0.
    \label{eq:remote_loading_surface_interaction}
\end{equation}

To provide a quantitative measure of the interaction effects, the SIFs obtained for the three-surface arrangements are normalized by the corresponding SIFs for a single isolated surface with the same geometry, material properties, surface parameters, and far-field loading. Thus, the interaction parameters are defined as
\begin{equation}
    \mathcal{I}_i^{(m,\eta)}
    =
    \frac{
        \left|K_i^{(m,\eta)}\right|
    }{
        \left|K_{i,\mathrm{single}}^{(\eta)}\right|
    },
    \qquad
    i=1,2,
    \qquad
    m=1,2,3,
    \qquad
    \eta = a,b\,,
    \label{eq:sif_interaction_factors}
\end{equation}
where \(\eta=a\) and \(\eta=b\) identify the tips \(\mathbf{a}^{(m)}\) and \(\mathbf{b}^{(m)}\), respectively, and \(K_{i,\mathrm{single}}^{(\eta)}\) denotes the corresponding SIF for the isolated surface. A value of \(\mathcal{I}_i^{(m,\eta)}>1\) indicates interaction-induced amplification while \(\mathcal{I}_i^{(m,\eta)}<1\) indicates shielding. In the absence of an appreciable interaction, \(\mathcal{I}_i^{(m,\eta)}\) approaches unity.

In the study of the effects of surface elasticity, the SIFs of isolated surfaces are recalculated for each value of \(\gamma\), while  \(\gamma\) is fixed in the study of the effects of surface separation distance.

\subsubsection{Effects of surface separation distance}
\label{subsubsec:effect_surface_separation}

In this study, the dimensionless parameter \(\gamma\) is fixed at \(\gamma=0.12\). The dimensionless parameter  $\widetilde{h}$ is defined in Eq.~\eqref{eq:dimensionless_surface_separation}
and the values of $ \widetilde{h}$ are chosen for both parallel and collinear arrangements as
\begin{equation}
    \widetilde{h}
    \in
    \left\{
        0.25,\,0.5,\,1,\,3,\,10
    \right\}.
    \label{eq:surface_separation_values}
\end{equation}

Additionally, to further highlight the curvature induced effects, the results for the problems with sinusoidal surfaces are compared with those for the corresponding problems with straight surfaces located on the horizontal lines that contain center-lines of the curved surfaces; the latter results  are obtained using the same problem parameters. In the following, we also identify the $m$-th straight surface as \(L^{(m)}\).

\begin{figure*}[!htbp]
    \centering
    \captionsetup[subfigure]{skip=2pt}

    \begin{subfigure}[t]{0.49\textwidth}
        \centering
        \includegraphics[width=\linewidth]
        {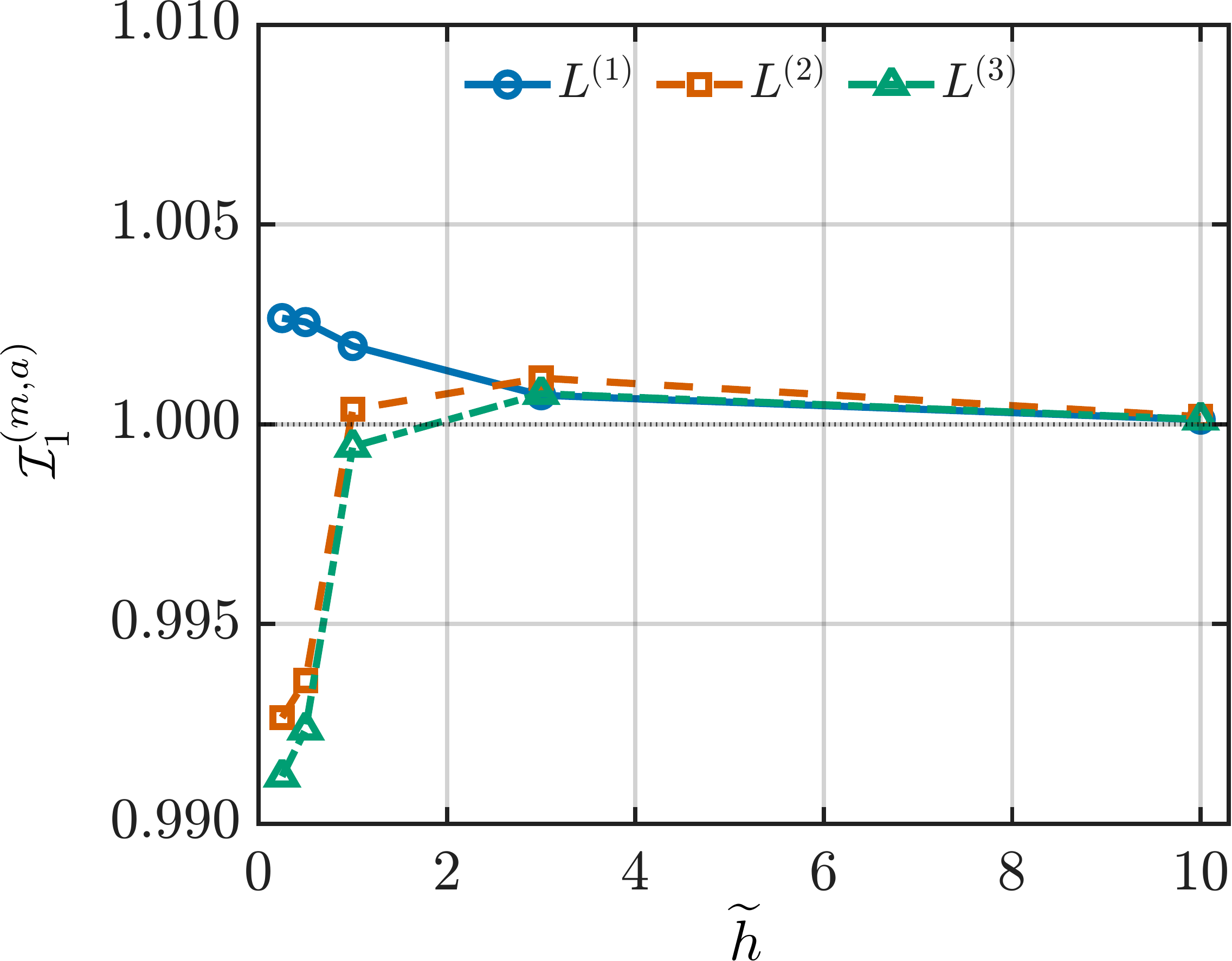}
        \caption{}
        \label{fig:collinear_sinusoidal_I1_a}
    \end{subfigure}
    \hfill
    \begin{subfigure}[t]{0.49\textwidth}
        \centering
        \includegraphics[width=\linewidth]
        {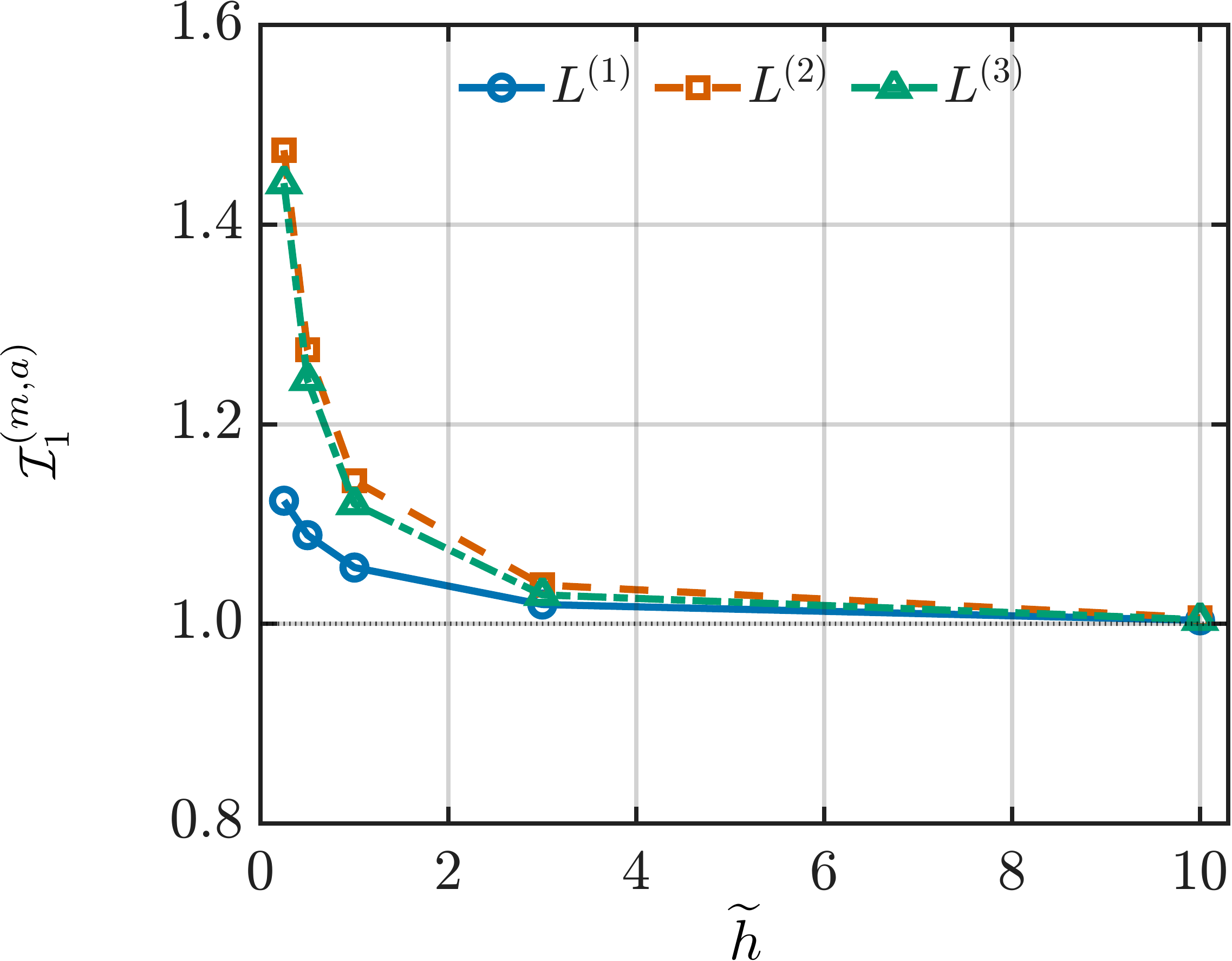}
        \caption{}
        \label{fig:collinear_straight_I1_a}
    \end{subfigure}

    \caption{
        Effect of \(\widetilde{h}\) on
        \(\mathcal{I}_1^{(m,a)}\) for three collinear material surfaces:
        (a) sinusoidal surfaces and (b) straight surfaces.
    }
    \label{fig:collinear_sif_interaction}
\end{figure*}

\begin{figure*}[!htbp]
    \centering
    \captionsetup[subfigure]{skip=2pt}

    \begin{subfigure}[t]{0.49\textwidth}
        \centering
        \includegraphics[width=\linewidth]
        {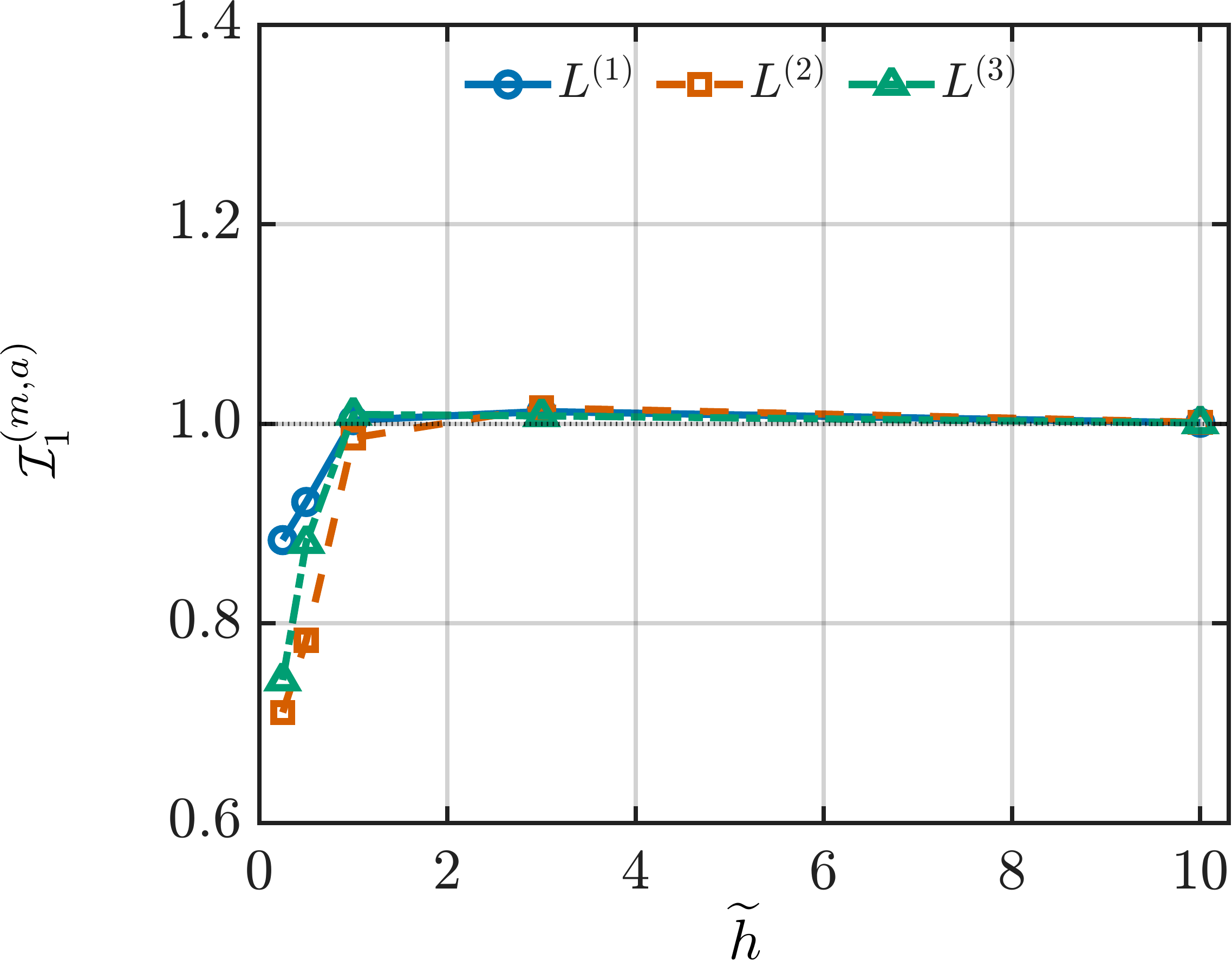}
        \caption{}
        \label{fig:parallel_sinusoidal_I1_a}
    \end{subfigure}
    \hfill
    \begin{subfigure}[t]{0.49\textwidth}
        \centering
        \includegraphics[width=\linewidth]
        {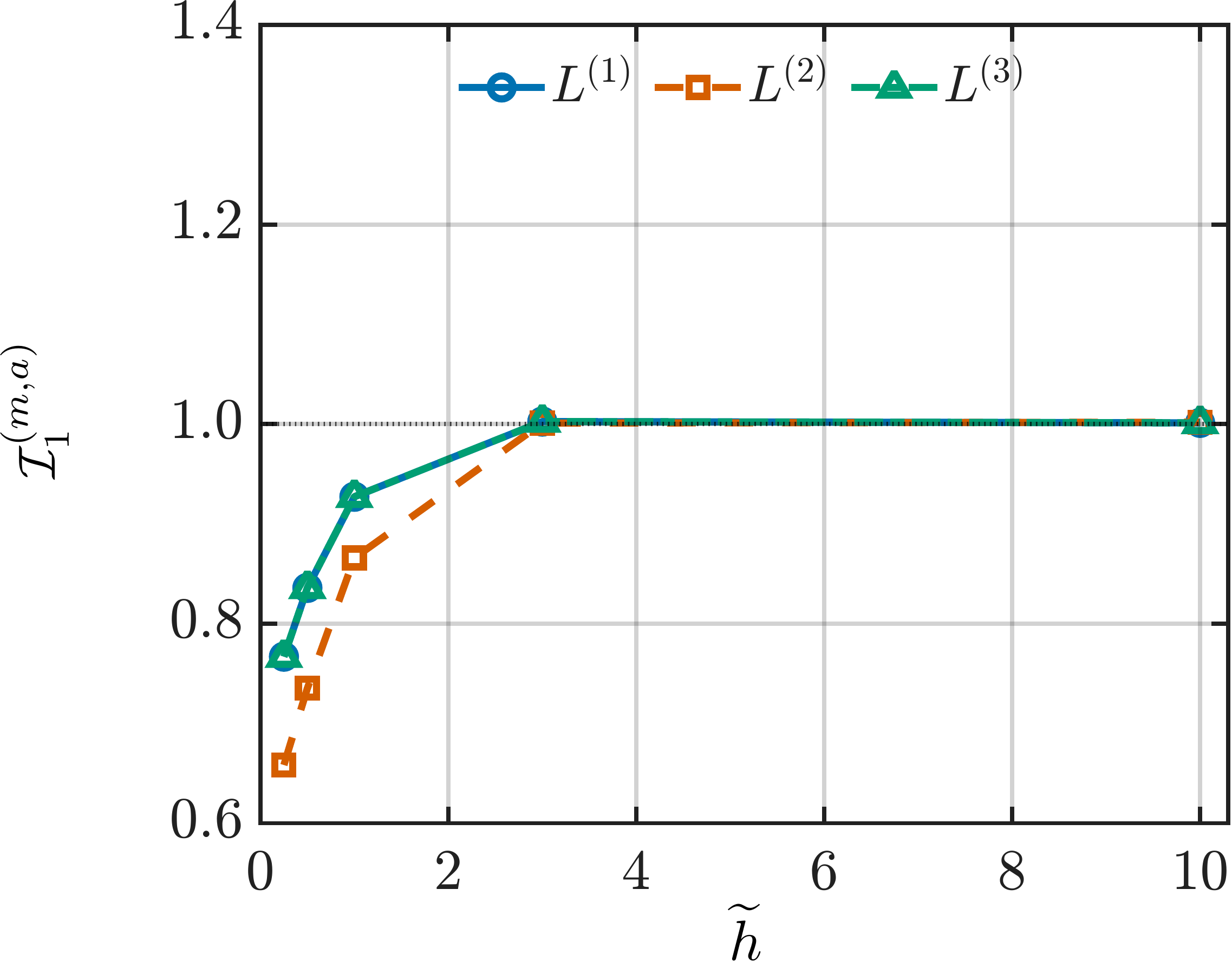}
        \caption{}
        \label{fig:parallel_straight_I1_a}
    \end{subfigure}

    \caption{
        Effect of \(\widetilde{h}\) on
        \(\mathcal{I}_1^{(m,a)}\) for three parallel material surfaces:
        (a) sinusoidal surfaces and (b) straight surfaces.
    }
    \label{fig:parallel_sif_interaction}
\end{figure*}

\begin{figure*}[!htbp]
    \centering
    \captionsetup[subfigure]{skip=2pt}

    \begin{subfigure}[t]{0.49\textwidth}
        \centering
        \includegraphics[width=\linewidth]
        {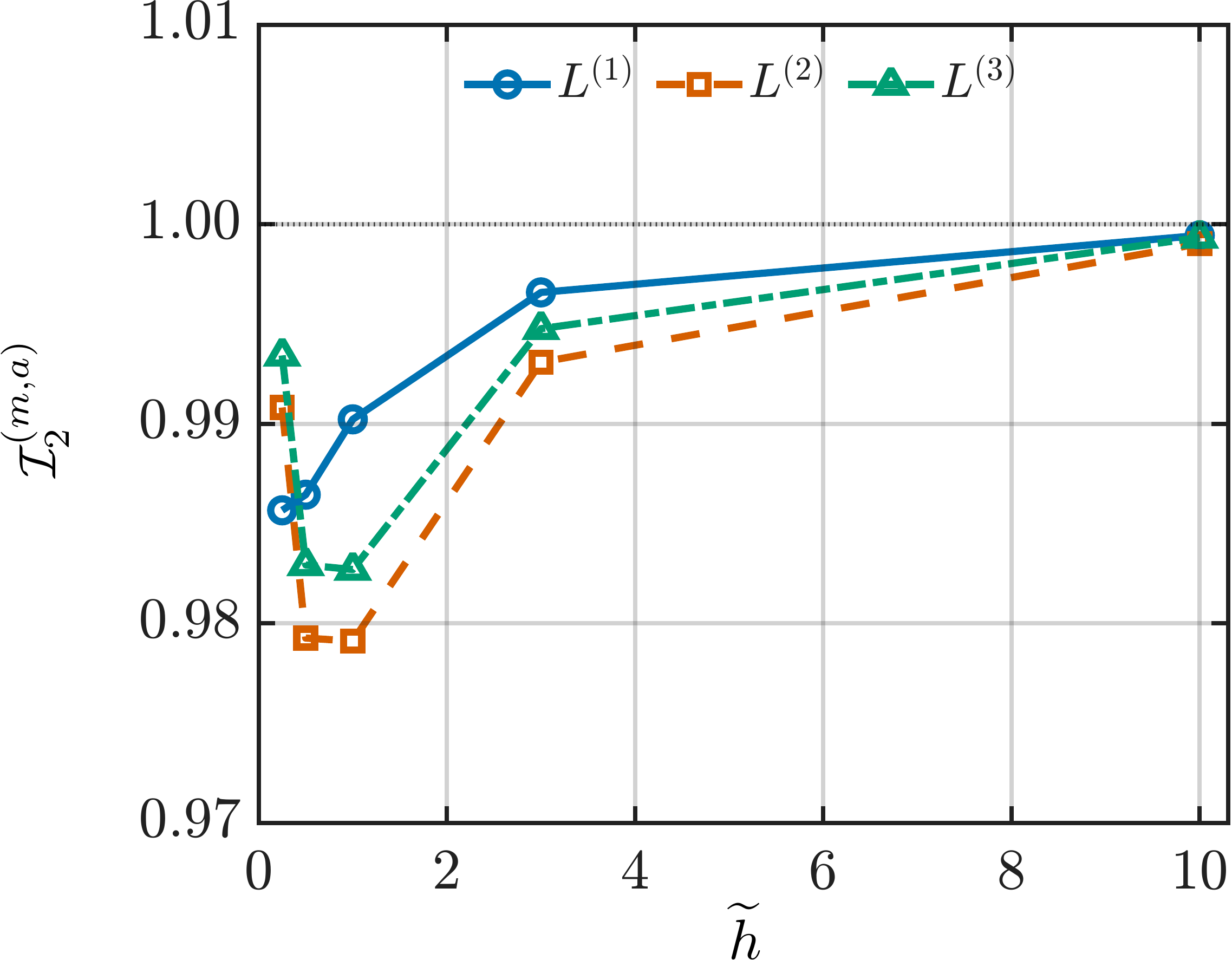}
        \caption{}
        \label{fig:collinear_sinusoidal_I2_a}
    \end{subfigure}
    \hfill
    \begin{subfigure}[t]{0.49\textwidth}
        \centering
        \includegraphics[width=\linewidth]
        {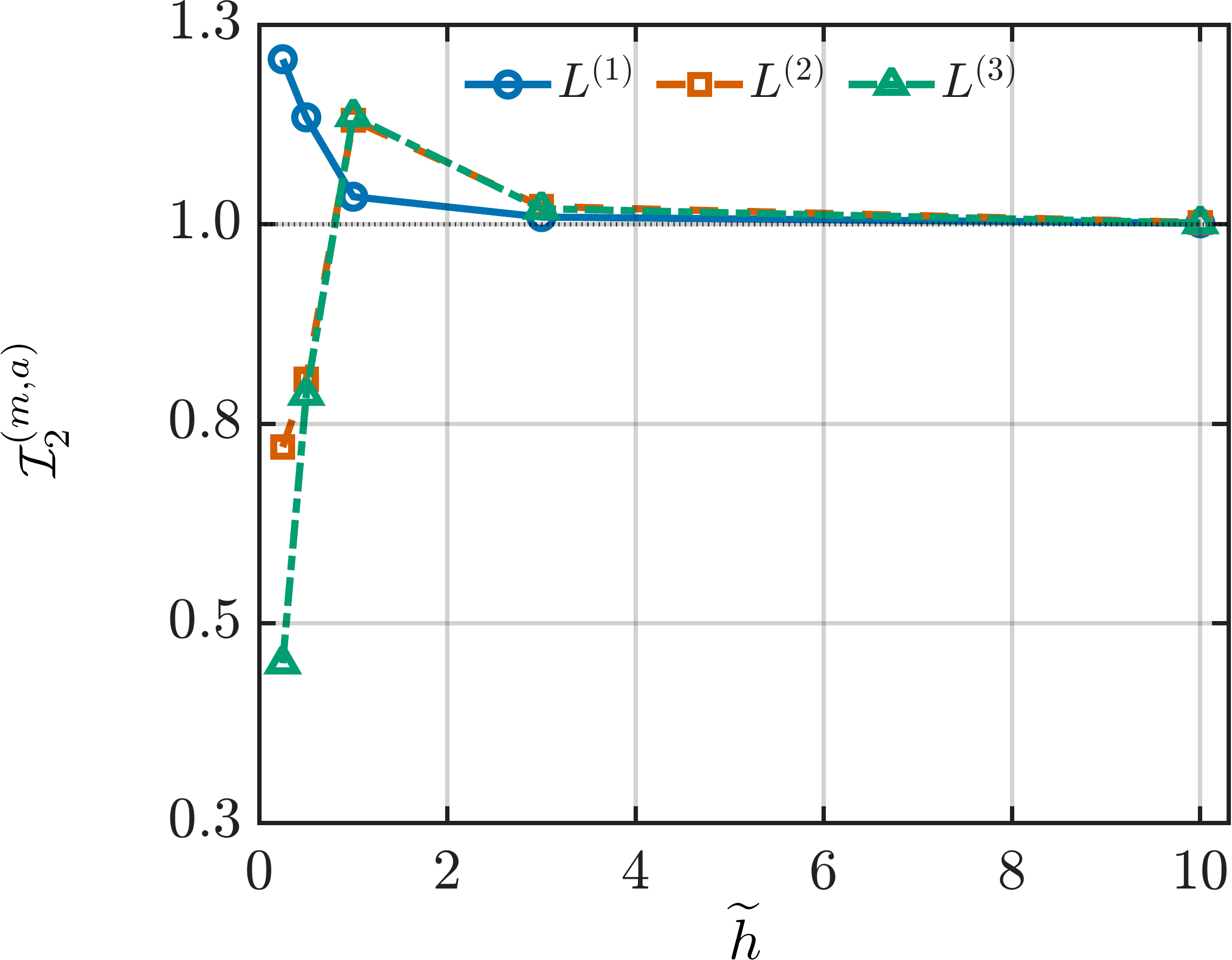}
        \caption{}
        \label{fig:parallel_sinusoidal_I2_a}
    \end{subfigure}

    \caption{
        Effect of 
        \(\widetilde{h}\) on \(\mathcal{I}_2^{(m,a)}\) for three sinusoidal material surfaces
        in the (a) collinear and (b) parallel configurations.
    }
    \label{fig:sinusoidal_I2_interaction_comparison}
\end{figure*}

For both arrangements, the three-surface configuration is invariant under a \(180^\circ\) rotation about its center, regardless of whether the surfaces are straight or sinusoidal. This transformation maps the surface \(L^{(1)}\) onto the surface \(L^{(3)}\), leaves the location of the surface \(L^{(2)}\) unchanged, and interchanges the tips \(\mathbf{a}\) and \(\mathbf{b}\). Because the prescribed far-field loading is also invariant under such a transformation, the following equalities hold:
\begin{equation}
\mathcal{I}_i^{(1,b)}
=
\mathcal{I}_i^{(3,a)},
\qquad
\mathcal{I}_i^{(2,b)}
=
\mathcal{I}_i^{(2,a)},
\qquad
\mathcal{I}_i^{(3,b)}
=
\mathcal{I}_i^{(1,a)},
\qquad i=1,2.
\label{eq:interaction_factor_tip_symmetry}
\end{equation}
Thus, in Figs.~\ref{fig:collinear_sif_interaction}--\ref{fig:sinusoidal_I2_interaction_comparison}, only the values of \(\mathcal{I}_i^{(m,a)}\) are plotted as functions of \(\widetilde{h}\). As these values are obtained as the result of normalization, Eq. (\ref{eq:sif_interaction_factors}), \(\mathcal{I}_i^{(m,a)}>1\) indicates amplification of the corresponding field near the tip compared to that of the single surface, while \(\mathcal{I}_i^{(m,a)}<1\) indicates shielding. For both arrangements \(\mathcal{I}_i^{(m,a)}\) approaches unity as \(\widetilde{h}\) increases, demonstrating the progressive convergence of the field to that of the isolated surface.

For the collinear arrangement, Fig.~\ref{fig:collinear_sif_interaction} shows that the values of \(\mathcal{I}_i^{(m,a)}\) strongly depend on the curvatures of the surfaces. For straight surfaces, the decrease in \(\widetilde{h}\) leads to substantial amplification of \(\mathcal{I}_i^{(m,a)}\), particularly at the tips facing adjacent surfaces.  At \(\widetilde{h}=0.25\), the corresponding values of \(\mathcal{I}_i^{(m,a)}\) for straight surfaces  \(L^{(2)}\)and \(L^{(3)}\) are approximately \(1.47\) and \(1.44\), respectively, while for surface \(L^{(1)}\) it is approximately \(1.12\).

In contrast, the values of \(\mathcal{I}_i^{(m,a)}\)  for the collinear sinusoidal surfaces remain within a roughly \(1\%\) interval of \(\mathcal{I}_i^{(m,a)}=1\) for  the entire range of \(\widetilde{h}\) considered. At small \(\widetilde{h}\), the values of \(\mathcal{I}_i^{(m,a)}\) for the surface \(L^{(1)}\) are slightly above \(\mathcal{I}_i^{(m,a)}=1\), while those for the surfaces \(L^{(2)}\) and \(L^{(3)}\) are slightly below that value.

For the parallel arrangement, Fig.~\ref{fig:parallel_sif_interaction} shows predominantly mode-I shielding at small \(\widetilde{h}\) for both straight and sinusoidal surfaces. When surfaces are closely spaced, their interaction reduces the effective opening action at the tips, resulting in stress shielding in their vicinities. The strongest shielding effects are observed near the tips of the middle surface. At \(\widetilde{h}=0.25\), \(\mathcal{I}_i^{(m,a)}\) ranges from approximately \(0.70\) to \(0.88\) for  sinusoidal surfaces and from \(0.65\) to \(0.76\) for straight surfaces, which constitutes the reduction of roughly \(12\%\)-\(35\%\) in the value of \(\mathcal{I}_1^{(m,a)}=1\). For straight surfaces, stronger shielding effects near the tips are observed, while for sinusoidal surfaces, the reduction in the values of \(\mathcal{I}_i^{(m,a)}\) is smaller. The results for isolated surfaces are recovered roughly at \(\widetilde{h}=1\) for sinusoidal surfaces and at \(\widetilde{h}=3\) for straight surfaces.

The distributions of \(\mathcal{I}_2^{(m,a)}\) for sinusoidal surfaces in both arrangements are presented in Fig.~\ref{fig:sinusoidal_I2_interaction_comparison}. The values of \(\mathcal{I}_2^{(m,a)}\) are not reported for straight surfaces because they vanish for an isolated surface due to symmetry in geometry and loading conditions. Therefore, normalization of Eq. (\ref{eq:sif_interaction_factors}) is simply not possible.

For the collinear arrangement of sinusoidal surfaces, the values of \(\mathcal{I}_2^{(m,a)}\) are slightly below unity, with deviations of only about \(2\%\). In contrast, for the parallel arrangement, strong mode-II interaction effects can be observed at small \(\widetilde{h}\), with \(\mathcal{I}_2^{(m,a)}=  1.25\) for \(L^{(1)}\), whereas those values for surfaces \(L^{(2)}\) and \(L^{(3)}\) are approximately \(0.77\) and \(0.45\), respectively. For both arrangements, the value of \(\mathcal{I}_2^{(m,a)}\) corresponding to an isolated surface is roughly achieved at \(\widetilde{h}=3\).


\begin{figure*}[!htbp]
    \centering
    \includegraphics[width=0.95\columnwidth]
    {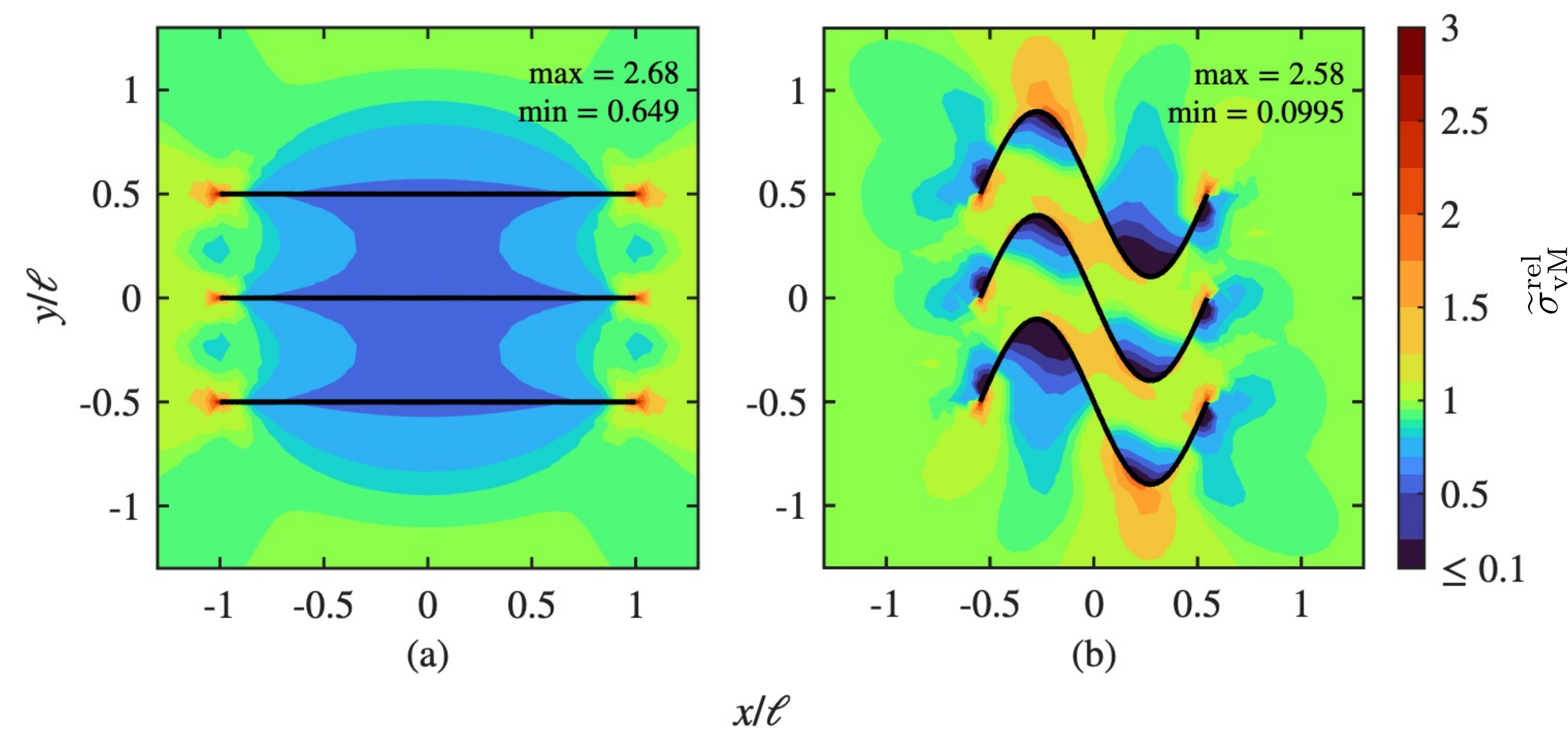}
    \caption{Distributions of the relative von Mises stress in the matrix surrounding three parallel surfaces for \(\widetilde{h}=0.5\) and \(\gamma=0.12\): (a) straight surfaces and (b) sinusoidal surfaces. The same color scale is used for both configurations.}
    \label{fig:parallel_relative_vm_straight_sinusoidal}
\end{figure*}

Figure~\ref{fig:parallel_relative_vm_straight_sinusoidal} compares the relative von Mises stress distributions around three parallel straight and sinusoidal surfaces separated by \(\widetilde{h}=0.5\). In both configurations, the relative von Mises stress approaches unity away from the surfaces, corresponding to the homogeneous matrix response to given loading. For straight surfaces, the stress field remains symmetric and exhibits localized amplification near the surface tips. Broad relaxation regions develop between the closely spaced surfaces. The maximum and minimum relative von Mises stress values are approximately \(2.68\) and \(0.649\), respectively.
For sinusoidal surfaces, the stress distribution becomes more spatially heterogeneous. The changing signed curvature of the surface alters the direction of load transfer along each surface, producing alternating regions of amplification and relaxation. The interaction among these regions generates several localized relaxation pockets around the curved segments. Although the maximum relative von Mises stress decreases slightly to approximately \(2.58\), the minimum decreases substantially to approximately \(0.0995\). Thus, curvature variation induces stronger changes in spatial distribution and depth of stress relaxation than in the maximum stress amplification. 

\subsubsection{Effect of surface elasticity parameter}
\label{subsubsec:effect_surface_elasticity}

To examine the effect of surface elasticity, the parameter \(\gamma\)
is varied for each prescribed \(\widetilde{h}\in
\left\{0.25,\,0.50,\,1.00,\,3.00\right\}\). The values of \(\gamma\) considered are
\begin{equation}
    \gamma
    \in
    \left\{
        0.1,\,1,\,10,\,100
    \right\}.
    \label{eq:surface_elasticity_values}
\end{equation}
Small values of \(\gamma\) identify stiffer surfaces, while the elastic stiffness of a surface is reduced when the value of \(\gamma\) increases. Under the assumption that \(\gamma\) remains the same on all three surfaces, the rotational symmetry of \(180^\circ\) of parallel and collinear arrangements is preserved and the interrelations of Eq.~\eqref{eq:interaction_factor_tip_symmetry} remain valid.  Therefore, only the results for 
\(\mathcal{I}_i^{(m,a)}\), for \(i=1,2\), are presented below. 



\begin{figure*}[!htbp]
    \centering
    \includegraphics[width=0.95\columnwidth]
    {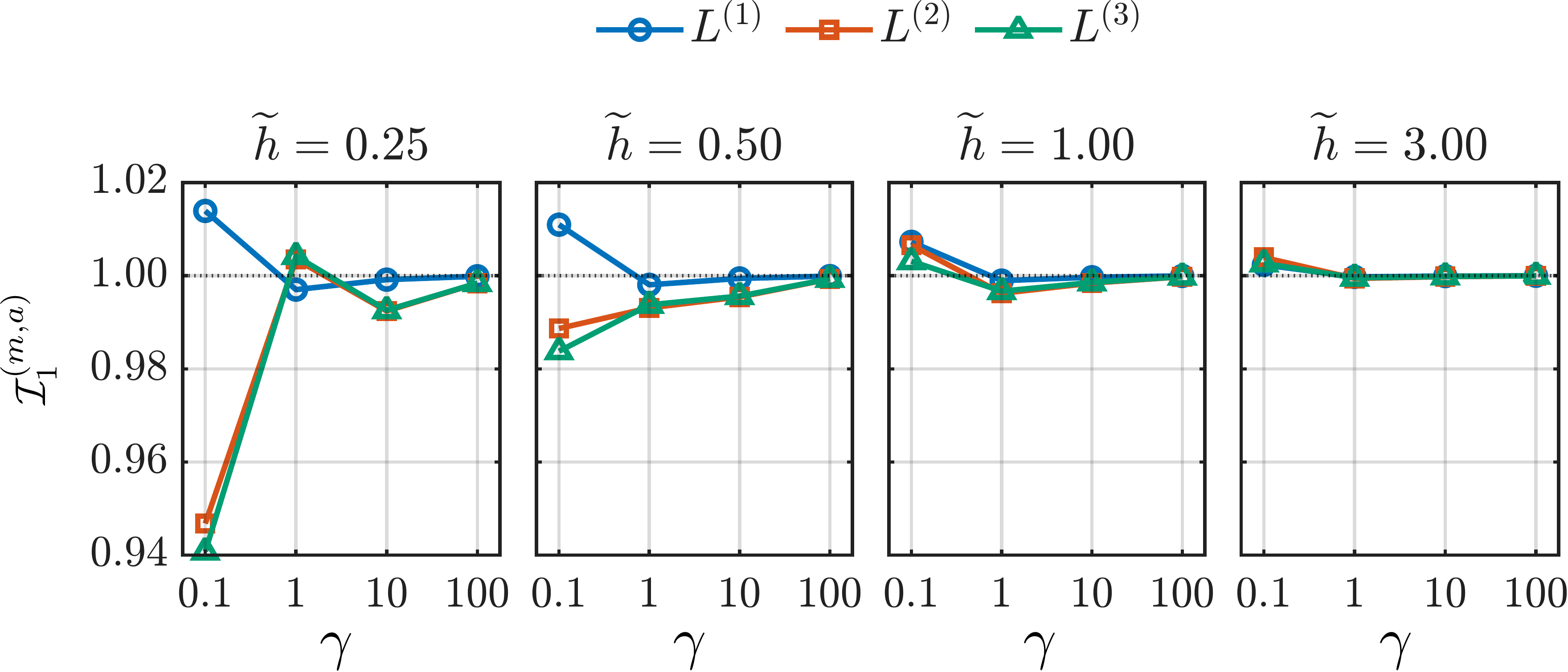}
    \caption{Effect of \(\gamma\) on \(\mathcal{I}_1^{(m,a)}\) for three sinusoidal surfaces in the collinear arrangement for selected values of \(\widetilde{h}\).}
    \label{fig:gamma_collinear_I1}
\end{figure*}

\begin{figure*}[!htbp]
    \centering
    \includegraphics[width=0.95\columnwidth]
    {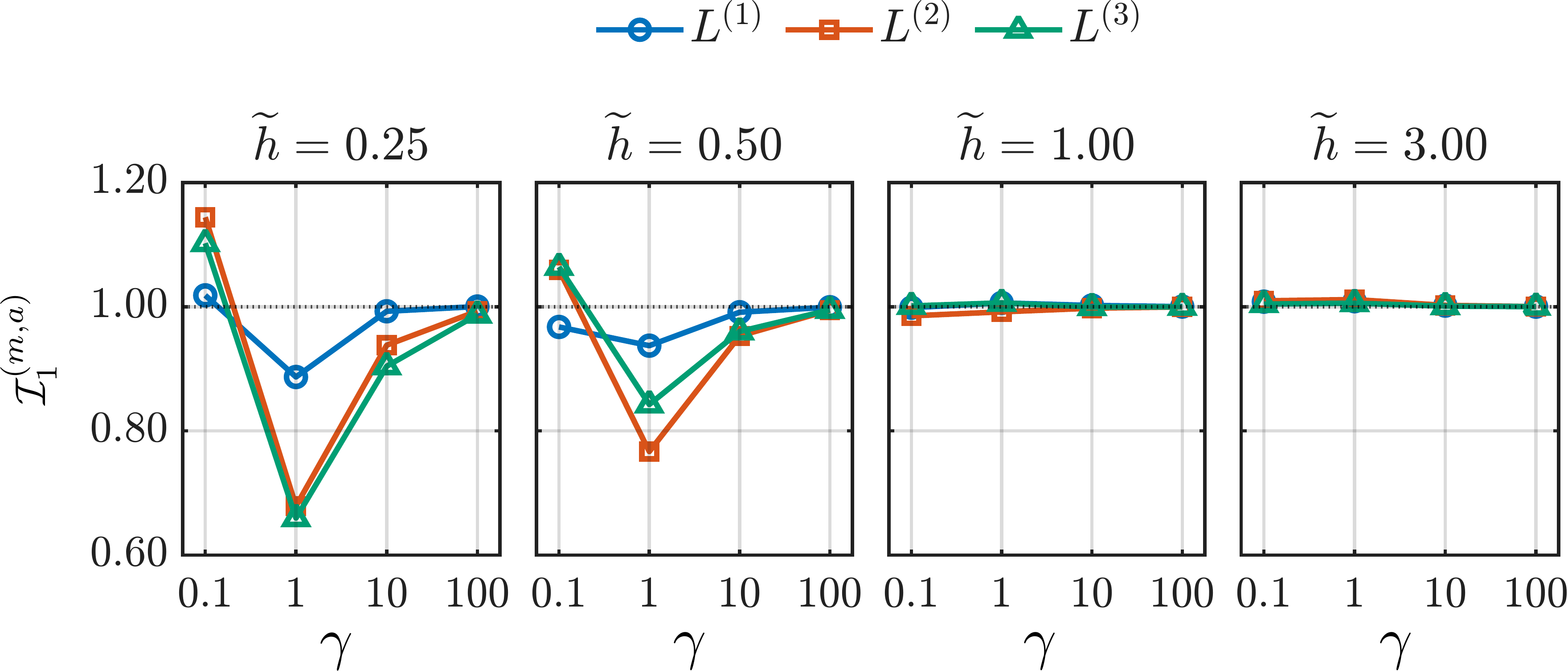}
    \caption{Effect of \(\gamma\) on \(\mathcal{I}_1^{(m,a)}\) for three sinusoidal surfaces in the parallel arrangement for selected values of \(\widetilde{h}\).}
    \label{fig:gamma_parallel_I1}
\end{figure*}

Figures~\ref{fig:gamma_collinear_I1} and \ref{fig:gamma_parallel_I1} illustrate the effects of \(\gamma\) on \(\mathcal{I}_1^{(m,a)}\) for collinear and parallel arrangements, respectively. For the collinear arrangement and \(\widetilde h=0.25\), the largest effects for all three surfaces are observed for \(\gamma=0.1\), where the values of \(\mathcal{I}_1^{(m,a)}\)  at the tips of \(L^{(1)}\) are slightly increased, indicating amplification, while those at the tips of \(L^{(2)}\) and \(L^{(3)}\) are decreased, indicating shielding. However, the values of \(\mathcal{I}_1^{(m,a)}\) at all tips remain within approximately \(6\%\) of \(\mathcal{I}_1^{(m,a)}=1\). The results for parallel arrangement are more sensitive to variations in \(\gamma\). At \(\widetilde h=0.25\), the near tip fields of all three surfaces exhibit amplification at \(\gamma=0.1\), with values of \(\mathcal{I}_1^{(m,a)}\) ranging from approximately \(1.02\) to \(1.14\). At \(\gamma=1\), the response changes to shielding, and the values of \(\mathcal{I}_1^{(m,a)}\)  decrease to approximately \(0.65\)--\(0.88\). As \(\gamma\) increases to \(100\), the interaction effects become insignificant and the values of \(\mathcal{I}_1^{(m,a)}\) approach unity.



\begin{figure*}[!htbp]
    \centering
    \includegraphics[width=0.95\columnwidth]
    {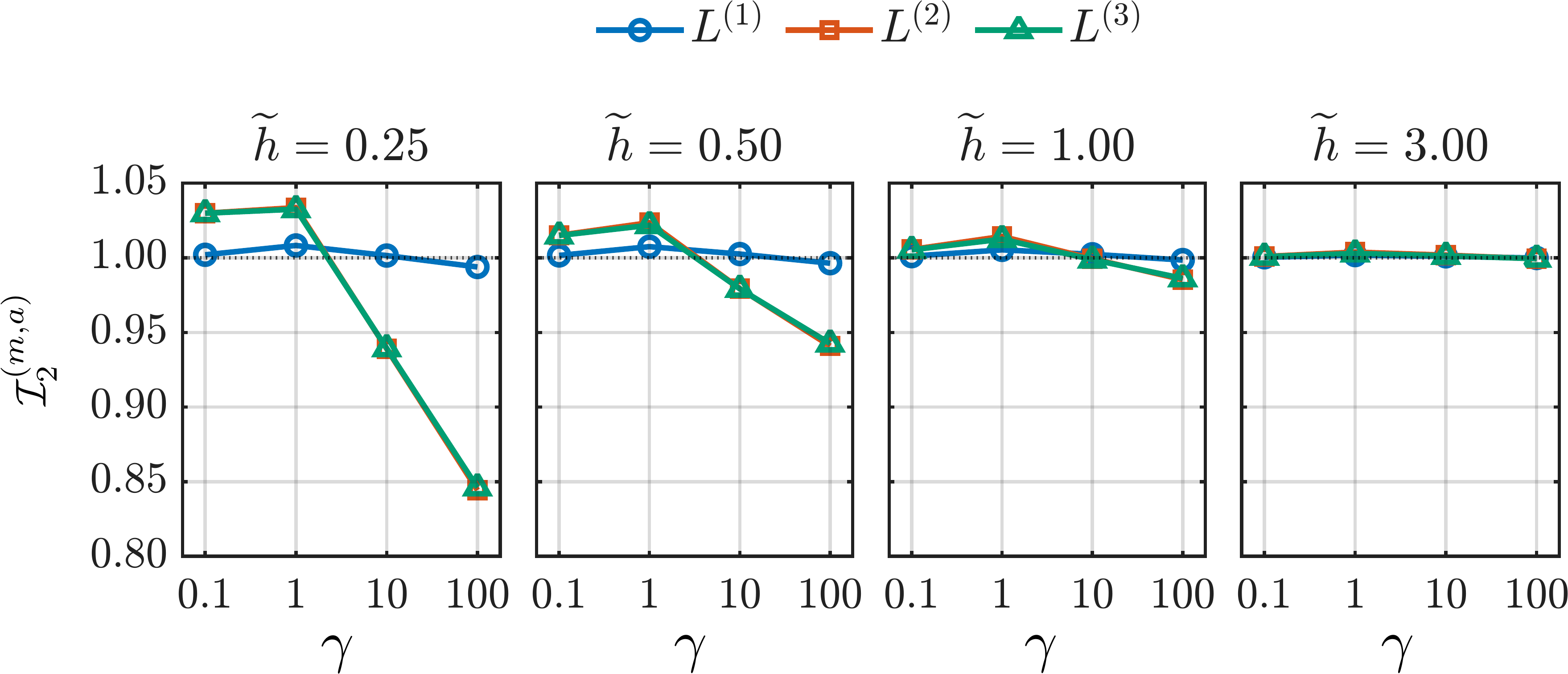}
    \caption{ Effect of \(\gamma\) on \(\mathcal{I}_2^{(m,a)}\) for three sinusoidal surfaces in the collinear arrangement for selected values of \(\widetilde{h}\).}
    \label{fig:gamma_collinear_I2}
\end{figure*}



\begin{figure*}[!htbp]
    \centering
    \includegraphics[width=0.95\columnwidth]
    {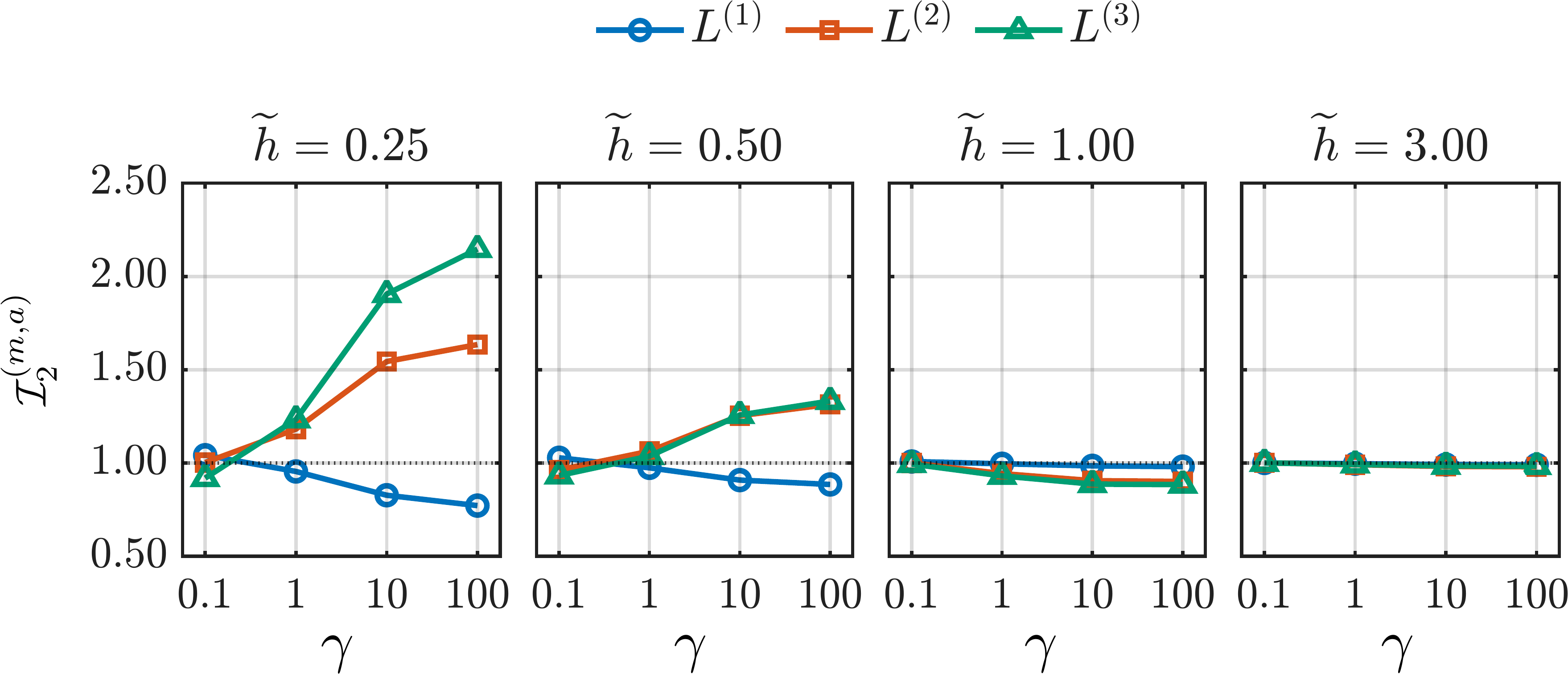}
    \caption{Effect of \(\gamma\) on \(\mathcal{I}_2^{(m,a)}\) for three sinusoidal surfaces in the parallel arrangement for selected values of \(\widetilde{h}\).}
    \label{fig:gamma_parallel_I2}
\end{figure*}

The corresponding distributions of \(\mathcal{I}_2^{(m,a)}\)are shown in Figs.~\ref{fig:gamma_collinear_I2} and \ref{fig:gamma_parallel_I2}. For the collinear arrangement and \(\widetilde h=0.25\), the values of \(\mathcal{I}_2^{(m,a)}\) at the tips of \(L^{(1)}\) remain close to unity, whereas those at the tips of \(L^{(2)}\) and \(L^{(3)}\) change from slight amplification at \(\gamma=0.1\) and \(1\) to shielding at larger \(\gamma\), reaching approximately \(0.85\) at \(\gamma=100\). Stronger effects are observed for the parallel arrangement. At \(\gamma=100\), the value of \(\mathcal{I}_2^{(m,a)}\) at the tip of \(L^{(1)}\) decreases to approximately \(0.75\), while the corresponding values for \(L^{(2)}\) and \(L^{(3)}\) increase to approximately \(1.62\) and \(2.14\), respectively. Thus, the stress fields at the tips of \(L^{(2)}\) and \(L^{(3)}\) are considerably more sensitive to changes in \(\gamma\) than those  for \(L^{(1)}\), particularly in the mode-II response.

For both arrangements, the effects  of \(\gamma\) decrease as \(\widetilde h\) increases, becoming less pronounced at \(\widetilde h=0.50\) and \(1.00\) and negligible at \(\widetilde h=3.00\).

    

    

    

\begin{figure*}[!htbp]
    \centering
    \includegraphics[width=0.95\columnwidth]
    {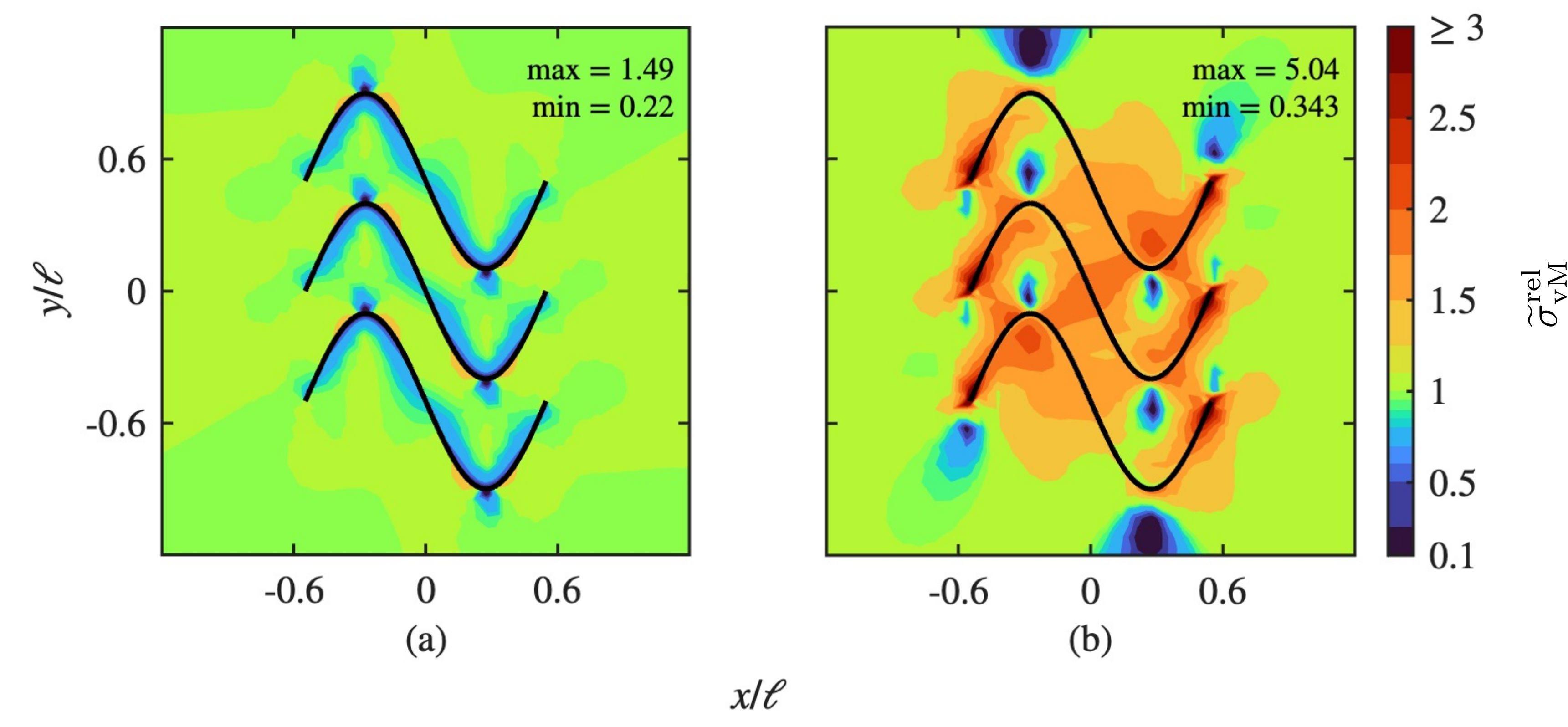}
    
    \caption{ Effect of \(\gamma\) on the relative von Mises stress distribution in the matrix surrounding three parallel sinusoidal surfaces for  \(\widetilde{h}=0.5\): (a) \(\gamma=0.1\) and (b) \(\gamma=10\). The same color scale is used for both cases.
}
\label{fig:parallel_relative_vm_gamma_comparison}
\end{figure*}

Figure~\ref{fig:parallel_relative_vm_gamma_comparison} presents the contour plots of the relative von Mises-stress for \(\gamma=0.1\) and \(\gamma=10\) for \(\widetilde{h}=0.5\). As for \(\gamma=0.1\), the surfaces are stiffer than for \(\gamma=10\), smaller surface strains are sufficient to generate the elastic surface stress required to oppose the residual surface tension and satisfy the tips conditions. The resulting matrix stress variations remain relatively localized, with the maximum and minimum relative von Mises stresses of approximately \(1.49\) and \(0.22\), respectively.

For \(\gamma=10\), the surfaces are less stiff and larger surface strains are required to produce the same change in surface stress, particularly near the tips. The associated variations of the surface stress components affect the tangential and normal traction jumps along the curved surfaces and produce wider stress-amplification regions and localized relaxation zones in the surrounding matrix. Consequently, the matrix stress field becomes more heterogeneous and the maximum relative von Mises stress increases substantially to approximately \(5.04\), while the minimum value is approximately \(0.343\). Thus, the contour plots demonstrate that changes in \(\gamma\) may strongly affect the load-transfer mechanism at a fixed surface separation distance.

\subsection{Example of modeling GNPs-reinforced composite }\label{Sub-Section:7.3}

In order to demonstrate the applicability of our approach to modeling GNPs-reinforced composite materials, we make use of experimental data reported in~\cite{Larsson2024}. There, the study of the overall properties of the epoxy material reinforced with layered graphene nanoplatelets was conducted. To understand the change in stiffness due to the reinforcements, the SEM images were digitized to describe the lengths, orientations, and spatial distribution of GNPs; the data were then used as input for FE-modeling of the overall response of the composite. Piecewise linear approximations were adopted to represent curved GNP morphologies. The epoxy matrix was characterized by \(\mu=983~\mathrm{MPa}\) and \(\nu=0.36\), while the Young's modulus  of the GNPs (with average thickness of \(h=5.57~\mathrm{nm}\) and average lateral dimension of \(L_I=0.88~\mu\mathrm{m}\)) was taken as \(E_I=876~\mathrm{GPa}\). However, some data have not been reported in~\cite{Larsson2024}, including  Poisson's ratio of the GNPs and the densities of the constituent materials. Information on densities is needed to estimate the number of GNPs required to maintain their weight fraction in the material, see Eqs. (1)-(3) of  ~\cite{Larsson2024}. Additionally, surface tension (prestress) was not included in the model reported in~\cite{Larsson2024}, despite the known fact that it exists on such a scale, see \cite{ChanHwang2023, Vassaux2022, Anagnostopoulos2015}. 

\begin{figure}[!hbpt]
    \centering
    \includegraphics[width=0.5\columnwidth]
    {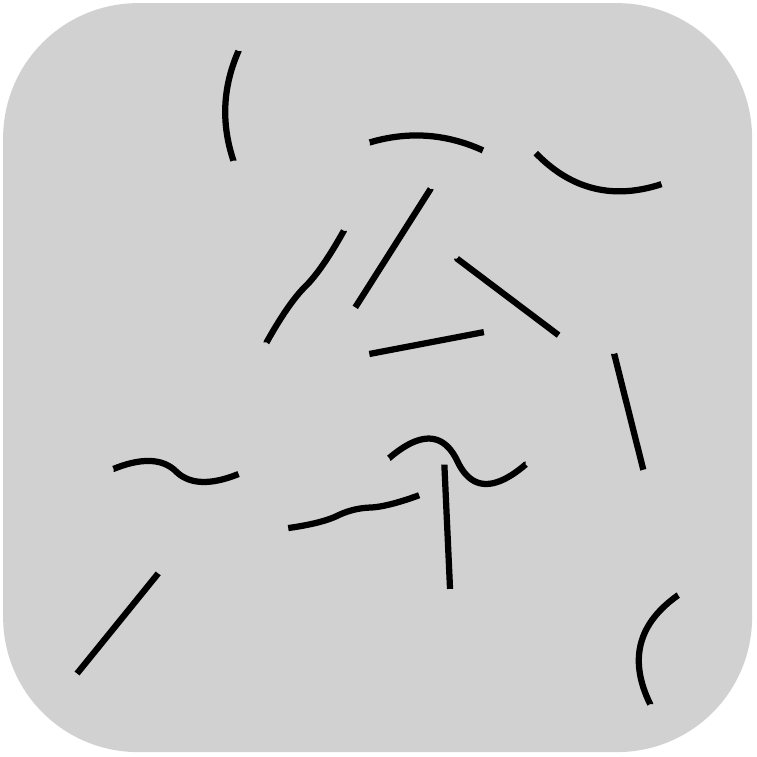}
    \caption{
        Schematic of the epoxy matrix containing the representative \(3~\mathrm{wt.}\%\) GNP morphology. The black curves represent the GNP material surfaces.
    }
    \label{fig:GNP_representative_morphology}
\end{figure}

In our example, we use one of the two digitized microstructure images presented in Fig.~8(c) of ~\cite{Larsson2024} for the epoxy with the GNPs content of \(3~\mathrm{wt.}\%\). We use that image as reference and create the representative morphology  consisting of \(14\) surfaces of different lengths, positions, and curvatures, each representing the GNP, Fig.~\ref{fig:GNP_representative_morphology}. Unlike in ~\cite{Larsson2024}, we treat each GNP as a smooth surface and assume that the surfaces do not intersect. The data missing in \cite{Larsson2024} were taken from \cite{Alvaredo-Atienza2020, Akter2024, Hadden2015} as follows: the value of the Poisson ratio for GNPs as \(\nu_I=0.19\), their densities as \(\rho_I=2.2~\mathrm{g/cm^3}\), and the density of the epoxy matrix as \(\rho=1.15~\mathrm{g/cm^3}\).

Using Eq.~\eqref{eq:surface_moduli_from_layer} and the data mentioned above, we obtained the following surface elasticity parameter for our model: 
\begin{equation}
    2\mu_S+\lambda_S=5062.06~\mathrm{N/m}.
    \label{eq:GNP_effective_surface_modulus}
\end{equation}
Using the aforementioned average lateral dimension \(L_I=0.88~\mu\mathrm{m}\), we obtain the chacteristic length of GNPs as \(\ell=L_I/2=0.44~\mu\mathrm{m}\), and hence
\begin{equation}
    \gamma
    =
    \frac{2\mu\ell}{2\mu_S+\lambda_S}
    \approx 0.171.
    \label{eq:GNP_dimensionless_surface_elasticity}
\end{equation}
As in ~\cite{Larsson2024}, we assume that
\begin{equation}
    \widetilde{\sigma}_0=0
    \label{eq:GNP_residual_surface_tension}
\end{equation}
and apply the unit macroscopic strain
\(\overline{\varepsilon}_{22}=1\),  \(\overline{\varepsilon}_{11}=\overline{\gamma}_{12}=0\). Under plane-strain conditions, the corresponding dimensionless far-field stresses are
\begin{equation}
    \widetilde{\sigma}_{11}^{\infty}=2.5714,
    \qquad
    \widetilde{\sigma}_{22}^{\infty}=4.5714,
    \qquad
    \widetilde{\sigma}_{12}^{\infty}=0.
    \label{eq:GNP_far_field_loading}
\end{equation}

\begin{figure*}[!htbp]
    \centering
    \includegraphics[width=\columnwidth]
    {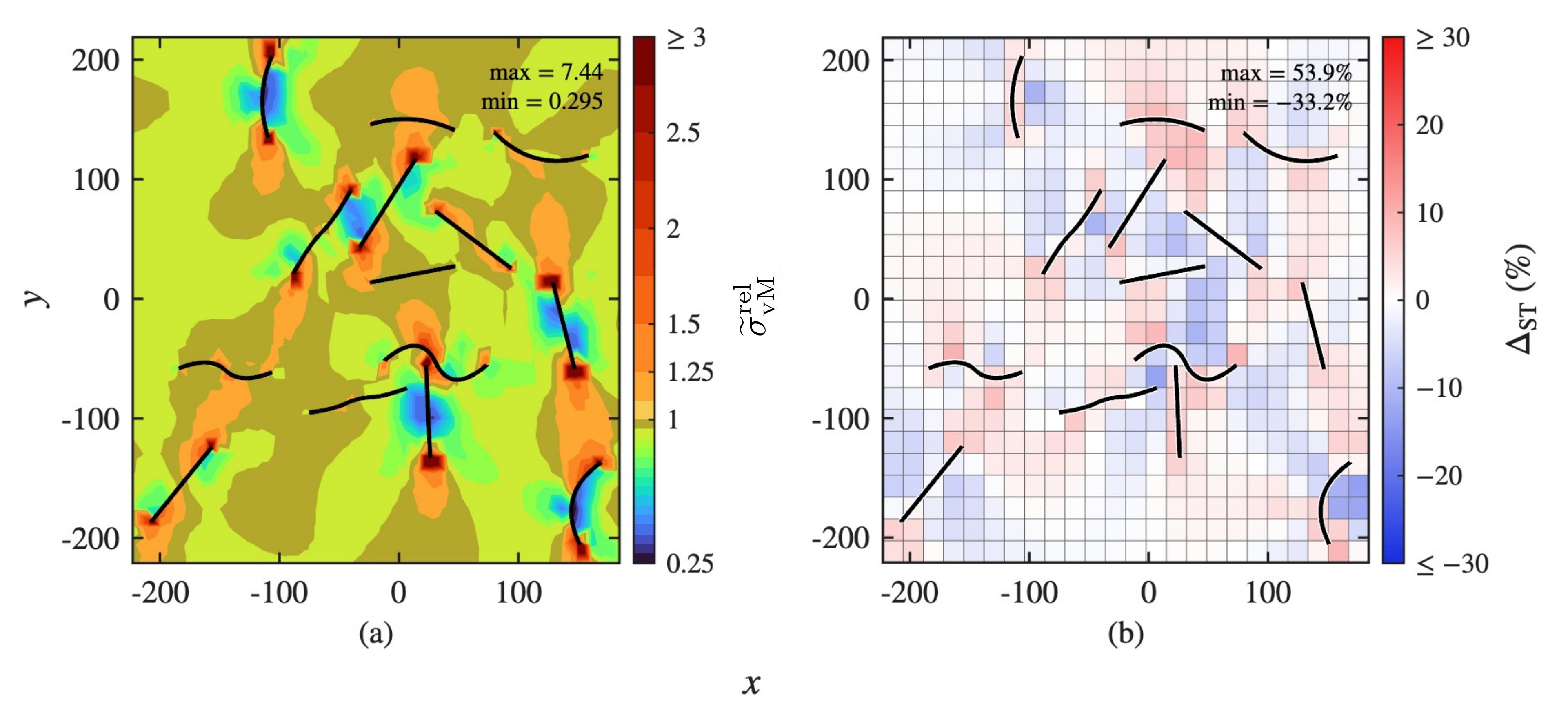}
    
    \caption{ Relative von Mises stress response of the matrix containing the representative \(3~\mathrm{wt.} \%\) GNP morphology: (a) distribution for \(\widetilde{\sigma}_0=0\) and (b) percentage change \(\Delta_{\mathrm{ST}}\) produced by the prescribed residual surface tension relative to case (a). The black curves represent the GNP material surfaces.}
\label{fig:relative_vm_GNP}
\end{figure*}

Figure~\ref{fig:relative_vm_GNP}(a) presents the relative von Mises stress contour plots for the adopted value \(\widetilde{\sigma}_0=0\). The combined effects of surface elasticity, variations in GNPs curvatures, orientations, and interactions between them produce a highly heterogeneous stress field, with the relative von Mises stress ranging from \(0.295\) to \(7.44\). Stress-amplification pockets occur primarily near the GNPs tips and along the surfaces, while adjacent regions exhibit stress relaxation. 

We also studied the effects of surface tension, which was not considered in ~\cite{Larsson2024}. As mentioned above, we used the same macroscopic uniaxial far-field strain of \(\overline{\varepsilon}_{22}=1\) as in ~\cite{Larsson2024}. However, under the latter assumption, the effects of surface tension of realistic magnitude would be very small. Therefore, to study the influence of surface tension under adopted loading, we used the "inflated"  value of \(\widetilde{\sigma}_0=1\). The effects due to the far-field load in such case would remain dominant, however, considerable local effects due to surface tension could be observed. To quantify them, we define the percentage change in the relative von Mises stress as
\begin{equation}
    \Delta_{\mathrm{ST}}(\mathbf{x})
    =
    \frac{
        \left.
        \widetilde{\sigma}_{\mathrm{vM}}^{\mathrm{rel}}(\mathbf{x})
        \right|_{\widetilde{\sigma}_0\neq 0}
        -
        \left.
        \widetilde{\sigma}_{\mathrm{vM}}^{\mathrm{rel}}(\mathbf{x})
        \right|_{\widetilde{\sigma}_0=0}
    }{
        \left.
        \widetilde{\sigma}_{\mathrm{vM}}^{\mathrm{rel}}(\mathbf{x})
        \right|_{\widetilde{\sigma}_0=0}
    }
    \times 100\%.
    \label{eq:surface_tension_percentage_change}
\end{equation}
Positive values of \(\Delta_{\mathrm{ST}}\) indicate that residual surface tension increases the local von Mises stress, whereas negative values indicate a reduction relative to the case without residual surface tension. Figure~\ref{fig:relative_vm_GNP}(b) presents the percentage change in the relative von Mises stress caused by residual surface tension. The value of \(\Delta_{\mathrm{ST}}\) ranges from \(-33.2\%\) to \(53.9\%\), indicating local stress reduction and amplification, respectively. Thus, residual surface tension redistributes the matrix stress, increasing it in some regions while reducing it in others.

\section{Conclusions}\label{Section:8} 

In this paper, a two-dimensional IGABEM framework was developed to model the load-transfer mechanism in composites reinforced by multiple thin, stiff, and prestressed reinforcements, e.g., nanosheets. For a given reinforcement morphology and mechanical properties of constituents, the framework produces accurate estimates of the local elastic fields everywhere in the material system and the SIFs at the reinforcement tips. Modeling the reinforcements as zero-thickness Gurtin--Murdoch material surfaces allows one to capture size-dependent effects without explicitly resolving the thicknesses of the reinforcements. The effects of residual surface tension and their contribution to local equilibrium and stress-transfer mechanisms are also taken into account. 

The framework is uniquely suited for the problems under study as it combines exact boundary-only integral representations for the fields with the Isogeometric Analysis capable of accurate representations of the geometrical features and unknown boundary fields. It enables the representation of multiple curved and/or wrinkled reinforcements whose traces are observed in SEM micrographs. Unlike the classical shear-lag model in which reinforcements are axially loaded through interfacial shear only, the present model accounts for coupled tangential and normal load transfer arising from surface curvature and surface tension.

The framework was validated against available benchmarks, and excellent agreement was obtained for the elastic fields and the SIFs. Investigation of the effects of selected governing parameters leads to the following conclusions. For the geometries studied, it was observed that, for the reinforcements located along the curves with mixed curvatures and non-zero residual tension, tangential components of surface stress tensors are reduced and their normal components are enhanced with increasing numbers of inflection points. With an increase in those numbers, the coupling of normal surface stress components and surface tension increases the influence of the latter. Although the maximum values of the dimensionless von-Mises stresses in the matrix are not changing much with increasing numbers of inflection points, zones of stress concentrations can be observed not just near the tips but also  around the reinforcements. The reinforcement's arrangements and relative orientations further influence the magnitude of SIFs producing either amplification of shielding. Thus, the shapes of the reinforcements and their spatial arrangement have a pronounced effect on a load transfer mechanism. It can also be concluded that for fixed geometrical arrangement, elastic properties of reinforcements significantly affect matrix stress distributions locally by amplifying or relieving these stresses.




This framework can be generalized to model problems involving finite domains and can be further used for Representative Volume Element-based homogenization procedures. The estimates of the effective mechanical properties obtained using the procedures could be compared with those measured experimentally. The framework can also be extended to three-dimensional problems by employing the
appropriate three-dimensional integral representations and NURBS approximations.

\section*{Acknowledgements}\label{Sec:Acknowledgement}
The first and third authors (R.S. Patil and S.G. Mogilevskaya) gratefully acknowledge the support from the U.S. National Science Foundation, award number NSF CMMI -2112894.
The second author (Z. Han) gratefully acknowledges the support provided by the Natural Science Foundation of Shanghai Basic Research Funding 25ZR1402003 and by the Fundamental Research Funds for the Central Universities 25D111907.






\appendix
\section{Singular Kernels and Boundary Integral Equations}
\label{app:singular_formulation}

\subsection{Kelvin Traction Kernel for Plane Strain}
\label{Appendix:Kelvin_traction_kernel}

The traction kernel, \(T_{kj}\), appearing in Eq.~\eqref{eq:SLEP_traction_general} is obtained by differentiating the Kelvin fundamental solution, \(G_{kj}\), in Eq.~\eqref{eq:Kelvin_displacement_kernel}, applying the strain--displacement relations and Hooke's law, and projecting the resulting stress tensor onto a line characterized by the unit normal
\(\mathbf{n}(\mathbf{x})\). Its Cartesian components are given by
\begin{equation}
T_{kj}(\mathbf{x},\boldsymbol{\xi})
=
-\frac{1}{\pi(\kappa+1)r}
\left\{
\frac{\kappa-1}{2}
\left[
r_{,k}n_j(\mathbf{x})
-
r_{,j}n_k(\mathbf{x})
+
\delta_{kj}
\frac{\partial r}{\partial n(\mathbf{x})}
\right]
+
2r_{,k}r_{,j}
\frac{\partial r}{\partial n(\mathbf{x})}
\right\},
\label{eq:Kelvin_traction_kernel}
\end{equation}
were, \(n_j(\mathbf{x})\) denotes the \(j\)-th Cartesian component of
the unit normal vector at the field point \(\mathbf{x}\), and
\(r_{,i}\) is defined in Eq.~\eqref{eq:Kelvin_kernel_notation}. The
quantity \(T_{kj}(\mathbf{x},\boldsymbol{\xi})\) represents the
\(k\)-th component of traction at \(\mathbf{x}\), acting on a line
with unit normal \(\mathbf{n}(\mathbf{x})\), due to a unit concentrated
force applied at \(\boldsymbol{\xi}\) in the \(j\)-th direction.

\subsection{Coupled Singular Boundary Integral Equations}
\label{app:coupled-singular-bies}

The systems of coupled singular boundary integral equations governing $\sigma^{S(q)}$ and $\omega^{S(q)}$ can be written at each point $\mathbf{x}_0\in L^{(q)}$, $q=1,\ldots,N$
as 

\begin{align}
\sigma^{S(q)}(\mathbf{x}_0)
={}&
\sigma_0^{(q)}
+
\left(\lambda_S^{(q)}+2\mu_S^{(q)}\right)
\Bigg\{
\varepsilon_\infty^{S(q)}(\mathbf{x}_0)
\nonumber\\
&+
\fint_{L^{(q)}}
\left[
K_{\sigma\ell}^{(qq)}(\mathbf{x}_0,\boldsymbol{\xi})
\Delta t_\ell^{(q)}(\boldsymbol{\xi})
+
K_{\sigma n}^{(qq)}(\mathbf{x}_0,\boldsymbol{\xi})
\Delta t_n^{(q)}(\boldsymbol{\xi})
\right]\mathrm{d}s_\xi
\nonumber\\
&+
\sum_{\substack{m=1\\m\neq q}}^{N}
\int_{L^{(m)}}
\left[
K_{\sigma\ell}^{(qm)}(\mathbf{x}_0,\boldsymbol{\xi})
\Delta t_\ell^{(m)}(\boldsymbol{\xi})
+
K_{\sigma n}^{(qm)}(\mathbf{x}_0,\boldsymbol{\xi})
\Delta t_n^{(m)}(\boldsymbol{\xi})
\right]\mathrm{d}s_\xi
\Bigg\},
\label{eq:appendix-stress-bie-traction}
\end{align}

\begin{align}
\omega^{S(q)}(\mathbf{x}_0)
={}&
\omega_\infty^{S(q)}(\mathbf{x}_0)
\nonumber\\
&+
\fint_{L^{(q)}}
\left[
K_{\omega\ell}^{(qq)}(\mathbf{x}_0,\boldsymbol{\xi})
\Delta t_\ell^{(q)}(\boldsymbol{\xi})
+
K_{\omega n}^{(qq)}(\mathbf{x}_0,\boldsymbol{\xi})
\Delta t_n^{(q)}(\boldsymbol{\xi})
\right]\mathrm{d}s_\xi
\nonumber\\
&+
\sum_{\substack{m=1\\m\neq q}}^{N}
\int_{L^{(m)}}
\left[
K_{\omega\ell}^{(qm)}(\mathbf{x}_0,\boldsymbol{\xi})
\Delta t_\ell^{(m)}(\boldsymbol{\xi})
+
K_{\omega n}^{(qm)}(\mathbf{x}_0,\boldsymbol{\xi})
\Delta t_n^{(m)}(\boldsymbol{\xi})
\right]\mathrm{d}s_\xi .
\label{eq:appendix-rotation-bie-traction}
\end{align}
In Eqs.~\eqref{eq:appendix-stress-bie-traction} and
\eqref{eq:appendix-rotation-bie-traction}, the first superscript of each kernel identifies the surface $L^{(q)}$ containing the evaluation point $\mathbf{x}_0$, whereas the second identifies the surface over which integration is performed. Thus, $(qq)$ indicates that $\mathbf{x}_0\in L^{(q)}$ and the integration is performed over the same surface $L^{(q)}$, while $(qm)$ indicates that $\mathbf{x}_0\in L^{(q)}$ and the integration is performed over $L^{(m)}$, $m\neq q$. The kernels appearing in
Eqs.~\eqref{eq:appendix-stress-bie-traction} and
\eqref{eq:appendix-rotation-bie-traction} are defined as
\begin{align}
K_{\sigma\ell}^{(qm)}
&=
-S_0^{(q)}K_{1\ell}^{(qm)}
+C_0^{(q)}K_{2\ell}^{(qm)},
&
K_{\sigma n}^{(qm)}
&=
-S_0^{(q)}K_{1n}^{(qm)}
+C_0^{(q)}K_{2n}^{(qm)},
\label{eq:appendix-surface-stress-kernels}
\\
K_{\omega\ell}^{(qm)}
&=
C_0^{(q)}K_{1\ell}^{(qm)}
+S_0^{(q)}K_{2\ell}^{(qm)},
&
K_{\omega n}^{(qm)}
&=
C_0^{(q)}K_{1n}^{(qm)}
+S_0^{(q)}K_{2n}^{(qm)},
\label{eq:appendix-surface-rotation-kernels}
\end{align}
in which, to distinguish the quantities evaluated at
$\mathbf{x}_0\in L^{(q)}$ from those evaluated at the integration point $\boldsymbol{\xi}\in L^{(m)}$, we introduced the following notations:

\begin{equation}
\begin{aligned}
S_0^{(q)}
&=
\sin\beta^{(q)}(\mathbf{x}_0),
&\qquad
C_0^{(q)}
&=
\cos\beta^{(q)}(\mathbf{x}_0),
\\
S_\xi^{(m)}
&=
\sin\beta^{(m)}(\boldsymbol{\xi}),
&\qquad
C_\xi^{(m)}
&=
\cos\beta^{(m)}(\boldsymbol{\xi}),
\end{aligned}
\label{eq:appendix-angular-quantities}
\end{equation}
with $\beta^{(m)}$ being the angle measured from the global
$x_1$-axis to the local unit normal on $L^{(m)}$ and 
\begin{align}
K_{1\ell}^{(qm)}
={}&
S_0^{(q)}s_\xi^{(m)}G_{11,1}
-S_0^{(q)}c_\xi^{(m)}G_{12,1}
-C_0^{(q)}s_\xi^{(m)}G_{11,2}
+C_0^{(q)}c_\xi^{(m)}G_{12,2},
\\
K_{1n}^{(qm)}
={}&
-S_0^{(q)}c_\xi^{(m)}G_{11,1}
-S_0^{(q)}s_\xi^{(m)}G_{12,1}
+C_0^{(q)}c_\xi^{(m)}G_{11,2}
+C_0^{(q)}s_\xi^{(m)}G_{12,2},
\\
K_{2\ell}^{(qm)}
={}&
S_0^{(q)}s_\xi^{(m)}G_{21,1}
-S_0^{(q)}c_\xi^{(m)}G_{22,1}
-C_0^{(q)}s_\xi^{(m)}G_{21,2}
+C_0^{(q)}c_\xi^{(m)}G_{22,2},
\\
K_{2n}^{(qm)}
={}&
-S_0^{(q)}c_\xi^{(m)}G_{21,1}
-S_0^{(q)}s_\xi^{(m)}G_{22,1}
+C_0^{(q)}c_\xi^{(m)}G_{21,2}
+C_0^{(q)}s_\xi^{(m)}G_{22,2},
\label{eq:appendix-displacement-kernels}
\end{align}
with \(
G_{ij,k}
=
\partial G_{ij}(\mathbf{x}_0,\boldsymbol{\xi})/\partial x_{0k}\,.
\)
The terms associated with the prescribed far-field loading are
\begin{align}
\varepsilon_\infty^{S(q)}(\mathbf{x}_0)
&=
-S_0^{(q)}
\frac{\partial u_1^\infty(\mathbf{x}_0)}
     {\partial s^{(q)}}
+C_0^{(q)}
\frac{\partial u_2^\infty(\mathbf{x}_0)}
     {\partial s^{(q)}},
\\
\omega_\infty^{S(q)}(\mathbf{x}_0)
&=
C_0^{(q)}
\frac{\partial u_1^\infty(\mathbf{x}_0)}
     {\partial s^{(q)}}
+S_0^{(q)}
\frac{\partial u_2^\infty(\mathbf{x}_0)}
     {\partial s^{(q)}}.
\label{eq:appendix-far-field-terms}
\end{align}
The local unit tangent and normal vectors at
$\mathbf{x}_0$ are
\begin{equation}
\boldsymbol{\ell}^{(q)}(\mathbf{x}_0)
=
\left(-S_0^{(q)},C_0^{(q)}\right),
\qquad
\mathbf{n}^{(q)}(\mathbf{x}_0)
=
\left(C_0^{(q)},S_0^{(q)}\right),
\label{eq:appendix-local-directions-x0}
\end{equation}
whereas those at $\boldsymbol{\xi}$ are
\begin{equation}
\boldsymbol{\ell}^{(m)}(\boldsymbol{\xi})
=
\left(-S_\xi^{(m)},C_\xi^{(m)}\right),
\qquad
\mathbf{n}^{(m)}(\boldsymbol{\xi})
=
\left(C_\xi^{(m)},S_\xi^{(m)}\right).
\label{eq:appendix-local-directions-xi}
\end{equation}


The kernel derivatives arise from differentiating the single-layer elastic-potential representation along the local tangent to $L^{(q)}$ at $\mathbf{x}_0$. The corresponding tangential differential operator is
\begin{equation}
\frac{\partial}{\partial s^{(q)}}
=
-S_0^{(q)}\frac{\partial}{\partial x_{01}}
+C_0^{(q)}\frac{\partial}{\partial x_{02}}.
\label{eq:appendix-tangential-operator}
\end{equation}

Local traction jumps
$\Delta t_\ell^{(m)}$ and $\Delta t_n^{(m)}$ are related to $\sigma^{S(m)}$, $\omega^{S(m)}$ and their first-order tangential derivatives through Eqs.~\eqref{eq:tangential_traction_jump_GM} and \eqref{eq:normal_traction_jump_GM}, respectively. Substitution of these relations into Eqs.~\eqref{eq:appendix-stress-bie-traction} and
\eqref{eq:appendix-rotation-bie-traction} leads to the system of coupled real-variable boundary integral equations written in terms of unknown $\sigma^{S(m)}$ and $\omega^{S(m)}$, $m=1,\ldots,N$. To obtain the final system of linear algebraic equations, all boundary integral equations are enforced by making the point $\mathbf{x}_0$ coincide with the collocation points $\mathbf{x}_k^{(m)}$, $k=1,\ldots,n^{(m)}$ at each $m$-th surface. As the equations for every surface are enforced for $\sigma^{S(m)}$ and $\omega^{S(m)}$, the final system of linear algebraic equations contains $2N \times n^{(m)}$ equations with $2N \times n^{(m)}$ unknowns.

\bibliographystyle{elsarticle-num}  
\bibliography{IGABEM_Framework_InfiniteDomain_arXiv}

\end{document}